\documentclass[article]{elsarticle}

\journal {Journal of Computational Physics}
\usepackage{amssymb}
\usepackage{amsmath}
\usepackage{subfigure}
\usepackage{array}
\usepackage{overcite}
\usepackage{lastpage}
\usepackage{fancyhdr}
\usepackage{booktabs}

\usepackage{geometry}
\usepackage{natbib}
\usepackage{hyperref}
\usepackage{color}

\usepackage{graphicx}
\usepackage{epsfig}
\usepackage{verbatim}
\usepackage{textcomp}
\usepackage{float}

\usepackage[normalem]{ulem}

\begin{document}

\begin{frontmatter}
\title{A Compressible High-Order Unstructured Spectral Difference
Code for Stratified Convection in Rotating Spherical Shells}
\author[gwu,ncar]{Junfeng Wang}
\ead{junfeng@gwmail.gwu.edu}
\author[gwu]{Chunlei Liang}
\ead{chliang@gwu.edu}
\author[ncar]{\corref{cor}Mark S. Miesch}


\ead{miesch@ucar.edu}
 \cortext[cor]{Corresponding author}

\address[gwu]{Department of Mechanical and Aerospace Engineering,
\\George Washington University, DC 20052}
\address[ncar]{High Altitude Observatory,
\\National Center for Atmospheric Research,
\\Boulder, CO 80301}

\begin{abstract}
We present a novel and powerful Compressible High-ORder Unstructured Spectral-difference (CHORUS) code for simulating thermal convection and related fluid dynamics in the interiors of stars and planets. The computational geometries are treated as rotating spherical shells filled with stratified gas. The hydrodynamic equations are discretized by a robust and efficient high-order Spectral Difference Method (SDM) on unstructured meshes. The computational stencil of the spectral difference method is compact and advantageous for parallel processing. CHORUS demonstrates excellent parallel performance for all test cases reported in this paper, scaling up to 12,000 cores on the Yellowstone High-Performance Computing cluster at NCAR.  The code is verified by defining two benchmark cases for global convection in Jupiter and the Sun.  CHORUS results are compared with results from the ASH code and good agreement is found. The CHORUS code creates new opportunities for simulating such varied phenomena as multi-scale solar convection, core convection, and convection in rapidly-rotating, oblate stars.
\end{abstract}

\end{frontmatter}

\section{Introduction}\label{sec:intro}

Turbulent convection is ubiquitous in stars and planets.  In intermediate-mass stars like the Sun, convection acts with radiation to transport the energy generated by fusion in the core to the surface where it is radiated into space.  In short, convection enables the Sun to shine.  It also redistributes angular momentum, establishing differential rotation (equator spinning about 30\% faster than the polar regions) and meridional circulations (with poleward flow near the surface).  Furthermore, turbulent solar convection and the mean flows it establishes act to amplify and organize magnetic fields, giving rise to patterns of magnetic activity such as the 11-year sunspot cycle.  Other stars similarly exhibit magnetic activity that is highly correlated with the presence of surface convection and differential rotation \citep{linsk85,hall08,saar10}.  Stars are hydromagnetic dynamos, generating vibrant, sometimes cyclic, magnetic activity from the kinetic energy of plasma motions.

Perhaps the biggest challenge in modeling solar and stellar convection is the vast range of spatial and temporal scales involved.  Solar observations reveal a network of convection cells on the surface of the Sun known as granulation \citep{sprui90}.  Each cell has a characteristic size of about 1000 km and a lifetime of 10-15 min.  However, in order to account for the differential rotation and cyclic magnetic activity of the Sun, larger-scale convective motions must also be present, occupying the bulk of the convection zone that extends from the surface down to 0.7 $R_\odot$ where $R_\odot$ is the solar radius \citep{miesc05}.  These so-called ``giant cells'' have characteristic length and time scales of order 100,000 and several weeks respectively.  Below the convective zone lies the convectively stable radiative zone where energy is transported by radiative diffusion.  The interface between the convective and radiative zones is a thin internal boundary layer that poses its own modeling challenges.  Here convection overshoots into the stable interior, exciting internal gravity waves and establishing a layer of strong radial shear in the differential rotation that is known as the {\em solar tachocline} \citep{miesc05}.

Another formidable modeling challenge is the geometry.  High-mass stars ($M \gtrsim 4 M_\odot$, where $M_\odot$ is the solar mass) are inverted Suns, with convectively stable (radiative) envelopes surrounding convectively unstable cores.  These are also expected to possess vigorous dynamo action but much of it is likely hidden from us, occurring deep below the surface where it cannot be observed with present techniques.  Stellar lifetimes are anticorrelated with mass, so all high-mass stars are significantly younger than the Sun.  Furthermore, stars spin down as they age due to torques exerted by magnetized stellar winds \citep{skuma72,charb93,soder01}. Thus, most high-mass stars spin much faster that the Sun, as much as one to two orders of magnitude.  The fastest rotators are significantly oblate.  For example, the star Regulus in the constellation of Leo has an equatorial diameter that is more than 30\% larger than its polar diameter \citep{mcali05}.

Other types of stars and planets pose their own set of challenges.  Low-mass main sequence stars ($M \lesssim M_\odot$) are convective throughout, from their core to their surface, so they require modeling strategies that can gracefully handle the coordinate singularity at the origin ($r=0$) as well as the the small-scale convection established by the steep density stratification and strong radiative cooling in the surface layers. Red giants have deep convective envelopes and dense, rapidly-rotating cores.  Jovian planets have both deep convective dynamos and shallow, electrically neutral atmospheric dynamics that drive strong zonal winds.

All of these systems have the common property that they are highly turbulent.  In other words, the Reynolds number $R_e = U L / \nu$ is very large, where $U$ and $L$ are velocity and length scales and $\nu$ is the kinematic viscosity of the plasma.  For example, in the solar convection zone $R_e > 10^{12}$ \citep{miesc05}.  Furthermore, though all of these systems are strongly stratified in density (compressible), most possess convective motions that are much slower than the sound speed $c_s$.  Thus, the Mach number $M_a << 1$.  Exceptions include surface convection in solar-like stars and low-mass stars and deep convection in the relatively cool red giants, where $M_a$ can approach unity.

Meeting this considerable list of challenges requires a flexible, accurate, robust, and efficient computational algorithm.  It is within this context that we here introduce the Compressible High-Order Unstructured Spectral difference (CHORUS) code.  CHORUS is the first numerical model of global solar, stellar, and planetary convection that uses an unstructured grid.   This valuable feature allows CHORUS to avoid the spherical coordinate singularities at the origin ($r=0$) and at the poles (colatitude $\theta = 0$, $\pi$) that plague codes based on structured grids in spherical coordinates.  Such coordinate singularities compromise computational efficiency as well as accuracy, due to the grid convergence that can place a disproportionate number of points
near the singularities and that can severely limit the time step through the Courant-Freidrichs-Lewy (CFL) condition.  Thus, CHORUS can handle a wide range of global geometries, from the convective envelopes of solar-like stars and red giants to the convective cores of high-mass stars.

The flexibility of the unstructured grid promotes maximal computational efficiency for capturing multi-scale nature of solar and stellar convection.  As noted above, boundary layers play an essential role in the internal dynamics of stars and planets.  In solar-like stars with convective envelopes, much of the convective driving occurs in the surface layers, producing granulation that transitions to giant cells through a hierarchical merging of downflow plumes \citep{miesc05}.  Meanwhile, the tachocline and overshoot region at the base of the convection zone play a crucial role in the solar dynamo.  The unstructured grid of CHORUS will enable us to locally enhance the resolution in these regions in order to capture the essential dynamics.  Similarly, optimal placing of grid points will allow CHORUS to efficiently model other phenomena such as core-envelope coupling in red giants (see above).  Furthermore, the unstructured grid is deformable.  So, it can handle the oblateness of rapidly-rotating stars as well as other steady and time-dependent distortions arising from radial pulsation modes or tidal forcing by stellar or planetary companions.

The spectral difference method (SDM) employed for CHORUS achieves high accuracy for a wide range of spatial scales, which is necessary to capture the highly turbulent nature of stellar and planetary convection.  Furthermore, its low intrinsic numerical dissipation enables it to be run in an inviscid mode (no explicit viscosity), thus maximizing the effective $R_e$ for a given spatial resolution.

One feature of CHORUS that is not optimal for modeling stars is the fully compressible nature of the governing equations.  The dynamics of stellar and planetary interiors typically operates at low Mach number such that acoustic time scales are orders of magnitude smaller than the time scales for convection or other large-scale instabilities.  Therefore, a fully compressible solver such as CHORUS is limited by the CFL constraint imposed by acoustic waves.  Many codes circumvent this problem by adopting anelastic or pseudo-compressible approximations that filter out sound waves.  We choose instead to define idealized problems by scaling up the luminosity to achieve higher Mach numbers while leaving other important dynamical measures such as the Rossby number unchanged.  This allows us to take advantage of the hyperbolic nature of the compressible equations, which is well suited for the SDM method and which promotes excellent scalability on massively parallel computing platforms.   Furthermore, the compressible nature of the equations will enable CHORUS to address problems that are inaccessible or at best challenging with anelastic and pseudo-compressible codes.  These include high-Mach number convection in Red Giants, the coupling of photospheric and deep convection, and the excitation of the radial and non-radial acoustic oscillations (p-modes) that form the basis of helio- and asteroseismology.

Currently CHORUS employs an explicit time-stepping method, which is not optimal for low Mach number flow, especially on non-uniform grids.  However, the SDM is well suited for split-timestepping and implicit time-stepping methods which we intend to implement in the future.

The purpose of this paper is to introduce the CHORUS code, to describe its numerical algorithm, and to verify it by comparing it to the well-established Anelastic Spherical Harmonic (ASH) code.  We begin with a discussion of the mathematical formulation and numerical algorithms of CHORUS in sections 2 and 3 respectively. In section 4 we generalize the anelastic benchmark simulations of \citep{Jones-2011} to provide initial conditions for compressible models like CHORUS.  We then address boundary conditions in section 5 and the conservation of angular momentum in section 6, which can be a challenge for global convection codes.  In section 7 we verify the CHORUS code by comparing its output to analogous ASH simulations for two illustrative benchmark cases, representative of Jovian planets and solar-like stars.  We summarize our results and future plans in section 8.

\section{Mathematical Formulation}
We consider a spherical shell of ideal gas, bounded by an inner spherical surface at $r=R_{i}$ and an outer
surface at $r=R_{o}$ where $r$ is the radius. We assume that the bulk of the mass is concentrated between both surfaces, and the gravity satisfies $\mathbf{g}=-g\mathbf{\hat{r}}=-\frac{GM}{r^{2}}\mathbf{\hat{r}}$ where $G$ is the gravitational constant, $M$ is the interior mass and $\mathbf{\hat{r}}$ is the radial unit vector. Consider a reference frame that is uniformly rotating about the $z$ axis with angular speed $\mathbf{\Omega_{o}} = \Omega_{o}\mathbf{\hat{e}_{z}}$ where $\mathbf{\hat{e}_{z}}$ is the unit vector in $z$ direction. In this rotating frame, the effect of Coriolis force is added to the momentum conservation equations. The Centrifugal forces are negligible as they have much less contribution in comparison with the gravity. However, the CHORUS code can handle oblate spheroids in rapidly-rotating objects and we will consider Centrifugal force effect in the future. The resulting system of hydrodynamic equations is

\begin{equation}
\frac{\partial \rho}{\partial t} = -\nabla \cdot (\rho \mathbf{u}),
\label{eqn:mass_convervation}
\end{equation}
\begin{equation}
\frac{\partial (\rho \mathbf{u})}{\partial{t}} = -\nabla \cdot \rho \mathbf{u} \mathbf{u} - \nabla p
+\nabla \cdot \bold{{\tau}} + \rho \mathbf{g} - 2\rho \mathbf{\Omega}_{0} \times \mathbf{u},
\label{eqn:monen_convervation}
\end{equation}
\begin{equation}
\frac{\partial E}{\partial{t}} = -\nabla \cdot ((E + p)\mathbf{u})
+ \nabla \cdot (\mathbf{u} \cdot \mathbf{{\tau}} - \mathbf{f}) + \rho \mathbf{u} \cdot \mathbf{g},
\label{eqn:energy}
\end{equation}
where $t$, $p$, $T$, $\rho$ and $\mathbf{u}$ are time, pressure, temperature, density and velocity vector respectively. $E$ is the total energy per unit volume and is defined as $E = \frac{p}{\gamma -1} + \frac{1}{2}\rho \mathbf{u}\cdot \mathbf{u}$ where $\gamma$ is the ratio of the specific heats. $\mathbf{\tau}$ is the viscous stress tensor for a Newtonian fluid.  The term $\mathbf{u}\cdot\mathbf{{\tau}}$ in Eq.(\ref{eqn:energy}) represents the viscous heating. The diffusive flux $\mathbf{f}$ is generally treated in the form of $\mathbf{f}=-\kappa \rho T \nabla S - \kappa_{r}\rho C_{p}\nabla T$ where $\kappa$ is the entropy diffusion coefficient, $S$ is the specific entropy, $\kappa_{r}$ is the thermal diffusivity (thermal conductivity and radiative conductivity), and $C_{p}$ is the specific heat at constant pressure. The entropy diffusion is a popular way of parameterizing the energy flux due to unresolved, subgrid-scale convective motions which tend to mix entropy \citep{Glatzmaier_1984,Miesch-2000,Jones-2011}. At the bottom of the convection zone, $\mathbf{f}$ is generally prescribed with a constant value which acts as the energy source of convection. The last term in Eq.(\ref{eqn:energy}) is the work done by buoyancy.\\

The governing equations for fully compressible model can be written in a conservative form as
\begin{equation}
\frac{\partial \mathbf{Q}}{\partial t}+ \frac{\partial \mathbf{F}}{\partial x} + \frac{\partial \mathbf{G}}{\partial y} +
\frac{\partial \mathbf{H}}{\partial z} + \mathbf{M} = 0,
\label{eqn:govn_conserve}
\end{equation}
where $\mathbf{Q}$ is the vector of conserved variables, $\mathbf{M}$ is the combination of the Coriolis force term and the gravitational force term,  and $\mathbf{F}$, $\mathbf{G}$, $\mathbf{H}$ are the total fluxes including both inviscid and viscous flux vectors in local Cartesian coordinates (transforming to an arbitrary geometry will be discussed in next section). We write these as $\mathbf{F}=\mathbf{F}_{inv}-\mathbf{F}_{v}$,
$\mathbf{G}=\mathbf{G}_{inv}-\mathbf{G}_{v}$, and $\mathbf{H}=\mathbf{H}_{inv}-\mathbf{H}_{v}$, where \\
\begin{eqnarray}
\mathbf{Q}=\left\{
\begin{array}{c}
\rho \\ \rho u \\ \rho v\\ \rho w \\ E
\end {array}
\right\}, &
\mathbf{M}=\left\{
\begin{array}{c}
0 \\ \rho g_{x} - 2\rho \Omega_{o}v \\ \rho g_{y}+2\rho \Omega_{o}u  \\ \rho g_{z} \\ \rho (ug_{x} + vg_{y} + wg_{z})
\end {array}
\right\},
\label{eqn:QL}
\end{eqnarray}

\begin{eqnarray}
\mathbf{F}_{inv}=\left\{
\begin{array}{c}
\rho u\\ \rho u^{2}+p\\ \rho uv\\ \rho uw \\u(E+p)
\end {array}
\right\}, &
\mathbf{G}_{inv}=\left\{
\begin{array}{c}
\rho v\\ \rho vu \\ \rho v^{2}+p \\ \rho vw \\v(E+p)
\end {array}
\right\}, &
 \mathbf{H}_{inv}=\left\{
\begin{array}{c}
\rho w\\ \rho wu \\ \rho wv \\ \rho w^{2}+p \\w(E+p)
\end {array}
\right\},
\label{eqn:FGH_inv}
\end{eqnarray}

\begin{equation}
 \mathbf{F}_{v}=\left\{
\begin{array}{c}
0 \\ \tau_{xx} \\ \tau_{yx} \\ \tau_{zx} \\ u\tau_{xx} + v\tau_{yx} + w\tau_{zx} + f_{x}
\end {array}
\right\},
 \mathbf{G}_{v}=\left\{
\begin{array}{c}
0 \\ \tau_{xy} \\ \tau_{yy} \\ \tau_{zy} \\ u\tau_{xy} + v\tau_{yy} + w\tau_{zy} + f_{y}
\end {array}
\right\}, \mbox{and}
\label{eqn:FGH_vis_1}
\end{equation}

\begin{equation}
\mathbf{H}_{v}=\left\{
\begin{array}{c}
0 \\ \tau_{xz} \\ \tau_{yz} \\ \tau_{zz} \\ u\tau_{xz} + v\tau_{yz} + w\tau_{zz} + f_{z}
\end {array}
\right\}.
\label{eqn:FGH_vis_2}
\end{equation}

In Eq.(\ref{eqn:QL})-(\ref{eqn:FGH_vis_2}), $u$, $v$, and $w$ are the velocity components in the $x$, $y$, and $z$ directions respectively.
The viscous stress tensor components in Eq.(\ref{eqn:FGH_vis_1}) and (\ref{eqn:FGH_vis_2}) can be written in the following form
\begin{equation}
    \begin{split}
\tau_{xy}&=\tau_{yx}=\mu(v_{x}+u_{y}), \\
\tau_{yz}&=\tau_{zy}=\mu(w_{y}+v_{z}), \\
\tau_{zx}&=\tau_{xz}=\mu(u_{z}+w_{x}),
    \end{split}
    \quad
    \begin{split}
\tau_{xx}&=2\mu(u_{x}-\frac{u_{x}+v_{y}+w_{z}}{3}),\\
\tau_{yy}&=2\mu(v_{y}-\frac{u_{x}+v_{y}+w_{z}}{3}),\\
\tau_{zz}&=2\mu(w_{z}-\frac{u_{x}+v_{y}+w_{z}}{3}),
    \end{split}
\end{equation}
where $\mu = \rho \nu$ is the dynamic viscosity and $\nu$ is the kinematic viscosity. \\

\section{Numerical Algorithm}\label{sec:algorithm}

The equations of motion (4)-(9) are solved using a Spectral Difference method (SDM).
The SDM is similar to the multi-domain staggered method originally proposed by Kopriva and his colleagues \citep{Kopriva-1996}. Liu et al \citep{Liu-2006} first formulated the SDM for wave equations by extending the multi-domain staggered method to triangular elements. The SDM was then employed by Wang et al for inviscid compressible Euler equations on simplex elements \citep{ZJWang-2007} and viscous compressible flow on unstructured grids \citep{wang-sun06}. The SDM is simpler than the traditional Discontinuous Galerkin (DG) method \citep{Cockburn-1998} since DG deals with the weak form of the equations and involves volume and surface integrals. The weak form is developed by integrating the product of a test function with the compressible Navier-Stokes equations. For the discontinuous Galerkin method, the integration is performed over the spatial coordinates of each finite element of the computational domain. The SDM is similar to the quadrature-free nodal discontinuous Galerkin method \citep{May-2011}. The SDM of CHORUS is designed for unstructured meshes of all hexahedral elements \citep{Liang-2009-3D}. This spectral difference approach employs high-order polynomials within each hexahedral element locally. In particular, we employ the roots of Legendre polynomials plus two end points for locating flux points \citep{huynh-07}. The stability for linear advection of this type of SDM has been proven by Jameson \citep{Jamson-2010}. Overall, high-order SDM is simple to formulate and CHORUS is very suitable for massively parallel processing.

\begin{figure}
\centering
\subfigure [Full spherical shell mesh]{\includegraphics[width=0.49 \textwidth, angle=0]{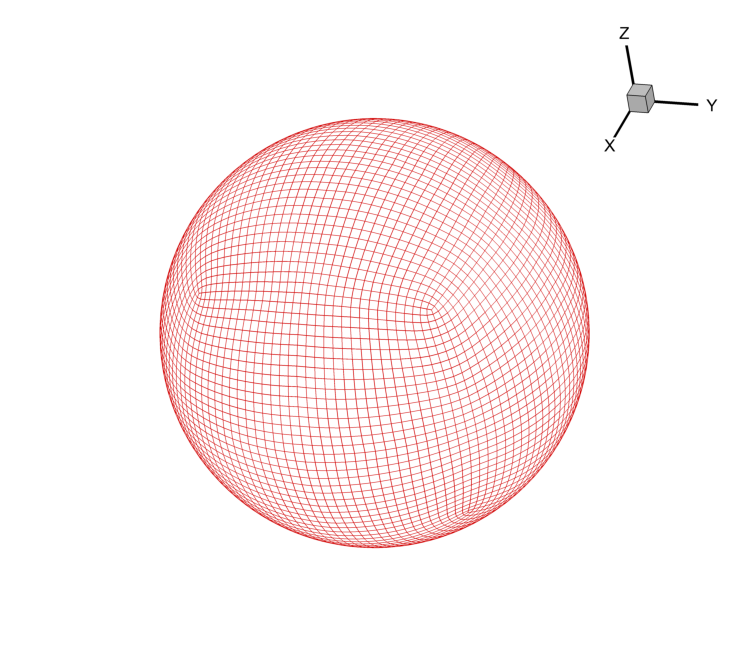}}
\subfigure [Spherical shell segment spanning 1/8 of the volume]{\includegraphics[width=0.49 \textwidth, angle=0]{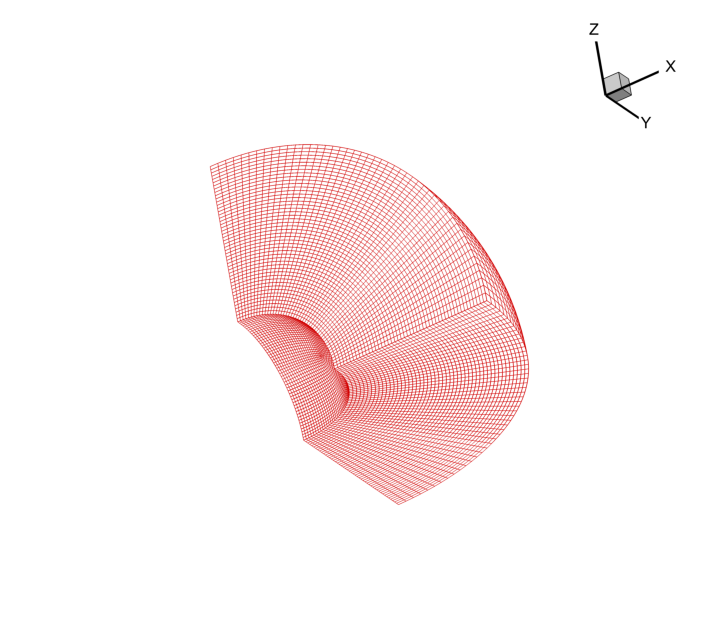}}
\caption{Unstructured mesh consisting of all hexahedral elements. The full spherical shell mesh (a) is generated by fusing eight segments as illustrated in (b) using the GAMBIT software \citep{GAMBIT}. At present, nodal points are uniformly distributed in the radial direction but GAMBIT also allows for variable radial resolution.}
\label{fig:mesh}
\end{figure}

The computational domain is divided into a collection of non-overlapping hexahedral elements as illustrated in Fig.\ref{fig:mesh}. These elements share similarities with the control volumes in the Finite Volume Method.
To achieve an efficient implementation, all hexahedral elements in the physical domain $(x,y,z)$ are transformed to a standard cube $(0\leq \xi \leq 1, 0\leq \eta \leq 1, 0\leq \zeta \leq 1)$. This mapping is achieved through a Jacobian matrix

\begin{equation}
\mathcal{J} = \frac{\partial (x,y,z)}{\partial (\xi, \eta, \zeta)} = \left[
\begin{array}{ccc}
 x_{\xi} & x_{\eta} & x_{\zeta} \\
 y_{\xi} & y_{\eta} & y_{\zeta} \\
 z_{\xi} & z_{\eta} & z_{\zeta}
\end {array}
\right].
\end{equation}

The governing equations in conservative form in the physical
domain as described by Eq.(\ref{eqn:govn_conserve}) are then transformed into the computational domain. The transformed equations are written as
\begin{equation}
\frac{\partial \mathbf{\tilde{Q}}}{\partial t} + \frac{\partial \mathbf{\tilde{F}}}{\partial \xi}
+ \frac{\partial \mathbf{\tilde{G}}}{\partial \eta} + \frac{\partial \mathbf{\tilde{H}}}{\partial \zeta} + \mathbf{\tilde{M}} = 0,
\end{equation}
where $\mathbf{\tilde{Q}} = |\mathcal{J}|\mathbf{Q}$ and $\mathbf{\tilde{M}} = |\mathcal{J}| \mathbf{M}$ using the determinant $|\mathcal{J}|$ of $\mathcal{J}$. The transformed flux components can be written as a combination of the physical flux components as

\begin{equation}
\left(
\begin{array}{c}
\mathbf{\tilde{F}} \\ \mathbf{\tilde{G}} \\ \mathbf{\tilde{H}}
\end {array}
\right) =
|\mathcal{J}|\mathcal{J}^{-1}
\left(
\begin{array}{c}
\mathbf{F} \\ \mathbf{G} \\ \mathbf{H}
\end{array}
\right).
\end{equation}

\begin{figure}[!htb]
\centering
\includegraphics[width=0.6 \textwidth, angle=0]{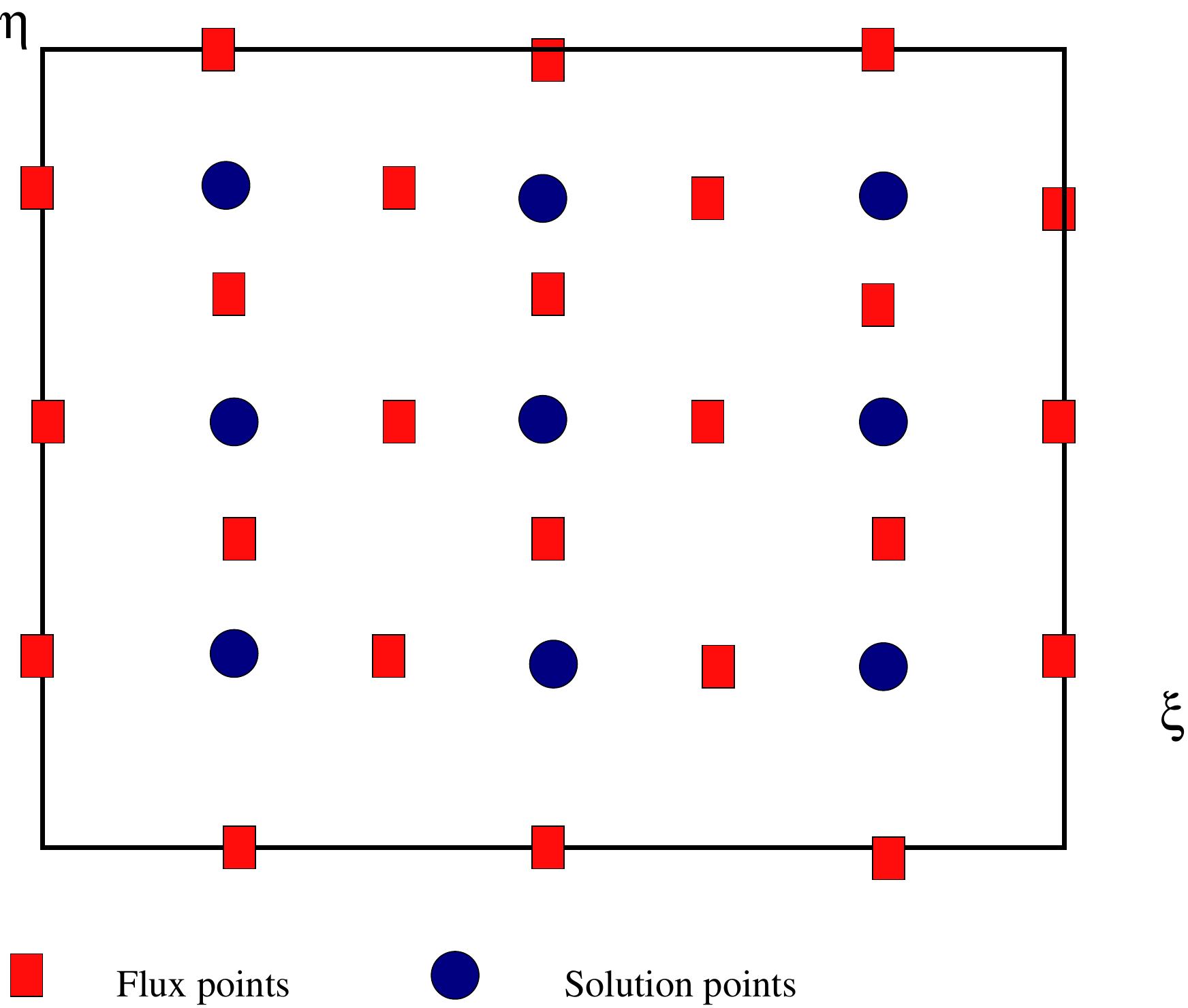}
\caption{Layout of a standard element for a 3rd order SDM in 2D}
\label{fig:thirdorder}.
\end{figure}

In the two-dimensional standard element as illustrated in Fig.\ref{fig:thirdorder}, two sets of points are defined, namely the solution and flux points for the SDM. A total of 9 solution points and 24 flux points are employed for the third-order SDM in 2D. A more detailed description of the SDM for quadrilateral elements can be found in \citep{Liang-2009}. The unknown conserved variables are stored at the solution points, while flux components $\mathbf{F}$, $\mathbf{G}$, and $\mathbf{H}$ are stored at the flux points in corresponding directions.

In order to construct a degree (N-1) polynomial in each coordinate direction, N solution points
are required (thus N is defined as the order the scheme). The solution points in 1D are chosen to be Chebyshev-Gauss-quadrature points defined by
\begin{equation}
X_{s} = \frac{1}{2}\left[1-\cos \left(\frac{2s-1}{2N}\cdot \pi \right)\right], \quad  s=1,2,\cdots,N.
\end{equation}

The flux points are selected to be the Legendre-Gauss-quadrature points plus the two end points, 0 and 1. Choosing $P_{-1}(\xi)=0$ and $P_{0}(\xi)=1$, we can determine the higher-degree Legendre polynomials as
\begin{equation}
P_{n}(\xi)=\frac{2n-1}{n}(2\xi-1)P_{n-1}(\xi)-\frac{n-1}{n}P_{n-2}(\xi), \quad n=1,\cdots,N-1.
\end{equation}

The locations of these Legendre-Gauss quadrature points for the N-th order SDM are the roots of equation $P_{n-1}(\xi)$ plus two end points.

Using the solutions at N solution points, a degree (N-1) polynomial can be built using the following
Lagrange basis defined as
\begin{equation}
h_{i}(X) = \prod_{s=1,s\neq i}^{N}\left(\frac{X-X_{s}}{X_{i}-X_{s}}\right).
\end{equation}

Similarly, using the (N+1) fluxes at the flux points, a degree N polynomial can be built for the
flux using a similar Lagrange basis defined as
\begin{equation}
l_{i+\frac{1}{2}}(X) = \prod_{s=0,i\neq i}^{N}\left(\frac{X-X_{s+\frac{1}{2}}}{X_{i+\frac{1}{2}}-X_{s+\frac{1}{2}}}\right).
\end{equation}

The reconstructed solution for the conserved variables in the standard element is just the tensor products of the three one-dimensional polynomials, i.e.,
\begin{equation}
Q(\xi,\eta,\zeta) = \sum_{k=1}^{N}\sum_{j=1}^{N}\sum_{i=1}^{N}\frac{\tilde Q_{i,j,k}}{|J_{i,j,k}|}h_{i}(\xi)\cdot h_{j}(\eta)\cdot h_{k}(\zeta).
\end{equation}

Similarly, the reconstructed flux polynomials take the following forms:
\begin{equation}
\begin{split}
\mathbf{\tilde F}(\xi,\eta,\zeta) = \sum_{k=1}^{N}\sum_{j=1}^{N}\sum_{i=1}^{N}\mathbf{\tilde F}_{i+\frac{1}{2},j,k}l_{i+\frac{1}{2}}(\xi)\cdot h_{j}(\eta)\cdot h_{k}(\zeta), \\
\mathbf{\tilde G}(\xi,\eta,\zeta) = \sum_{k=1}^{N}\sum_{j=1}^{N}\sum_{i=1}^{N}\mathbf{\tilde G}_{i,j+\frac{1}{2},k}
h_{i}(\xi)\cdot l_{j+\frac{1}{2}}(\eta)\cdot h_{k}(\zeta), \\
\mathbf{\tilde H}(\xi,\eta,\zeta) = \sum_{k=1}^{N}\sum_{j=1}^{N}\sum_{i=1}^{N}\mathbf{\tilde H}_{i,j,k+\frac{1}{2}}
h_{i}(\xi)\cdot h_{j}(\eta)\cdot l_{k+\frac{1}{2}}(\zeta).
\end{split}
\end{equation}

The flux polynomials are element-wise continuous, but discontinuous across element interfaces. For computing the inviscid fluxes, an approximate Riemann solver \citep{Rusanov-1961,Roe-1981} is employed to compute a common flux at interfaces and to ensure conservation and stability. Here the Rusanov flux treatment for the $\xi$ direction  is formulated as
$\mathbf{\tilde F}_{rus} = \frac{1}{2}({\mathbf{\tilde F}_{L}^{inv} + \mathbf{\tilde F}_{R}^{inv}}- sgn(\mathbf{n}\cdot \nabla \xi)(|V_{n}|+c_{s})(Q_{R}-Q_{L})|J\nabla \xi|)$, where $\mathbf{n}$ is the interface normal direction, $V_{n}$ is the fluid velocity normal to the interface and $c_{s}$ is the speed of sound.
If the normal direction of the cell interface is mapped to either the $\eta$ or $\zeta$ direction, the Riemann fluxes can be formulated similarly.

For calculating the viscous fluxes, a simple averaging procedure is used for evaluating fluxes at interfaces \citep{Liang-2009}. This procedure is similar to the BR1 scheme \citep{Bassi1997}. For future implementation of implicit time stepping methods, we can extend the CHORUS code to use BR2 scheme \citep{Bassi05}.

The number of Degrees of Freedom (DOFs) for CHORUS simulations in this paper is computed as
\begin{equation}
DOFs = N_{element} \times N^{3},
\end{equation}
where $N_{element}$ is the total number of elements in the spherical shell and $N$ is the order of the scheme, which is equal to the number of solution points in each direction within one standard element.

\begin{figure}[h!]
\centering
\includegraphics[width=0.7 \textwidth, angle=0]{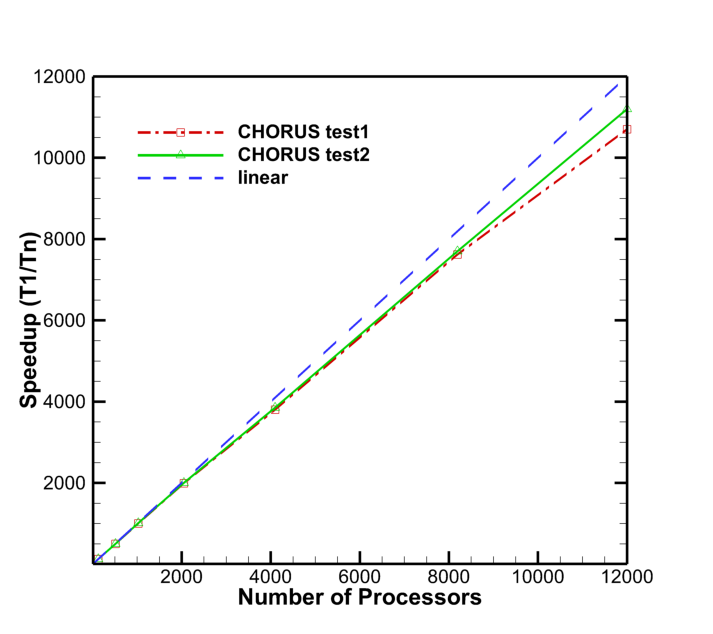}
\caption{Strong scaling of the CHORUS code on Yellowstone for 18.9M (test1) and 70.8M (test2) DOFs.  The latter test in particular achieves over 92\% effeciency for 12k processors.}
\label{fig:scalability}.
\end{figure}

The CHORUS code is written in FORTRAN 90 and efficient parallel performance is achieved by using the Message Passing Interface (MPI) for interprocessor communication.  The ParMetis package \citep{Karypis-1998} is used to partition the unstructured mesh by means of a graph partitioning method.  The parallel scalability of the CHORUS code is shown in Fig.\ref{fig:scalability} using the Yellowstone High-Performance computing cluster at the National Center for Atmospheric Research (NCAR), which is built on IBM's iDataPlex architecture with Intel Sandy Bridge processors.  These numerical experiments demonstrate strong scaling, with $T_{1}$ denoting the execution time of the sequential CHORUS code and $T_{n}$ denoting the execution time of the parallel CHORUS code with $n$ processors. Two sets of simulations using the 4th order SDM are shown. The total numbers of elements in physical domain for test1 and test2 are 294,912 and 1,105,920 respectively, which correspond to 18,874,368 and 70,778,880 DOFs in the 4th-order SDM simulations.

\section{Initial Conditions and Stratification}\label{sec:IC}

Though the equations that CHORUS solves are general, we are intersted in simulating global-scale convection in stellar and planetary interiors for reasons discussed in Section 1.  Thermal convection is a classical fluid instability in the sense that it can develop from a static equilibrium state that satisfies certain instability criteria \citep{chand61}.  The first is the Schwarzschild criterion which requires a negative (superadiabatic) radial entropy gradient $\partial S/\partial r< 0$.  The second is that the buoyancy force must be sufficiently strong to overcome viscous and thermal diffusion.  This is typically quantified in terms of the Rayleigh number, which must exceed a critical value in order for convection to ensue.

Though linear theory is generally concerned with static, equilibrium initial conditions, numerical simulations can tolerate initial conditions that are not in equilibrium.  However, it is still desirable to initiate nonlinear simulations with states that are close to equilibrium in order to mitigate violent initial transients and minimize nonlinear (dynamic) equilibration times.

In the sections that follow we describe how we set up the initial conditions for spherical shells of convection.  CHORUS can also handle convective cores and fully convective geometries but we defer these applications to future papers.  Note that these initial conditions are static relative to the uniform rotation of the coordinate system ($\mathbf{u} = 0$). Thus, the differential rotation, meridional circulation, and convective motions are in no way imposed; they are zero initially. After specifying the static initial conditions, CHORUS automatically introduces random thermal perturbations through the non-axisymmetric distribution of unstructured grid points to excite the convection which in turn establishes the mean flows.  Note also that the stratification in a stellar or planetary convection zone is nearly hydrostatic.  Convection will modify this but only slightly.  So, the initial conditions not only excite the convection but they also establish the basic background stratification including crucial simulation properties such as the density contrast across the convection zone.

\subsection{Static Equilibrium Equations}
To establish the static initial conditions we first consider a steady state in the absence of motions ($\partial /\partial t = 0$, $\mathbf{u} = 0$).  Then the governing equations (\ref{eqn:mass_convervation}) - (\ref{eqn:energy}) reduce to the equation of hydrostatic balance,
\begin{equation}
\frac{dp_{0}}{dr} = -\rho_{0} g,
\label{eqn:hydrostatic}
\end{equation}
and the equation of thermal energy balance,
\begin{equation}
\frac{d}{dr}(r^{2} \kappa \rho_0 T_{0} \frac{dS_{0}}{dr} +  r^{2}\kappa_{r}\rho_0 C_{p} \frac{dT_{0}}{dr}) = 0,
\label{eqn:thermal_equilibrium}
\end{equation}
where the subscript $0$ denotes the initial state. These are supplemented with the ideal gas law
\begin{equation}
p_{0} = {\mathfrak{R} \rho_{0} T_{0}},
\label{eqn:gas_relation}
\end{equation}
where $\mathfrak{R}$ is the specific gas constant, and the equation for specific entropy,
\begin{equation}
S_{0} = C_{p}\ln(\frac{p_{0}^{1/\gamma}}{\rho_{0}}).
\label{eqn:entropy}
\end{equation}

\subsection{Polytropic, Adiabatic Reference State}\label{sec:polytrope}
Though we solve the equations in dimensional form, it is useful to define several nondimensional numbers that characterize the parameter regime of the solution:
\begin{equation}\label{eqn:numbers}
R_{a} = \frac{GMd\Delta S}{\nu \kappa C_{p}},
P_{r} = \frac{\nu}{\kappa},
E_{k} = \frac{\nu}{\Omega_{o}d^{2}},
N_{\rho} = \ln(\frac{\rho_{i}}{\rho_{o}}),
\beta = \frac{R_{i}}{R_{o}} .
\end{equation}
Here $R_{a}$ is the Rayleigh number, $P_{r}$ is the fluid Prandtl number, $E_{k}$ is the
Ekman number, $\exp(N_{\rho})$ is the density ratio across the layer, with $\rho_{i}$ and $\rho_{o}$
as the the densities at the inner and outer boundaries, $d=R_{o}-R_{i}$,
$\beta$ is the aspect ratio, and $\Delta S$ is the entropy difference across the layer,
averaged over latitude and longitude.

The hydrostatic balance equation (\ref{eqn:hydrostatic}), along with the constitutive equations (\ref{eqn:gas_relation}) and (\ref{eqn:entropy}), can be satisfied by introducing a polytropic stratification as described by Jones et al.\ \citep{Jones-2011}:
\begin{equation}
\rho_{a} = \rho_{c}\chi^{n}, \mbox{\hspace{.1in}}
T_{a} = T_{c}\chi,\mbox{\hspace{.1in} and } p_{a} = p_{c}\chi^{n+1},
\end{equation}
where
\begin{equation}
\chi = c+\frac{\alpha d}{r}
\end{equation}
and
\begin{equation}
c=\frac{2\chi_{o}-\beta-1}{1-\beta}, \alpha = \frac{(1+\beta)(1-\chi_{o})}{(1-\beta)^{2}}, \\
\chi_{o}=\frac{\beta+1}{\beta\exp(N_{\rho}/n)+1}, \chi_{i}=\frac{1+\beta-\chi_{o}}{\beta} ~~.
\end{equation}
The subscripts $i$, $o$, and $c$ refer to the bottom, top, and middle of the layer respectively. Once the dimensionless numbers together with other physical input values are determined, the profiles of $\rho_{a}$, $T_{a}$, $p_{a}$, and $S_a$ can be evaluated.

It follows from (\ref{eqn:entropy}) that setting $n=1/(\gamma-1)$ yields an adiabatic stratification ($\partial S_a / \partial r = 0$).  We do so here so that the $a$ subscript denotes a polytropic, adiabatic stratification.  This provides an excellent first approximation to a stellar or planetary convection zone which is very nearly adiabatic due to the high efficiency of the convection.  Though they cannot be strictly adiabatic because they must satisfy the Schwarzschild criterion ($\partial S/\partial r < 0$), they are {\em nearly adiabatic} in the sense that \citep{Gilman_and_Glatzmaier}
\begin{equation}
\epsilon \equiv -\frac{d}{C_{p}}(\frac{d\overline{S}}{dr}) << 1.\label{eqn:epsilon}
\end{equation}
The bar denotes an average over latitude and longitude.
Equation (\ref{eqn:epsilon}) is the basis of the so-called {\em anelastic approximation} in which the equations of motion are derived as perturbations about a static, often (but not necessarily) adiabatic, reference state \citep{gough69,Gilman_and_Glatzmaier,lantz99}.

Although we do not employ the anelastic approximation here, an adiabatic, polytropic reference provides a useful starting point for setting up the initial conditions. However, the process cannot end here because this polytropic solution does not in general satisfy Eq.(\ref{eqn:thermal_equilibrium}) and, since it does not satisfy the Schwarzschild criterion, it will not excite convection.

\subsection{Almost Flux Balance Approach}\label{sec:almost}
In anelastic systems, the superadiabatic component of the entropy gradient ($\partial S/\partial r < 0$) is assumed to be small so it can be specified indepedently of the adiabatic reference state \citep{Jones-2011}.  This amounts to setting $\rho_0 = \rho_a$, $T_0 = T_a$, and $P_0 = P_a$, and then solving equation (\ref{eqn:thermal_equilibrium}) for $dS_0/dr$:
\begin{equation}
\frac{d S_0}{dr} \equiv \Gamma = \frac{1}{\rho_{a}T_{a}\kappa}(\frac{L}{4\pi r^{2}}
+ \rho_{a}C_{p}\kappa_{r}\frac{dT_{a}}{dr}),
\label{eqn:target_entropy}
\end{equation}
where $L$ is the luminosity.

This procedure breaks down in fully compressible systems because equation (\ref{eqn:entropy}) will only be satisfied to lowest order in $\epsilon$.  This is a high price to pay merely to satisfy equation (\ref{eqn:thermal_equilibrium}) which should have little bearing on the final dynamical equilibrium achieved after the onset of convection.  It is more essential to satisfy the constitutive equations (\ref{eqn:gas_relation}) and (\ref{eqn:entropy}) precisely, together with the hydrostatic balance equation (\ref{eqn:hydrostatic}) to avoid a rapid initial restratification.

We achieve this by introducing an extra step in the initialization procedure.  As in anelastic systems, we compute the polytropic, adiabatic stratification as described in section \ref{sec:polytrope} and we calculate $\Gamma$ as defined in Eq.(\ref{eqn:target_entropy}).  However, unlike anelastic systems, we treat $\Gamma$ as a target entropy gradient and then solve equations (\ref{eqn:hydrostatic}), (\ref{eqn:gas_relation}), and (\ref{eqn:entropy}) precisely using a separate finite difference code. In particular, we solve the following two equations for $\rho_0$ and $P_0$ using $\rho_a$ and $P_a$ as an initial guess
\begin{equation}
\frac{\Gamma}{C_{p}} = -(\frac{1}{\rho_{0}}\frac{d\rho_{0}}{dr} + \frac{g\rho_{0}}{\gamma p_{0}}), \quad and \quad
\frac{dp_{0}}{dr} = -\rho_{0} g.
\label{eqn:state_equations}
\end{equation}

The temperature $T_0$ is then given by (\ref{eqn:gas_relation}).  This process produces a superadiabatic, hydrostatic, spherically-symmetric initial state that satisfies equations (\ref{eqn:hydrostatic}), (\ref{eqn:gas_relation}), and (\ref{eqn:entropy}).  The entropy gradient will be equal to $\Gamma$ but the thermal energy balance equation (\ref{eqn:thermal_equilibrium}) will only be satisfied to lowest order in $\epsilon$.

An alternative approach would be to solve all four equations (\ref{eqn:hydrostatic})-(\ref{eqn:entropy}) simultaneously for the four unknowns $P_0$, $\rho_0$, $T_0$, and $S_0$.  Our \emph{Almost Flux Balance} approach is much easier to implement and provides an effective way to initiate convection.

\section{Boundary Conditions}\label{sec:BCs}
The inner and outer boundaries are assumed to be impenetrable and free of viscous stresses
\begin{equation}
V_{r} = \frac{\partial}{\partial r}(\frac{V_{\theta}}{r})
    = \frac{\partial}{\partial r}(\frac{V_{\phi}}{r}) = 0,
\end{equation}
where $V_{r}, V_{\theta}, V_{\phi}$ are the velocity components in spherical coordinates $(r,\theta,\phi)$.

In addition, a constant heat flux $\mathbf{f \cdot \hat{r}} = L/(4\pi R_i^2$) is imposed at the bottom boundary and the temperature is fixed at the top boundary.

The present CHORUS code employs 20 nodes including 8 corner points and 12 mid-edge points for each hexahedral element in the physical domain.  A precise treatment of the curved top and bottom boundaries of the spherical shells is then assured by the iso-parametric mapping procedure mentioned in Section 3. For calculating inviscid fluxes on element interfaces, an approximate Riemann solver \citep{Rusanov-1961} is generally used. However, exact inviscid fluxes are employed on top and bottom boundaries by using the fact that $V_r$ is precisely zero on spherical shell boundaries. A careful transformation between Cartesian and Spherical coordinate systems is conducted for computing viscous fluxes on two boundaries in order to ensure the stress-free conditions.

\section{Angular Momentum Conservation}

The equations of motion (1)-(3) express the conservation of mass, energy, and linear momentum.  The conservation of angular momentum follows from these equations and the impenetrable, stress-free boundary conditions discussed in section \ref{sec:BCs}.  These are hyperbolic equations and we express them in conservative form when implementing the numerical algorithm, as disscussed in section \ref{sec:algorithm}.  This, together with the spectral accuracy within the elements and the approximate Riemann solver employed at cell edges, ensures that the mass, energy, and linear momentum are well conserved as the simulation proceeds.  However, we do not explicitly solve a conservation equation for angular momentum.  Numerical errors including both truncation and round-off errors can result in a small change in angular momentum over each time step. Though these changes may be small, even a highly accurate algorithm can accummulate errors over thousands or millions of time steps that can compromise the validity of a simulation.  This can be an issue in particular for unstructured grid codes like ours that solve the conservation equations in Cartesian geometries that are mapped to conform to the spherical boundaries.  However, conservation of angular momentum can be violated even in highly accurate pseudo-spectral simulations, as reported by Jones et al.\ \citep{Jones-2011}.  For this reason, we introduce an angular momentum correction scheme similar to one of the schemes described by Jones et al \citep{Jones-2011}.

Two correction steps are taken in the CHORUS code to maintain constant angular momentum over long simulation intervals including:

\begin{enumerate}[Step 1.]
\item Calculate three Cartesian components of angular momentum explicitly, namely $(L_{x},L_{y},L_{z})$.
\item Add a commensurate rigid body with rotating rate $(\delta \Omega_{x}, \delta \Omega_{y}, \delta \Omega_{z})$ to remove the angular momentum discrepancy.
\end{enumerate}

In Step 1, three Cartesian components are evaluated over the full spherical shell $V$ as
\begin{equation}
\begin{split}
L_{x} = \int_{V}\rho r(w\sin \theta \sin \phi - v\cos \theta)dV , \\
L_{y} = \int_{V}\rho r(u\cos \theta - w\sin \theta \cos \phi)dV , \\
L_{z} = \int_{V} \rho r(v\sin \theta \cos \phi - u\sin \theta \sin \phi)dV .
\end{split}
\label{eqn:angular_momentum}
\end{equation}
Note that all three components are initially zero relative to the rotating coordinate system.

The introduced rigid body rotating rate $(\delta \Omega_{x}, \delta \Omega_{y}, \delta \Omega_{z})$ in Step 2 is
determined by
\begin{equation}\label{eq:domega}
\begin{split}
\delta \Omega_{x} = L_{x}/I_{x},\quad
\delta \Omega_{y} = L_{y}/I_{y}, \quad
\delta \Omega_{z} = L_{z}/I_{z},
\end{split}
\end{equation}
where $I_{x}, I_{y}, I_{z}$ are the moment of inertia of the the spherical shell in the $x$, $y$ and $z$ directions respectively, and $I_{x}=\int_{V} \rho (y^{2}+z^{2})dV$, $I_{y}=\int_{V} \rho (x^{2}+z^{2})dV$, and $I_{z}=\int_{V} \rho (x^{2}+y^{2})dV$. Once $(\delta \Omega_{x}, \delta \Omega_{y}, \delta \Omega_{z})$ is obtained, the Cartesian velocity components $u$, $v$, and $w$ will be updated at each solution point using
\begin{equation}
\begin{split}
u_{new}=u_{old}+\delta \Omega_{z} \sqrt{x^{2}+y^{2}}\sin ({\rm atan2} (y,x)) - \delta \Omega_{y}
            \sqrt{x^{2}+z^{2}} \cos ({\rm atan2} (x,z)), \\
v_{new}=v_{old}-\delta \Omega_{z} \sqrt{x^{2}+y^{2}}\cos ({\rm atan2} (y,x)) + \delta \Omega_{x}
            \sqrt{y^{2}+z^{2}} \sin ({\rm atan2} (z,y)), \\
w_{new}=w_{old}+\delta \Omega_{y} \sqrt{x^{2}+z^{2}}\sin ({\rm atan2} (x,z)) - \delta \Omega_{x}
            \sqrt{y^{2}+z^{2}} \cos ({\rm atan2} (z,y)),
\end{split}
\label{eqn:velocity_update}
\end{equation}
where function ${\rm atan2}(m_{1},m_{2}) = 2 \arctan (\frac{m_{1}}{\sqrt{m_{1}^{2} + m_{2}^{2}}+m_{2}}).$

We note that this correction procedure applies equally well in the case of an oblate, rapidly-rotating star.  The total moment of inertia in each direction is numerically calculated by summing up the moment of inertia in each element of the unstructured grid so it works regardless of the shape of the star.

\begin{figure}[!htb]
\centering
\includegraphics[width=0.9 \textwidth, angle=0]{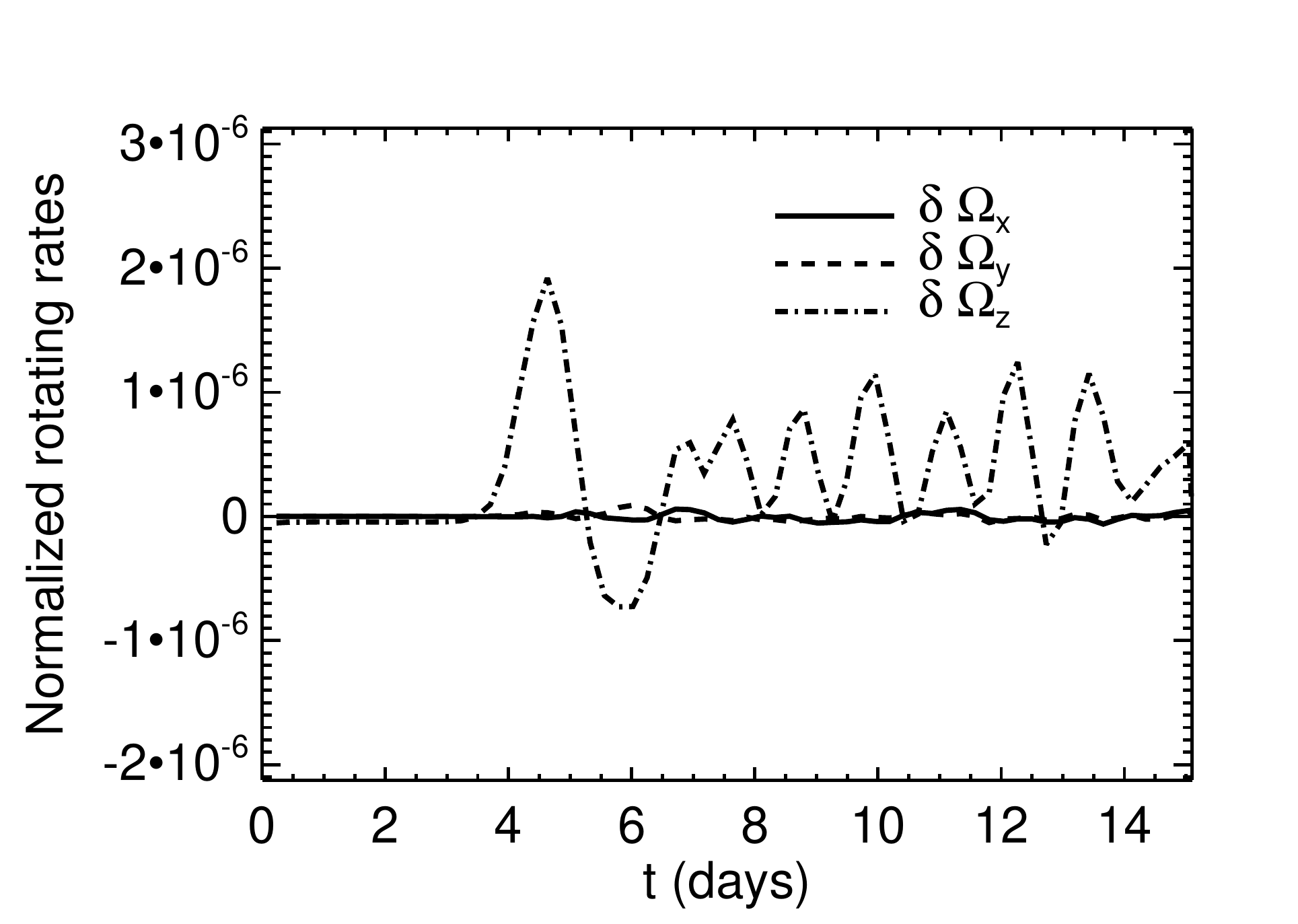}
\caption{Deviations from angular momentum conservation expressed in terms of angular velocity variations according to Eq.\ (\ref{eq:domega}). $\delta \Omega_{x}, \delta \Omega_{y}$, and $\delta \Omega_{z}$ are sampled every 5000 iterations just before the correction procedure and normalized to the rotation rate of the coordinate system $\Omega_0$.}
\label{fig:omega}
\end{figure}

This correction procedure is computationally expensive if performed at every time step. Thus, in all CHORUS simulations, we only do correction every 5000 time steps. Fig.\ref{fig:omega} shows the time evolution of $\delta \Omega_{x}$, $\delta \Omega_{y}$ and $\delta \Omega_{z}$ from a representative CHORUS simulation. This demonstrates the high accuracy of the numerical algorithm since the relative angular momentum error never exceeds $2\times 10^{-6}$ even at this 5000-step correction interval.  Furthermore, it demonstrates that the cumulative long-term errors in the angular momentum components are well controlled.

\section{Code Verification}

We verify the CHORUS code by comparing its results to the well-established Anelastic Spherical Harmonic (ASH) code \citep{Clune-1999,Miesch-2000,Brun-2004}.  However, we acknowledge that this comparison is not ideal since the two codes solve different equations.  As mentioned in section \ref{sec:polytrope}, the equations of motion in the anelastic approximation are obtained by linearizing the fully compressible equations (1)-(3) about a hydrostatic, spherically-symmetric reference state.  So we would only expect CHORUS and ASH to agree in the limit $\epsilon \rightarrow 0$, where $\epsilon$ is the normalized radial entropy gradient defined in Eq.\ (\ref{eqn:epsilon}).  A thorough comparison between the two systems would involve a linear and nonlinear analysis demonstrating convergence as $\epsilon \rightarrow 0$.  This lies outside the scope of the present paper.  Here we focus on defining two benchmark simulations, patterned after the gaseous atmosphere of Jupiter and the convective envelope of the Sun, and compare the results from CHORUS and ASH.  Despite subtle differences in the model equations and substantial differences in the numerical method, we demonstrate good agreement between the two codes.  This serves to verify CHORUS and to pave the way for future applications.

\subsection{The ASH Code and Anelastic Benchmarks}\label{sec:ash}

ASH is a multi-purpose code designed to simulate the hydrodynamics (HD) and magnetohydrodynamics (MHD) of solar and stellar interiors in global spherical geometries.  It was first developed over 15 years ago and has remained at the leading edge of the field ever since, continually improving in its physical sophistication and parallel efficiency on high-performance computing platforms.  ASH results have appeared in over 100 publications with applications ranging from convection and dynamo action in solar-like stars and fully-convective low-mass stars, to core convection and dynamo action in massive stars,  to MHD instabilities and stably-stratified turbulence, to the generation of and transport by internal gravity waves, to tachocline confinement, to flux emergence, to the HD and MHD of red giants \citep{miesc05,miesc07,brun10,strug11}.

ASH is based on the anelastic approximation (sec.\ \ref{sec:polytrope}) and uses a poloidal-toroidal decomposition of the mass flux to ensure that the anelastic form of the mass continuity equation is satisfied identically ($\mathbf{\nabla \cdot} (\rho_0 \mathbf{u}) = 0$).  It is a pseudo-spectral code that uses triangularly-truncated spherical harmonic basis functions in the horizontal dimensions.  Although earlier versions of ASH employed Chebyshev basis functions in the radial dimension, the version presented here uses a centered, fourth-order finite difference scheme that was introduced to improve parallel efficiency.  The radial grid is uniformly spaced for the simulations presented here and the boundary conditions are as specified in section \ref{sec:BCs}.   Time stepping is accomplished using an explicit Adams-Bashforth scheme for the nonlinear terms and a semi-implicit Crank-Nicolson scheme for linear terms, both second order.

ASH is one of four global anelastic codes that were validated using a series of three carefully defined benchmark simulations presented by Jones et al.\citep{Jones-2011}.  All three benchmarks had shell geometries and dimensional parameters that were chosen to represent deep convection in Jupiter's extended atmosphere but they differed in their degree of magnetism (one was non-magnetic) and turbulent intensity (spanning laminar and turbulent dynamo solutions).  In order to concisely represent the effective parameter space, the benchmark simulations were specified through a series of non-dimensional parameters as defined in Eq.(\ref{eqn:numbers}), namely $R_a$, $P_r$, $E_k$, and $\beta$.  These benchmarks made use of a hydrostatic, adiabatic, polytropic reference state as described in sec.\ \ref{sec:polytrope}, specified through the additional nondimensional parameters $N_\rho$ and $n$ (the polytropic index).  All four codes used similar numerical methods (pseudospectral, spherical harmonic) and agreed to within a few percent for a variety of different metrics of physical quantities.

The benchmarks we define here are inspired by the anelastic convection-driven dynamo bechmarks of Jones et al.\ \citep{Jones-2011}.  However, some modifications to the anelastic benchmarks are necessary in order to ensure that they are consistent with the fully compressible equations solved by CHORUS.  We already discussed one example of this when defining the initial conditions in sec.\ \ref{sec:almost}.

Another significant modification of the anelastic benchmarks that we introduce here concerns the Mach number.  An implicit requirement of the anelastic approximation is that the Mach number of the flow is much less than unity $M_a = U/c_s << 1$, where $c_s$ is the sound speed.  This is well justified in stellar convection zones where $M_a$ is typically less than $10^{-3}$.  In a fully compressible code this places a severe constraint on the allowable time step permitted by the Courant-Freidrichs-Lewy (CFL) condition $\Delta t < \delta / c_s$ where $\delta$ is some measure of the minimum grid spacing.  Anelastic codes are not subject to this constraint.  If the Mach number is low, the CFL constraint imposed by the sound speed is much more stringent than that imposed by the flow field $\Delta t  < \delta / U$.  In the future we will mitigate this constraint through the use of implicit and local time stepping.  Here we address it by defining benchmark problems for which the Mach number is low (justifying the anelastic approximation in ASH) but not too low (mitigating the CFL constraints of CHORUS).

The CFL constraint arising from the sound speed can be a major issue for global, rotating convection simulations where the equilibration time scale is much longer than the dynamical time scale.  If one neglects structural stellar evolution, the longest time scale in the system is the thermal relaxation timescale $T_{r}=E/L$, which can exceed $10^5$ years in stars; by comparison the dynamical time scale is of order one month.  However, a more relevant time scale for equilibration of the convection is the thermal diffusion time scale $T_{d}=d^{2}/\kappa$, which is $\approx 65.74$ days for the Jupiter benchmark and $\approx 58.44$ days for the solar benchmark ($T_{r}\approx 3.55\times 10^{4}$ days and $5.45\times 10^{4}$ days respectively).

\subsection{Metrics of Physical Quantities}\label{sec:metrics}

Before proceeding to the simulation results, we first define several important metrics that provide a means to compare CHORUS and ASH.  These include the mean kinetic energy relative to the rotating reference frame and its mean-flow components, namely the the differential rotation (DRKE) and the meridional circulation (MCKE):

\begin{equation}
KE = \frac{1}{V}\int_{V} \frac{1}{2}\rho (\mathbf{u}\cdot \mathbf{u}) dV,
\end{equation}
\begin{equation}
DRKE = \frac{1}{V}\int_{V}\frac{1}{2}\rho \langle V_{\phi} \rangle^{2}r^{2}\sin \theta dr d\theta d\phi,
\end{equation}
\begin{equation}
MCKE = \frac{1}{V}\int_{V}\frac{1}{2}\rho(\langle V_{r} \rangle^{2} + \langle V_{\theta} \rangle^{2})r^{2}\sin \theta dr d\theta d\phi ~~,
\end{equation}
where $V$ denotes the volume of the computational domain and angular brackets denote averages over longitude.

The growth rate of the kinetic energy is defined as
\begin{equation}
\sigma = \frac{d(\ln KE)}{dt}.
\end{equation}

From Eq.(\ref{eqn:energy}), four components of the energy flux are involved in transporting energy in the radial direction, namely the enthalpy flux $F_{e}$, kinetic energy flux $F_{k}$, radiative flux $F_{r}$ and entropy flux $F_{u}$. In a statistically steady state, these four fluxes together must account for the full luminosity imposed at the bottom boundary:
\begin{equation}
F_{e} + F_{k} + F_{r} + F_{u} = F_{*} = \frac{L}{4\pi r^{2}},
\end{equation}
where
\begin{eqnarray}
F_{e} &=& \overline{\rho}C_{p}V_{r}(T - \overline{T}), \\
F_{k} &=& \frac{1}{2}\overline{\rho}V_{r}(\mathbf{u}\cdot \mathbf{u}), \\
F_{r} &=& -\kappa_{r}\overline{\rho}C_{p}\frac{\partial \overline{T}}{\partial r},
\end{eqnarray}
and
\begin{equation}
F_{u} = -\kappa \overline{\rho}\overline{T}\frac{\partial \overline{S}}{\partial r}.
\end{equation}
$\overline{\rho}$, $\overline{T}$, and $\overline{S}$ denote the mean density, temperature and entropy,
averaged over horizontal surfaces.

On each horizontal surface, the mean Mach number is defined as
\begin{equation}
Ma = \frac{V_{rms}}{C_{s}},
\end{equation}
where
\begin{equation}
V_{rms} = \sqrt{\frac{1}{4\pi}\int_{0}^{\pi}\int_{0}^{2\pi}(\mathbf{u}\cdot \mathbf{u})\sin\theta d\theta d\phi},
\end{equation}
and the mean sound speed is
\begin{equation}
C_{s} = \sqrt{\gamma \frac{\overline{p}}{\overline{\rho}}} .
\end{equation}

\subsection{Jupiter Benchmark}

The deep, extended outer atmosphere of Jupiter is thought to be convectively
unstable \citep{hubba02}.  This has motivated substantial work on the internal dynamics of
giant planets and inspired the parameter regimes chosen for the anelastic benchmarks of
Jones et al.\ \citep{Jones-2011}.  Our first benchmark for comparing CHORUS and ASH is
similar to the hydrodynamic benchmark of Jones et al.\ \citep{Jones-2011} apart from
the thermal boundary conditions.  Whereas we impose a fixed heat flux on the lower boundary
and a fixed temperature on the upper (sec.\ \ref{sec:BCs}), Jones et al.\ fix the specific
entropy $S$ on both boundaries.  Note that this means that the Rayleigh number $R_a$,
defined in terms of $\Delta S$ in Eq. (\ref{eqn:numbers}), does change somewhat
in our simulations as the convection modifies the entropy stratification.  This is in
contrast to Jones et al.\ where it was held fixed.

The parameters for this case are specified in
Table \ref{tab:parameter_Jupiter}.   The value of $R_a$ listed is the initial
value, before convection ensues.  The number of DOFs used for the CHORUS simulation
is about a factor of five larger than that used for the ASH simulation, as indicated
in Table \ref{tab:computational_effort}.  Both simulations use the same boundary conditions
and initial conditions, apart from the random perturbations needed to excite convection
which are generated independently by each code.

\begin{table}
    \begin{tabular}{l}
    \toprule
    Dimensionless parameters \\
    $E_{k} = 10^{-3}$, $N_{\rho}=5$, $\beta = 0.35$, $Ra = 351,806$, $Pr = 1$, $n = 2.0$ \\ \\
    Defining physical input values \\
    $R_{o}$ = 7 $\times$ $10^{9}$ $cm$, $\Omega_{o}$ = 1.76 $\times$ $10^{-4}$ $s^{-1}$, $M$ = 1.9 $\times$ $10^{30}$ $g$, $\rho_{i}$ = 1.1 $g$ $cm^{-3}$ \\
    $\mathfrak{R}$ = 3.503 $\times$ $10^{7}$ $erg$ $g^{-1}$ $K^{-1}$, $G$ = 6.67 $\times$ $10^{-8}$ $g^{-1}$ $cm^{3}$ $s^{-2}$, $\kappa_{r}=0$  \\
    \\
    Derived physical input values \\
    $R_{i}$ = 2.45 $\times$ $10^{9}$ $cm$, $d$ = 4.55 $\times$ $10^{9}$ $cm$, $\nu$ = 3.64364 $\times$ $10^{12}$ $cm^{2}$ $s^{-1}$,\\ $\kappa$ = 3.64364 $\times$ $10^{12}$ $cm^{2}$ $s^{-1}$, $\gamma$ = 1.5  \\
    \\
    Other thermodynamic quantities \\
    $L$ = 7.014464 $\times$ $10^{32}$ $erg$ $s^{-1}$, $C_{p}$ = 1.0509 $\times$ $10^{8}$ $erg$ $g^{-1}$ $K^{-1}$  \\
    \bottomrule
    \end{tabular}
    \caption{Parameters for the Jupiter benchmark}
    \label{tab:parameter_Jupiter}
\end{table}

\begin{figure}[!htp]
\centering
\subfigure[]{\label{fig:KE_density_Jupiter}\includegraphics[width=0.49 \textwidth, angle=0]{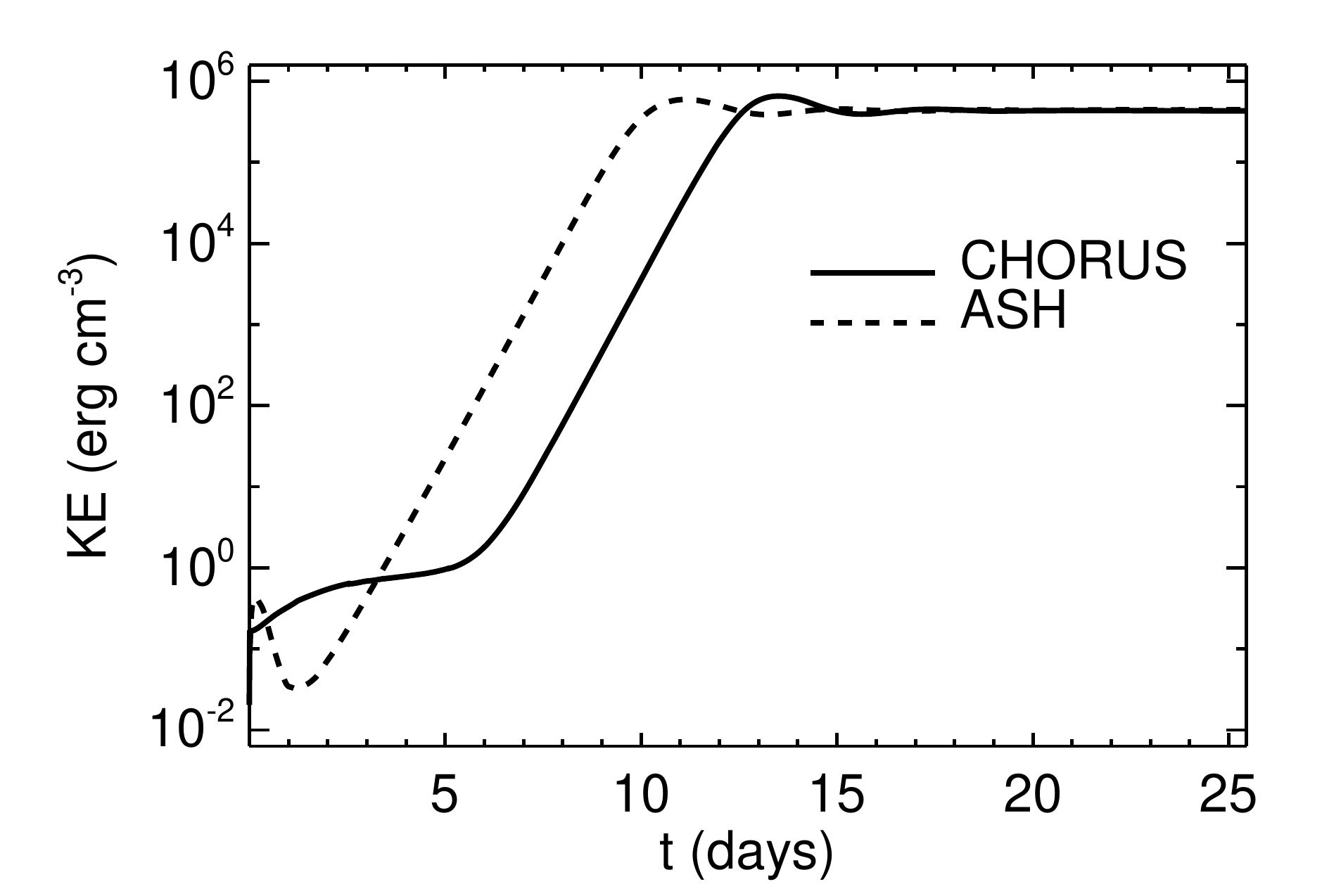}}
\subfigure[]{\label{fig:Growth_Jupiter}\includegraphics[width=0.49 \textwidth, angle=0]{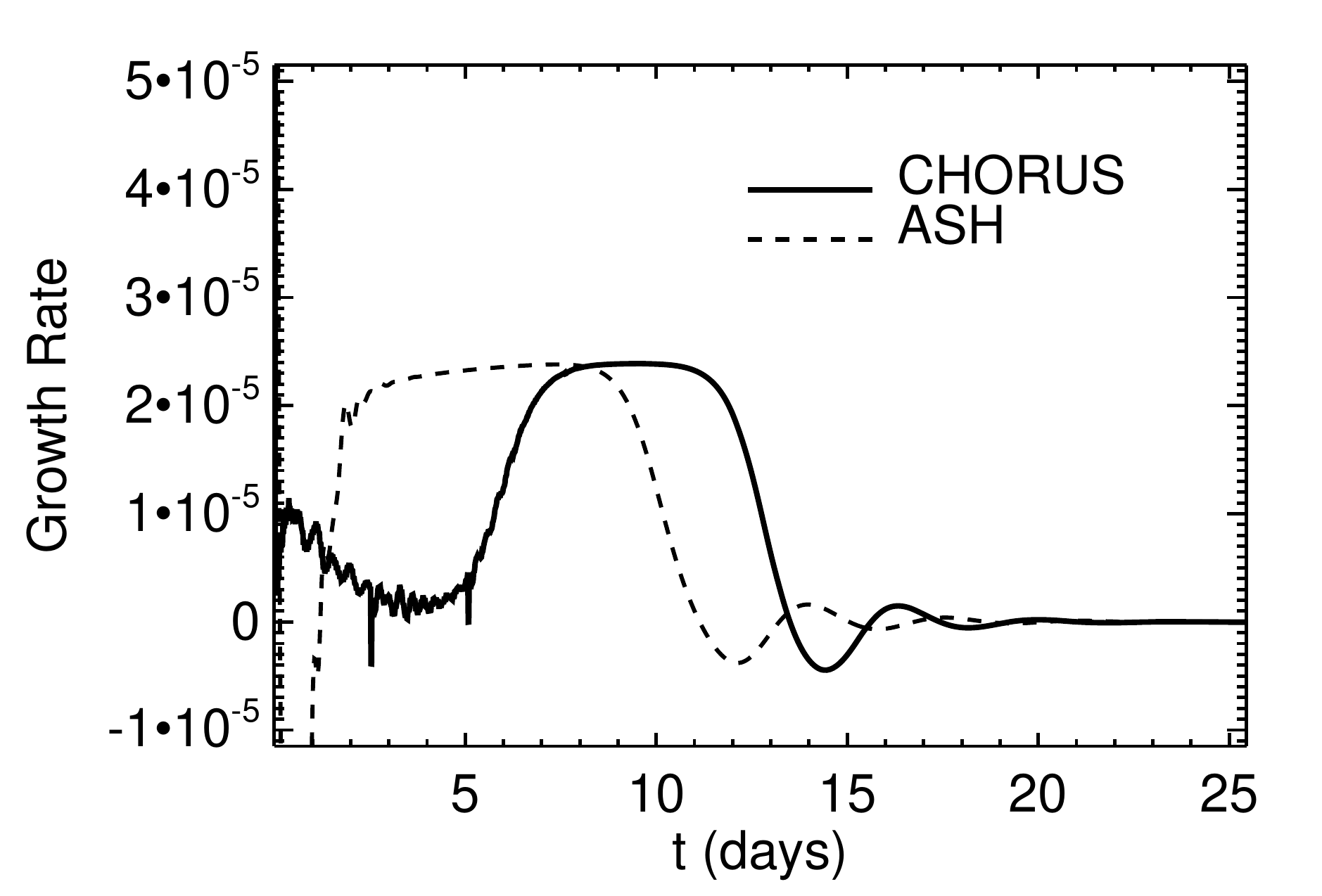}}
\caption{(\textit{a}) Kinetic energy density and (\textit{b}) growth rate for the Jupiter benchmark.}
\label{fig:_jupiter}.
\end{figure}

\subsubsection{Exponential Growth and Nonlinear Saturation}\label{sec:growth}
The evolution of the kinetic energy densities for both CHORUS and ASH simulations are illustrated in Fig.\ref{fig:KE_density_Jupiter}. As mentioned above, both simulations start with the same background stratification but with different random perturbations.  The small amplitude of the initial perturbations ensures that each simulation begins in the linear regime.  For each simulation there is an initial adjustment period before the flow locks on to the fastest-growing eigenmode which then grows exponentially.

The initial establishment period is different for each simulation, lasting roughly 8 days for the CHORUS simulation and 2 days for the ASH simulation.  However, this is to be expected from the different mix of random perturbations.  A meaningful comparison can only be made between the two codes after they reach the exponential growth phase, which is followed by nonlinear saturation and a subsequent equilibrium.  In other words, the two features of Fig.\ \ref{fig:KE_density_Jupiter} that should be compared are the slope of the line in the linear growth phase and the value of the kinetic energy density after each simulation has saturated and equilibrated.

This first point of comparison, namely the growth rate, is highlighted in
Fig.\ref{fig:Growth_Jupiter}.  The peak values achieved by each simulation in the linear regime reflect the growth rate of the preferred linear eigenmode and agree to within about 2\%;
$\sigma = 2.38\times 10^{-5}$ for CHORUS and $ \sigma = 2.33 \times 10^{-5}$ for ASH.  We define the nonlinear saturation time $T_n$ as the time at which the growth rate first crosses zero after the exponential growth phase.

Again, the saturation time is different between the two simualtions because of the random nature of the initial conditions.  Thus, in order to compare the two cases in the nonlinear regime we define a sampling time $T_{s}$ to be 10 days after the saturation time, $T_n$. This is $T_{s}=23.51$ days for the CHORUS simulation and $T_{s}=21.15$ days for the ASH simulation.  Averaged between ($T_{s}-2$) days and $T_{s}$ days, $KE=4.33\times 10^{5}$ $erg$ $cm^{-3}$ for the CHORUS simulation and $KE=4.15\times 10^{5}$ $erg$ $cm^{-3}$ for the ASH simulation. The difference is about $4\%$.   Another point of comparison is the relative magnitude of the mean flows, as quantified by the DRKE and MCKE defined in sec.\ \ref{sec:metrics}.  At the sampling time $T_{s}$, DRKE/KE $ = 0.113$ and MCKE/KE $ = 2.45\times 10^{-4}$ for the CHORUS simulation while DRKE/KE $ = 0.134$ and MCKE/KE $ = 2.58\times 10^{-4}$ for the ASH simulation.  This corresponds to a difference of about 16\% and 5\% for the DR and MC respectively.  Small differences between the two codes of order several percent are to be expected due to differences between the compressible and anelastic equations.  Furthermore, the relatively large difference in the DRKE/KE likely comes about because the simulations are not fully equilibrated.  We address these issue further in the following section (sec.\ \ref{sec:mach}).

\subsubsection{Mach Number, $\epsilon$ and Flux Balance}\label{sec:mach}
\begin{figure}[!htp]
\centering
\subfigure[]{\label{fig:Ma_JUPITER}\includegraphics[width=0.49 \textwidth, angle=0]{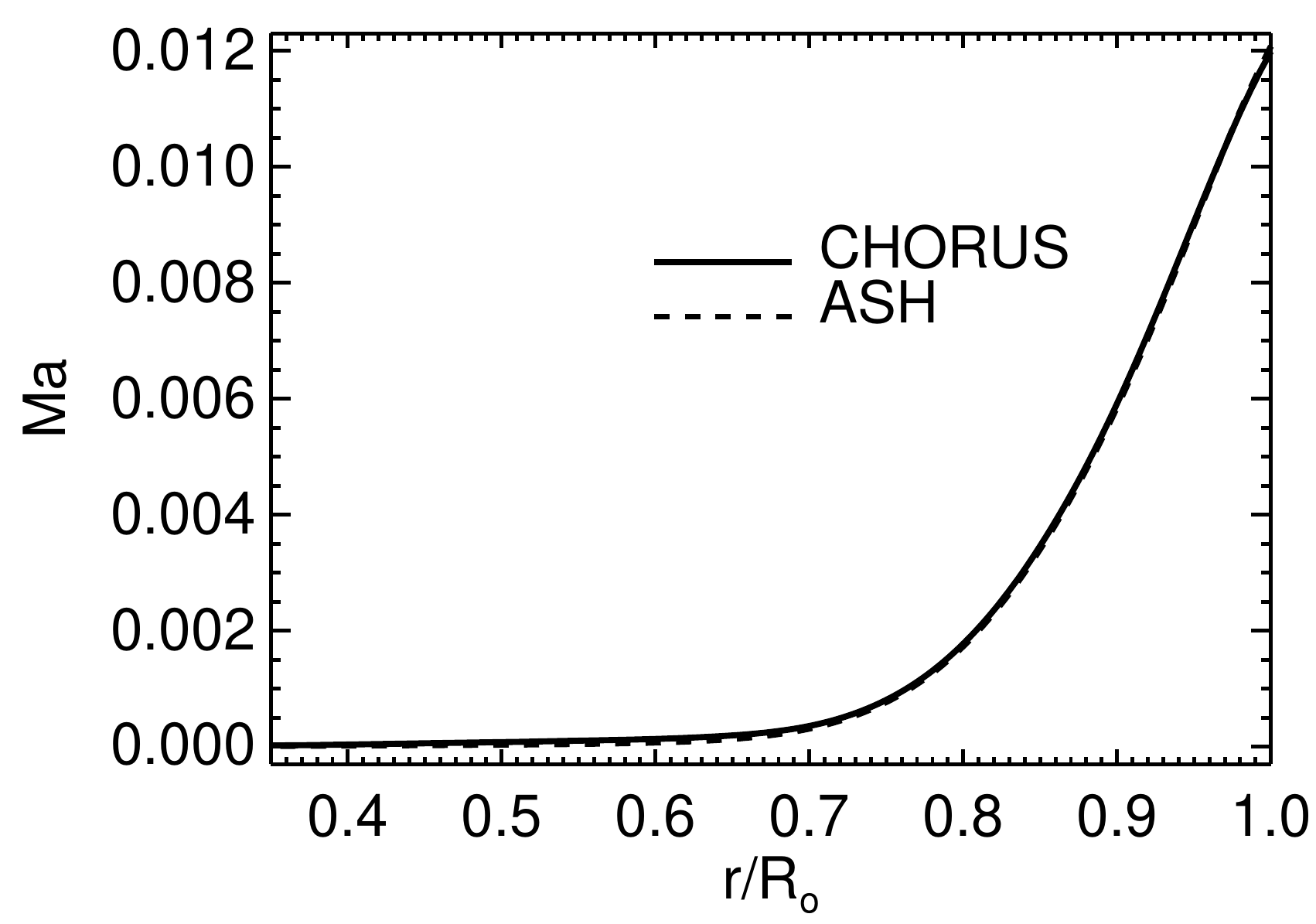}}
\subfigure[]{\label{fig:epsilon_JUPITER}\includegraphics[width=0.49 \textwidth, angle=0]{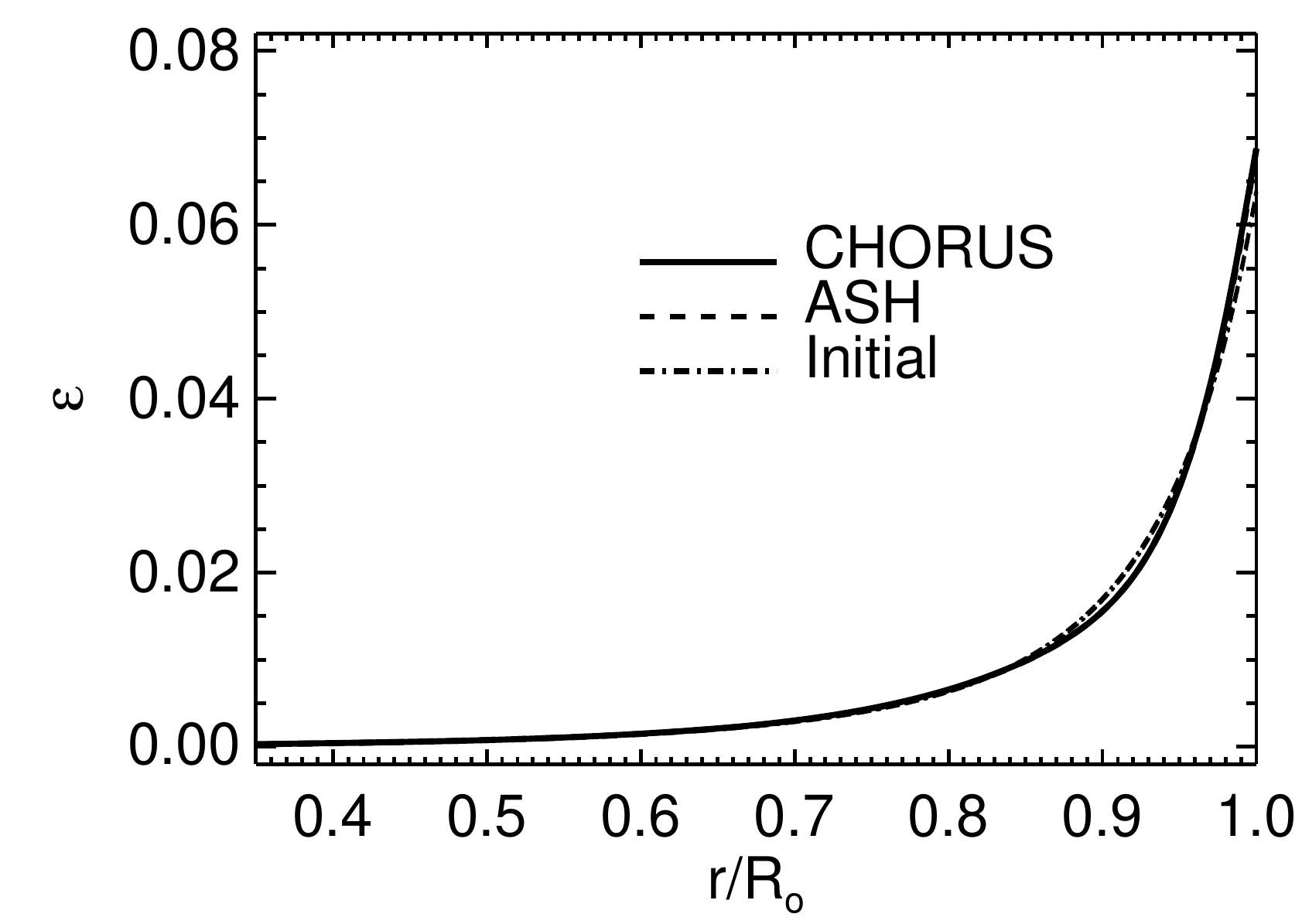}}
\caption{(\textit{a}) Mach number and (\textit{b}) $\epsilon$ for the Jupiter benchmark, plotted at the sampling time $T_{s}$. In (\textit{b}), the initial condition is also plotted for comparison (dash-dotted line). The CHORUS and ASH curves are nearly indistinguishable.}
\end{figure}

As mentioned in secs.\ \ref{sec:polytrope} and \ref{sec:ash}, the comparisons between ASH and CHORUS are only meaningful if the stratification is nearly adiabatic ($\epsilon << 1$) and the Mach number is small.  These conditions are required for the validity of the anelastic approximation.  Fig.\ \ref{fig:Ma_JUPITER} demonstrates that these conditions are met, but that departures are significant.  In particular, we expect that the anelastic and compressible equations are only equivalent to lowest order in $\epsilon$, which reaches a value as high as 0.07 in the upper convection zone (Fig.\ref{fig:epsilon_JUPITER}).

In Fig.\ref{fig:Ma_JUPITER}, the mean Mach number is minimum at the bottom and increases with radius and reaches the maximum (about 0.012) at the top in both the CHORUS and ASH simulations. As defined in Eq.(\ref{eqn:epsilon}), $\epsilon$ is proportional to the mean entropy gradient $\frac{d\overline{S}}{dr}$. Identical initial entropy gradients for the CHORUS and ASH simulations implies that they have the same initial degrees of adiabaticity. When convection is present, the associated energy flux leads to a redistribution of entropy, tending to smooth out the entropy gradient. This is the reason that $\epsilon$ in Fig.\ref{fig:epsilon_JUPITER} becomes smaller than the initial near $r=0.92R_{o}$.

\begin{figure}[!htp]
\centering
\subfigure[CHORUS]{\includegraphics[width=0.49 \textwidth, angle=0]{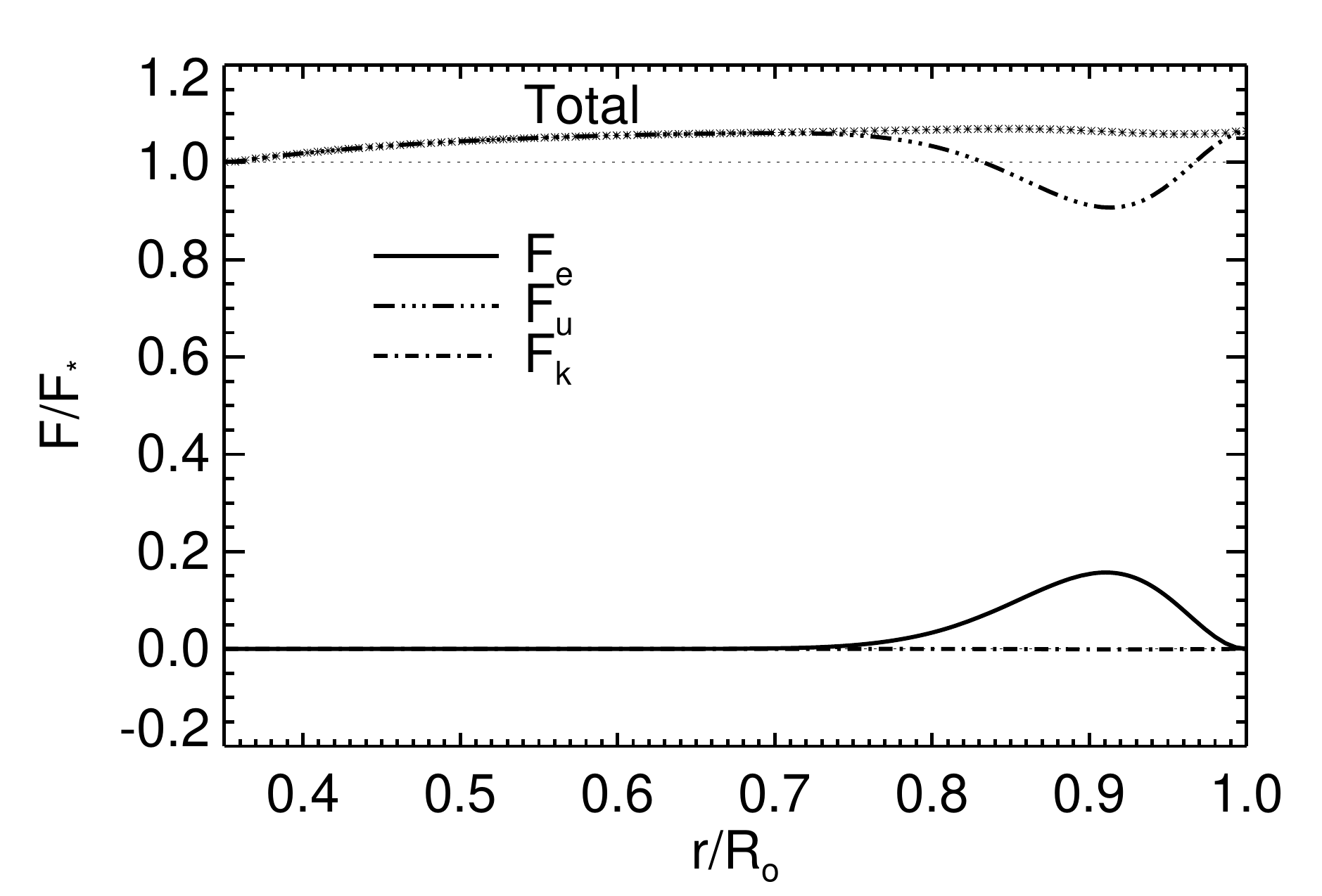}}
\subfigure[ASH]{\includegraphics[width=0.49 \textwidth, angle=0]{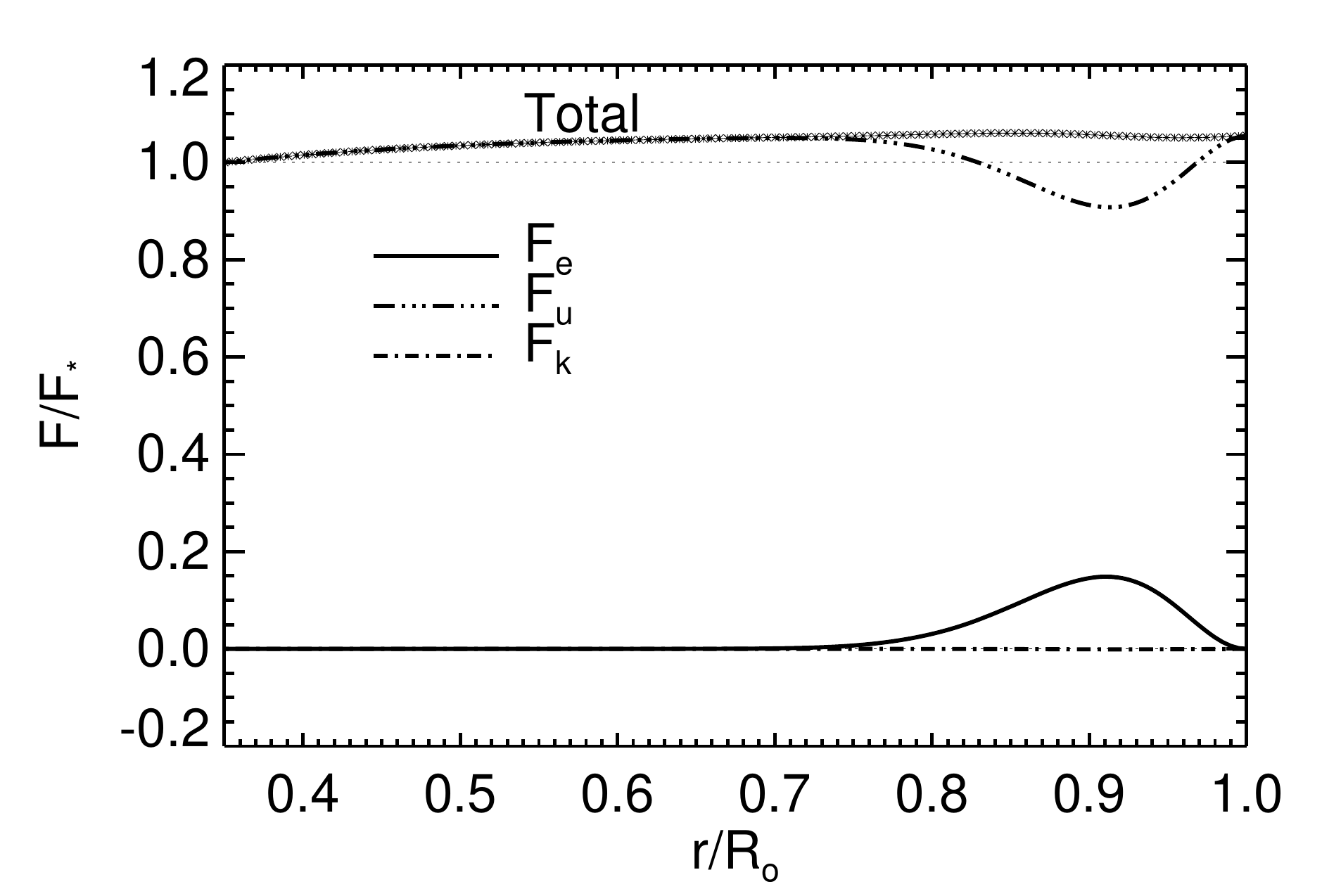}}
\caption{Components of the radial energy flux for the Jupiter benchmark integrated over horizontal surfaces at the sampling time $T_{s}$ for (\textit{a}) CHORUS and (\textit{b}) ASH.  All values are normalized by the total flux $F_{*}$ imposed at the bottom boundary.}
\label{fig:flux_jupiter}.
\end{figure}

The location $r=0.92R_{o}$ corresponds to where the efficiency of the convection peaks.  This is demonstrated in Fig.\ref{fig:flux_jupiter} which shows the components of the energy flux defined in sec.\ \ref{sec:metrics}.  The {\em almost flux balance} initialization described in sec.\ \ref{sec:almost} establishes an entropy stratification that carries most of the energy flux through entropy diffusion $F_{u} \approx F_{*}$, with a slight over-luminosity of about 7\% in the upper convection zone.  Once convection is established, it carries roughly 15\% of this flux, flattening the entropy gradient and reducing $F_{u}$.  The kinetic energy flux $F_{k}$ for this case is negligible.  The maximum values of $F_{e}$ at $r=0.92R_{o}$ for the CHORUS and ASH simulations are 0.1536 and 0.1484 respectively, showing the difference of $3.5\%$.

\begin{figure}[!htp]
\centering
\includegraphics[width=0.80 \textwidth, angle=0]{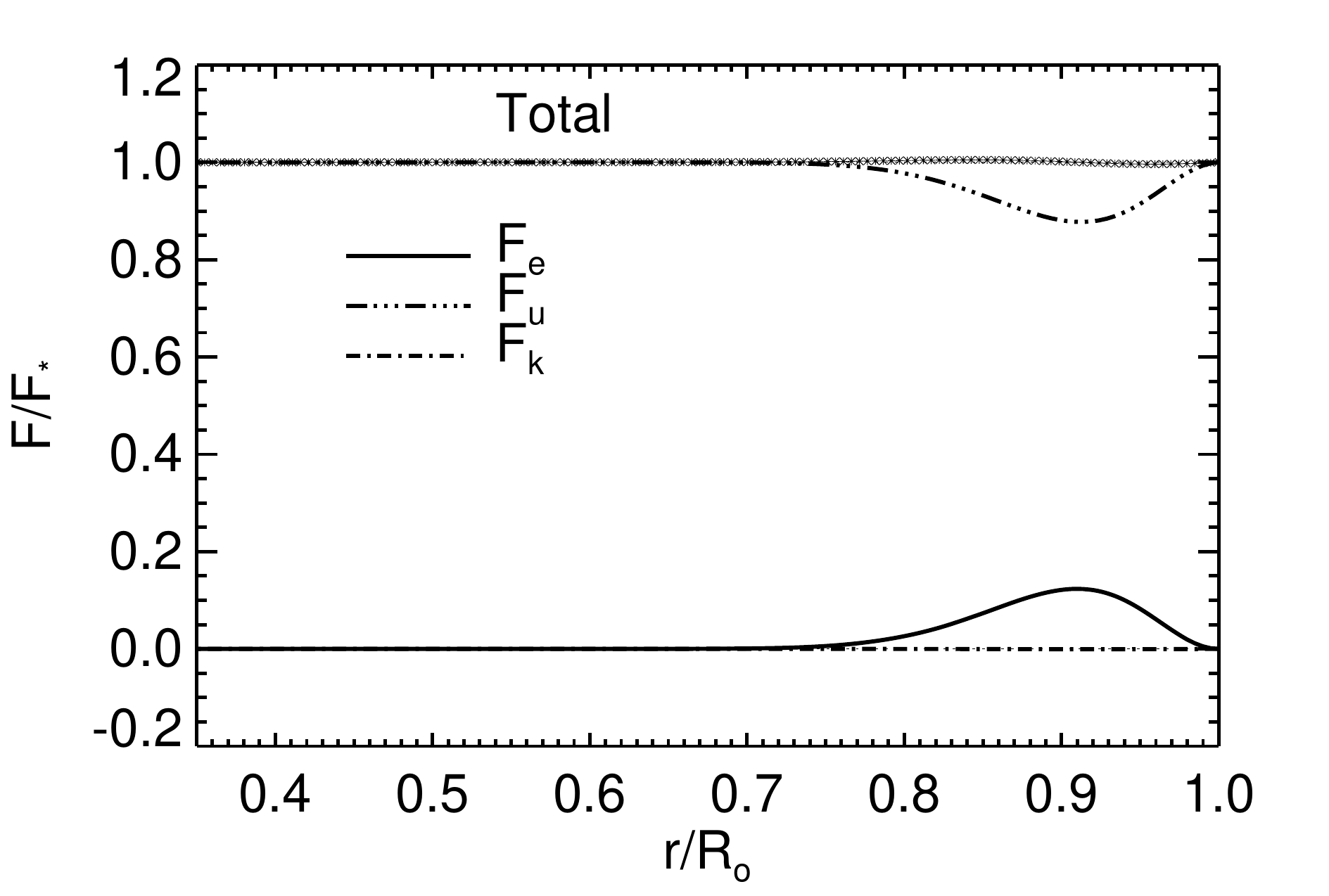}
\caption{Components of the radial energy flux as in Fig.\ \ref{fig:flux_jupiter}
but at a much later time, $t \sim $ 1800 days, computed with the ASH code, showing an equilibrium solution.}
\label{fig:flux_1800}.
\end{figure}

In a nonlinear equilibrium state, the sum of the normalized fluxes in Fig.\ \ref{fig:flux_jupiter} should be unity.  This is clearly not the case; as mentioned above both simulations are over-luminous by about 7\% due to the initial entropy stratification.  This will eventually subside but the process is slow, occurring gradually over a time scale that is longer than the thermal diffusion time scale of $T_{d} \sim 65.74$ days but shorter than the thermal relaxation time scale of $T_{r} \sim 3.55\times 10^4$ days (see sec.\ \ref{sec:ash}).  This is demonstrated in Fig.\ref{fig:flux_1800} which shows the flux balance in the ASH simulation after 1800 days.  Given the greater computational cost of CHORUS (sec.\ \ref{sec:performance}), and the satisfactory agreement at the sampling time (within the expected order $\epsilon$), we choose not to run the CHORUS simulation to full equilibration.

\subsubsection{Convection Structure}

\begin{figure}[!htp]
\centering
\subfigure[CHORUS simulation]{\includegraphics[width=0.95 \textwidth, angle=0]{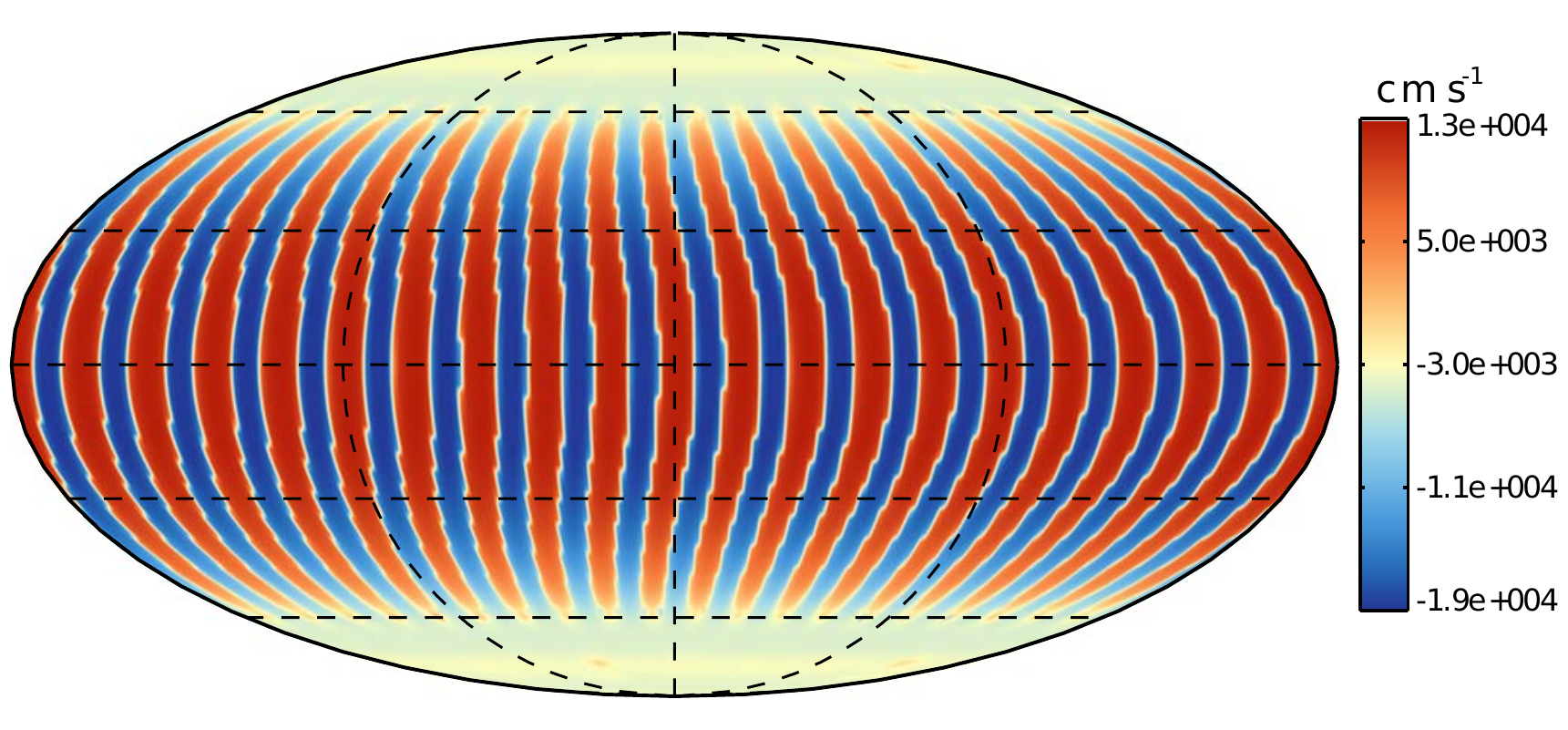}}
\subfigure[ASH simulation]{\includegraphics[width=0.95 \textwidth, angle=0]{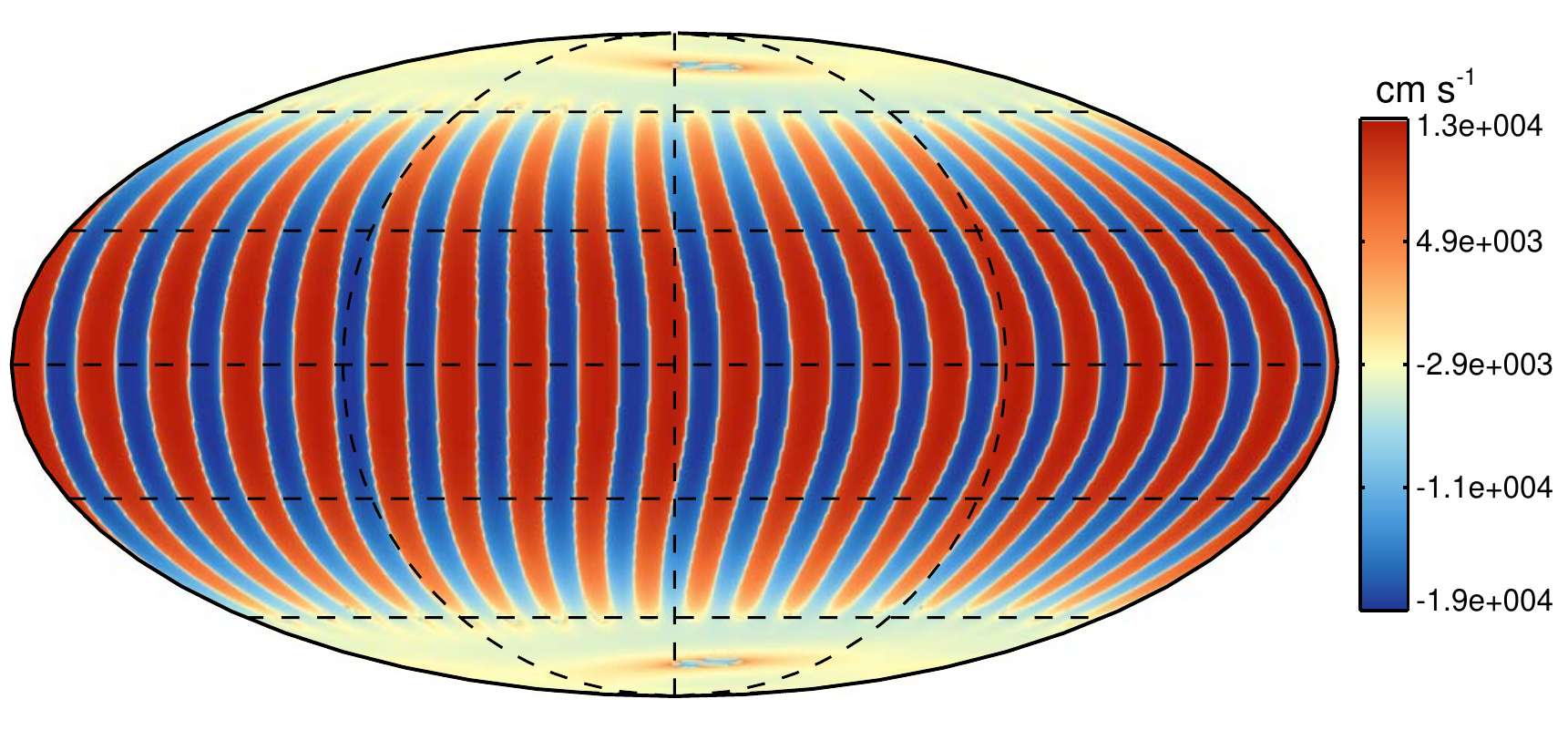}}
\caption{Mollweide projections of radial velocity $V_{r}$ for the Jupiter benchmark at the horizontal surface $r=0.95R_{o}$, taken for (\textit{a}) CHORUS and (\textit{b}) ASH at the sampling time $T_{s}$. Red and blue tones denote the upflow and downflow as indicated by the color bar.}
\label{fig:Vr_Jupiter}.
\end{figure}

The structure of the convection at the sampling time is illustrated in Fig.\ \ref{fig:Vr_Jupiter}.  Though this is well into the nonlinear regime, both simulations are dominated by a series of columnar convective rolls approximately aligned with the rotation axis but sheared slightly in the prograde direction at low latitudes by the differential rotation.  These are the well-known `banana cells' characteristics of convection in rotating spherical shells and most apparent for laminar parameter regimes \citep{miesc05}.  Though this is well into the nonlinear regime, they reflect the preferred linear eigenmodes and are well described by a single sectoral spherical harmonic mode with $\ell = m$, where $\ell$ and $m$ are the spherical harmonic degree and order.  The degree and order $\ell$ and $m$ can also be interpreted as the total wavenumber and the longitudinal wavenumber respectively.

\begin{figure}[!htp]
\centering
\subfigure[]{\includegraphics[width=0.325 \textwidth, angle=0]{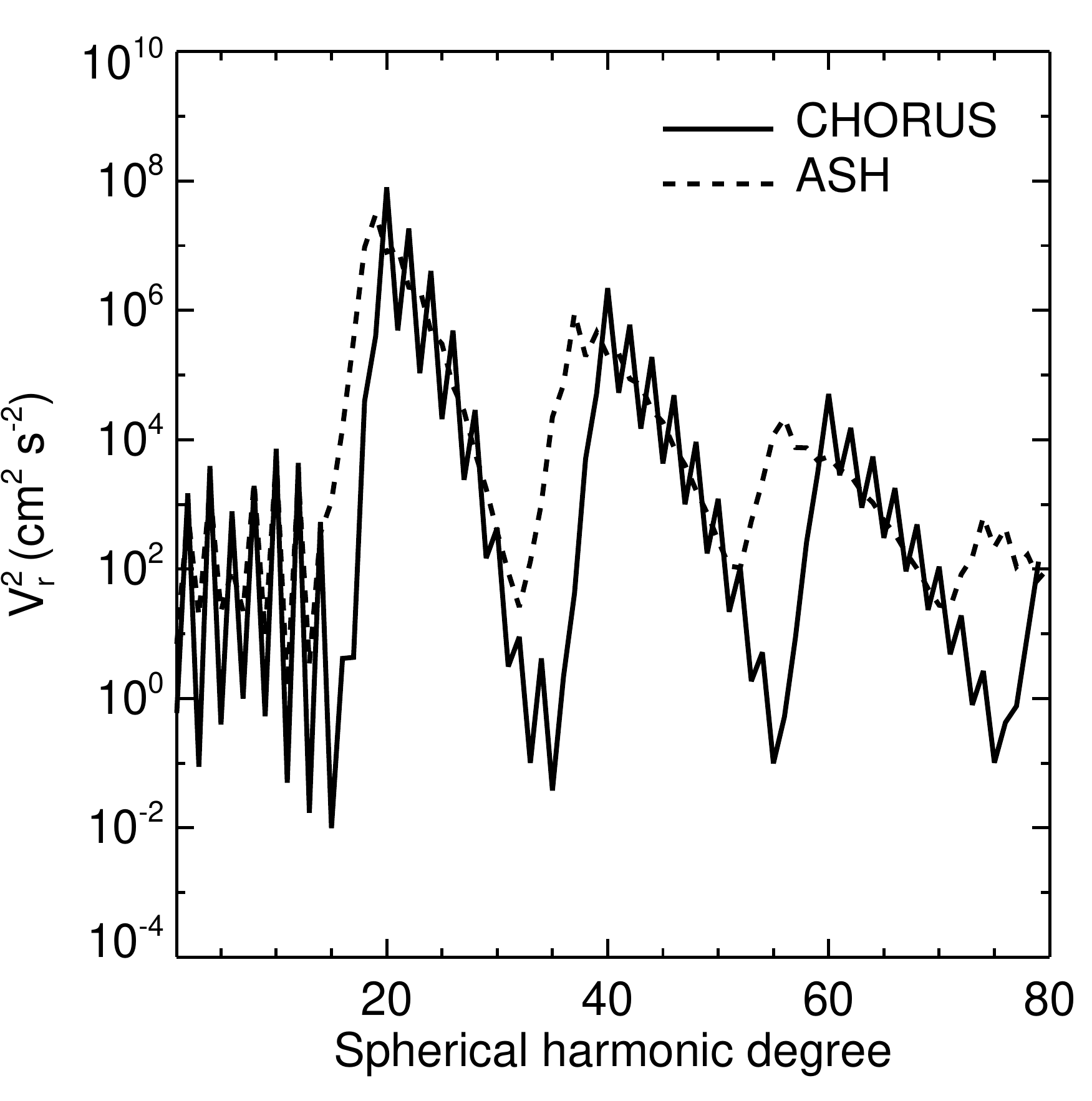}}
\subfigure[]{\includegraphics[width=0.325 \textwidth, angle=0]{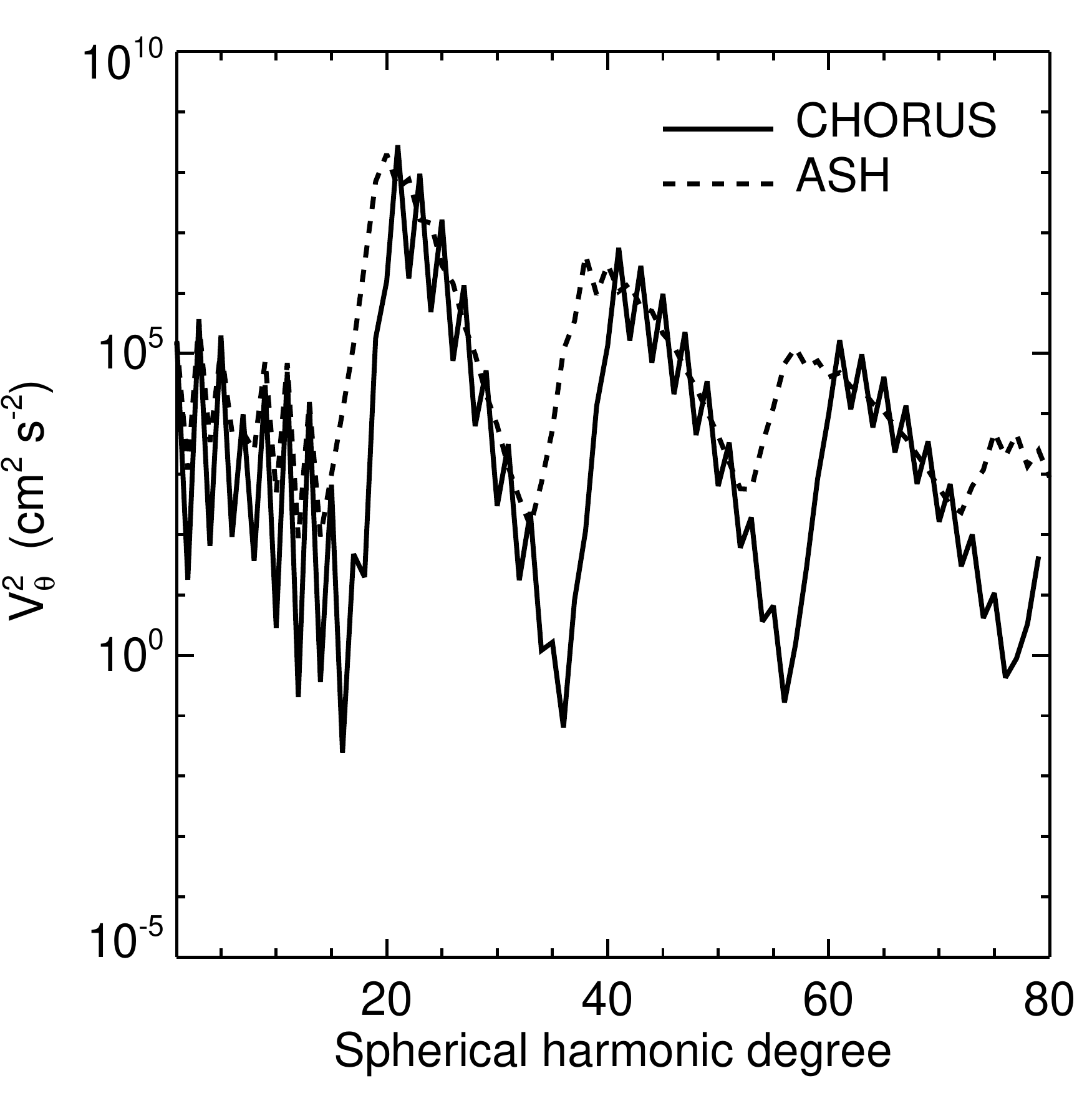}}
\subfigure[]{\includegraphics[width=0.325 \textwidth, angle=0]{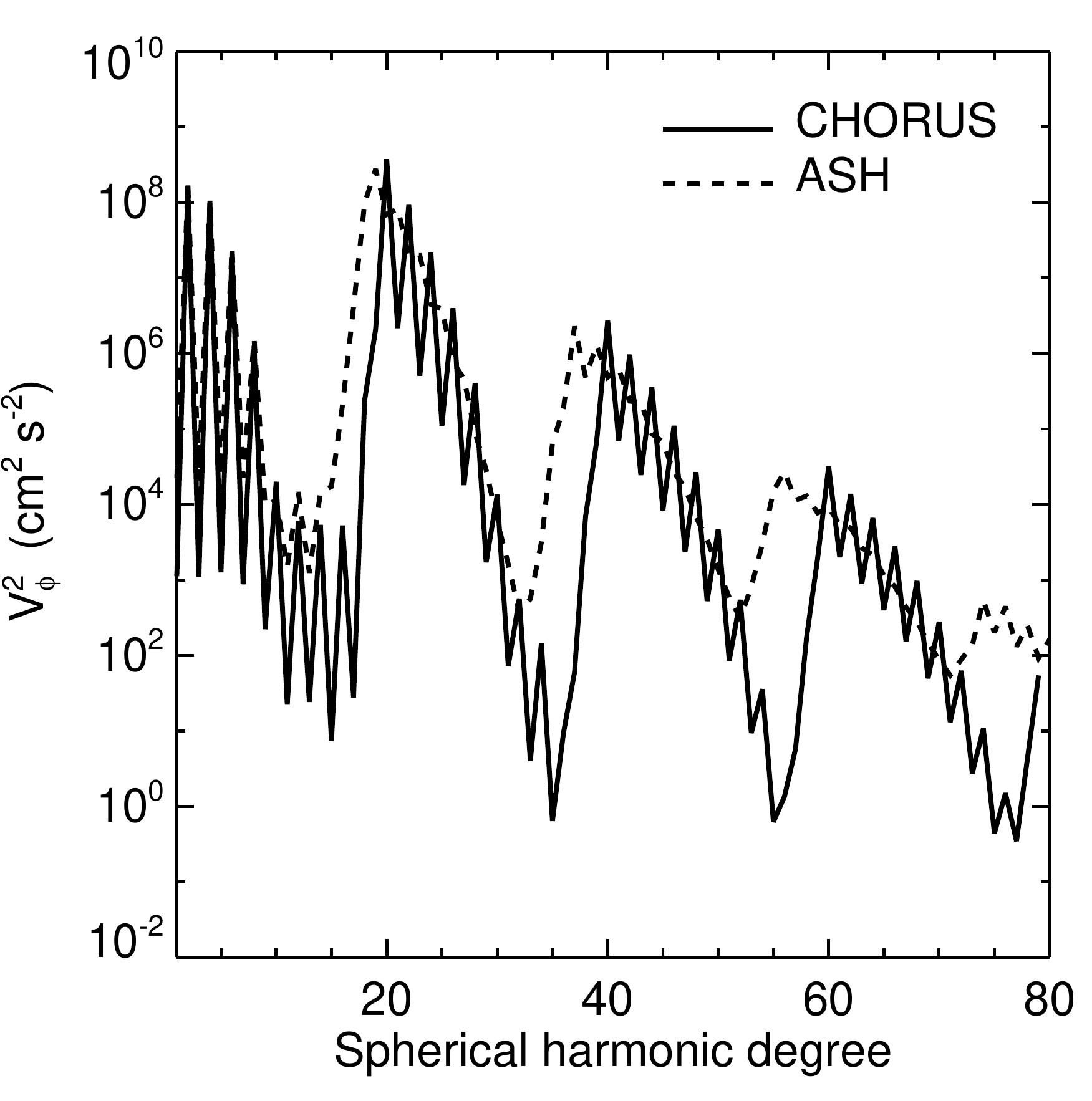}}
\caption{Shown are the power spectra of (\textit{a}) radial velocity $V_{r}$,
(\textit{b}) meridional velocity $V_{\theta}$, and (\textit{c}) zonal velocity $V_{\phi}$
as a function of spherical harmonic degree $\ell$ for the Jupiter benchmark at $r = 0.98 R_o$ and $t = T_{s}$.}
\label{fig:Spectrum_Jupiter}.
\end{figure}

Close scrutiny of Fig.\ \ref{fig:Vr_Jupiter} reveals that the two simulations exhibit a slightly different mode structure, with CHORUS selecting an $m=20$ mode and ASH selecting an $m=19$ mode.  This is demonstrated more quantitatively in Fig.\ref{fig:Spectrum_Jupiter} which shows the spherical harmonic spectra of the velocity field on the horizontal surface $r = 0.98R_{o}$ as a function of spherical harmonic degree $\ell$ (summed over $m$) at the sampling time $T_{s}$. In the CHORUS simulation, the spectra of radial velocity $V_{r}$, meridional velocity $V_{\theta}$, and zonal velocity $V_{\phi}$ peak at $\ell = 20$ and its higher harmonics, $\ell = 40$, $60$ and $80$. In comparison, the velocity spectra in the ASH simulation peak at $\ell = 19$, $38$, $57$ and $76$.

This level of agreement is consistent with the expected accuracy of the anelastic and compressible systems.  As discussed by Jones et al.\ \citep{Jones-2011}, the linear growth rates of the $m = 19$ and $m = 20$ modes in this hydrodynamic benchmark are the same to order $\epsilon$ and even a single anelastic code may choose one or the other depending on the random initial perturbations.  The laminar nature of this benchmark highlights these small differences; in more turbulent parameter regimes where the Rayleigh number far exceeds the critical value, a broad spectrum of modes is excited and the results are less sensitive to the details of the initial conditions and the linear eigenmodes.  This is demonstrated by the solar benchmark described in sec.\ \ref{sec:solarbm}, which is in a more turbulent parameter regime and which exhibits closer agreement between the velocity spectra.

\subsubsection{Mean Flows}
\begin{figure}[!htp]
\centering
\subfigure{\includegraphics[width=0.99 \textwidth, angle=0]{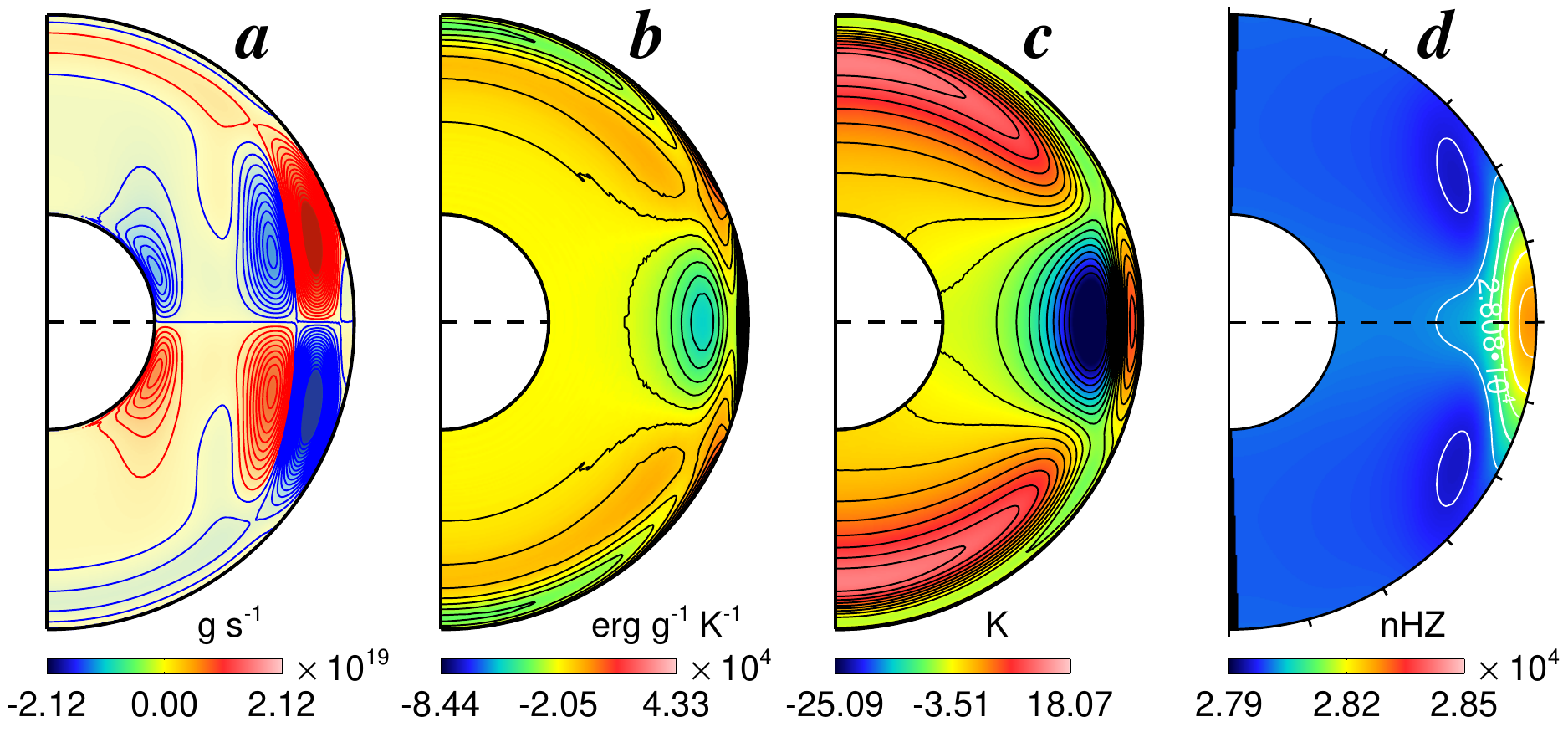}}
\subfigure{\includegraphics[width=0.99 \textwidth, angle=0]{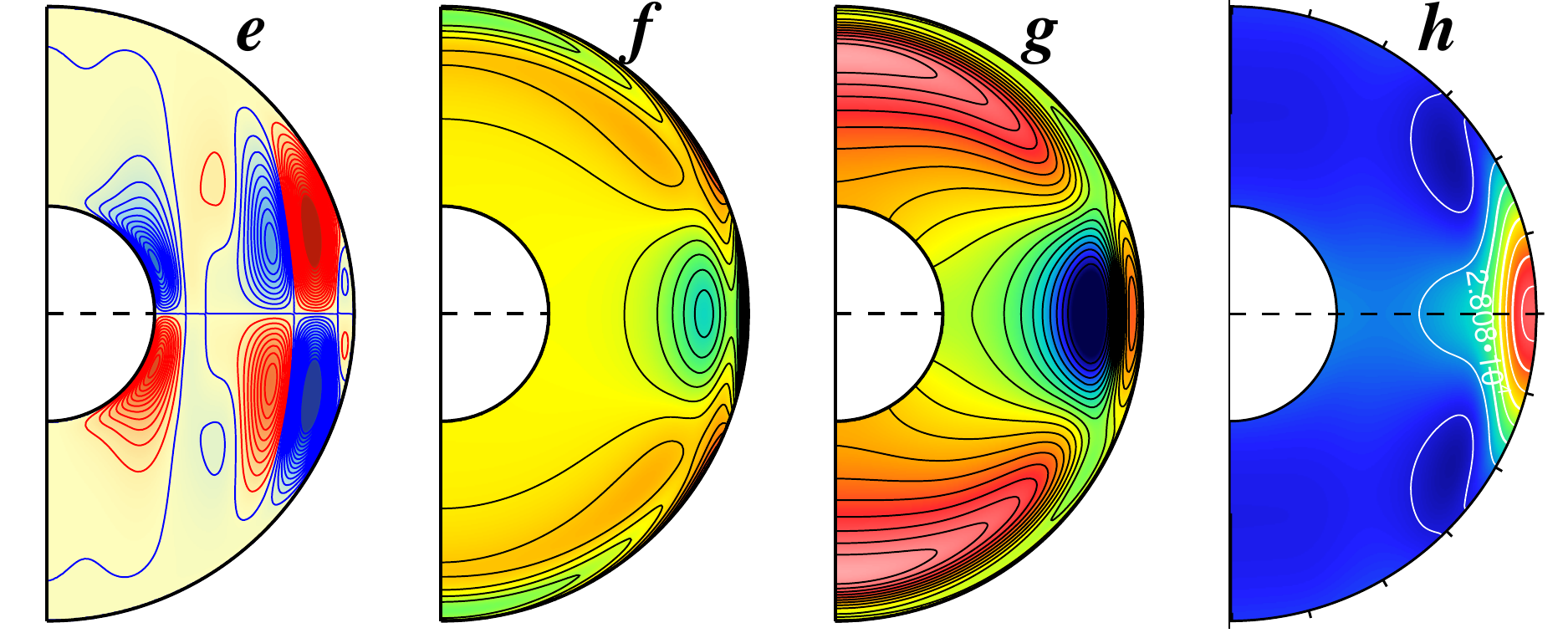}}
\caption{Mean flows in the Jupiter benchmark from CHORUS (top row) and ASH (bottom row).  Shown are (\textit{a}, \textit{e}) meridional circulation expressed in terms of the stream function $\Psi$, with red and blue denoting clockwise and counterclockwise circulations respectively, (\textit{b}, \textit{f}) specific entropy perturbation $S^{'}$, (\textit{c}, \textit{g}) temperature perturbation $T^{'}$, and (\textit{d}, \textit{h}) differential rotation, expressed in terms of the angular velocity $\Omega$.  The top and bottom plots in each column share the same color bar.}
\label{fig:MCDR_Jupiter}.
\end{figure}

The mean (averaged) flows for the Jupiter benchmark are
shown in Fig.\ \ref{fig:MCDR_Jupiter}.  All averages span 2 days (about
4.8 rotation periods), starting at the sampling time $T_{s}$.

The meridional circulation is expressed in terms of a stream function, $\Psi$, defined as
\begin{eqnarray}
r\sin \theta \langle \overline{\rho}V_{r} \rangle= -\frac{1}{r}\frac{\partial\Psi}{\partial
\theta}, \quad \mbox{and} \quad r \sin \theta\langle \overline{\rho}V_{\theta} \rangle = \frac{\partial \Psi}{\partial r} ~ ,
\end{eqnarray}
and the differential rotation is expressed in terms of the angular velocity
\begin{equation}
\Omega = \frac{1}{2\pi}(\frac{\langle V_{\phi} \rangle}{r\sin \theta} + \Omega_{o}).
\end{equation}
We define thermal variations $S^{'}$ and $T^{'}$ by averaging over longitude and time and then subtracting the spherically-symmetric component ($\ell = m = 0$) in order to highlight variations relative to the mean stratification.

The CHORUS and ASH results in Fig.\ \ref{fig:MCDR_Jupiter} correspond closely with a few notable exceptions.  Near the equator in the upper convection zone, $\Omega$ in Fig.\ref{fig:MCDR_Jupiter}(\textit{d}) is somewhat smaller than that in Fig.\ref{fig:MCDR_Jupiter}(\textit{h}).  This is also reflected by the lower DRKE/KE noted in sec.\ \ref{sec:growth}.  As mentioned there, this discrepancy may in part be because the simulations are not strictly in equilibrium at the sampling time (Fig.\ \ref{fig:flux_jupiter}).  We would expect the correspondence to improve if we were to run CHORUS for several thousand days, giving the mean flows ample time to equilibrate along with the stratification.  The wiggles in the $S^\prime$ plot for CHORUS (Fig.\ \ref{fig:MCDR_Jupiter}(\textit{b})) can be attributed to two factors.  First, unlike ASH, the relevant state variable in CHORUS is the total energy $E$.  The specific entropy must be obtained from $E$ by subtracting out contributions from the kinetic energy and the mean stratification to obtain the pressure variations, from which $S^{'}$ is computed (using also the density variation).  This residual nature of $S^\prime$ along with the post-processing step of interpolating the CHORUS results onto a structured, spherical grid both contribute numerical errors.  In ASH, by contrast, $S$ is a state variable.  The correspondence of Figs.\ \ref{fig:MCDR_Jupiter}(\textit{b}) and (\textit{f}) despite these numerical errors is a testament to the accuracy of CHORUS.

\subsection{Solar Benchmark}\label{sec:solarbm}

Having gained confidence in simulating the Jupiter benchmark, we now look into
defining a benchmark that has more in common with the Sun.  This serves
two purposes.  First, it helps verify the CHORUS code by probing a different
region in parameter space, this one more turbulent.  Second, it makes explicit
contact with solar and stellar convection which is the primary application we
had in mind when developing CHORUS.

The most significant differences between the solar and the Jupiter benchmarks include the Rayleigh number $R_a$, the density stratification $N_\rho$, and the radiative heat flux $F_r$.  The value of $R_a$ is about four times larger in the solar benchmark.  This, combined with the larger value of $E_k$ implies that the flow is more turbulent and less rotationally constrained (larger Rossby number). In the middle of the convection zone, the Rossby number $\sim 1.56\times 10^{-2}$ for the Sun benchmark and it is larger than the Jupiter benchmark $\sim 3.81\times 10^{-4}$. As in the Sun, the radiative flux $F_r$ (which was set to zero in the Jupiter benchmark) carries energy through the bottom boundary, extending into the lower convection zone (see Fig.\ \ref{fig:flux_Sun}).  This allows us to set the radial entropy gradient $\partial S/\partial r$ to zero at the lower boundary, which is what marks the base of the convection zone in the Sun.  Another significant difference is the aspect ratio $\beta$, for which we use a solar-like value of $0.73$.

As mentioned in secs.\ \ref{sec:intro} and \ref{sec:ash}, a challenge in simulating solar convection
with a compressible code like CHORUS is the low value of the Mach number.  We address this challenge by scaling up the luminosity $L$ by a factor of 1000 relative to the actual luminosity of the Sun.  According to the mixing-length theory of convection, the rms velocity should scale as $L^{1/3}$.  Thus, to preserve a solar-like value of the Rossby number (important for achieving reasonable mean flows), this suggests that we would need to scale up the rotation rate $\Omega_0$ by a factor of approximately 10 relative to the actual Sun. We then chose values of of $\kappa$ and $\nu$ to give practical values for $R_{a}$ and $E_{k}$.  Table \ref{tab:parameter_Sun} gives the parameters for the solar benchmark.

The radiative flux $F_{r}$ is parameterized by expressing the radiative diffusion as $\kappa_{r} = \lambda (c_{0} + c_{1}\omega + c_{2}\omega^{2})$, where $c_{0} = 1.5600975\times 10^{8}$, $c_{1} = -4.5631718\times 10^{7}$, $c_{2} = 3.3370368 \times 10^{6}$, and $\omega = r \times 10^{-10}$. The parameter $\lambda$ is chosen so that $F_{r} = L/(4\pi r^{2})$ on the bottom boundary. As for the Jupiter benchmark, we define a sampling time $T_{s}$ that is 10 days after the nonlinear saturation time.  The sampling times are $T_{s}=15.10$ days and $T_{s}=14.59$ days for CHORUS and ASH respectively.  The number of DOFs used for the CHORUS simulation is about a factor of 6.8 larger than that used for the ASH simulation, as indicated in Table \ref{tab:computational_effort}.

\begin{table}
    \begin{tabular}{l}
    \toprule
    Dimensionless parameters \\
    $E_{k}$ = 2.447 $\times$ $10^{-3}$, $N_{\rho}$ = 3, $\beta$ = 0.736762, $R_{a}$ = 1,428,567, $Pr$ = 1, $n$ = 1.5 \\ \\
    Defining physical input values \\
    $R_{o}$ = 6.61 $\times$ $10^{10}$ $cm$, $\Omega_{o}$ = 8.1 $\times$ $10^{-5}$ $s^{-1}$, $M$ = 1.98891 $\times$ $10^{33}$ $g$, $\rho_{i}$ = 0.21 $g$ $cm^{-3}$ \\
    $\mathfrak{R}$ = 1.4 $\times$ $10^{8}$ $erg$ $g^{-1}$ $K^{-1}$, $G$ = 6.67 $\times$ $10^{-8}$ $g^{-1}$ $cm^{3}$ $s^{-2}$ \\
    \\
    Derived physical input values \\
    $R_{i}$ = 4.87 $\times$ $10^{10}$ $cm$, $d$ = 1.74 $\times$ $10^{10}$ $cm$, $\nu$=6.0 $\times$ $10^{13}$ $cm^{2}$\ $s^{-1}$,\\ $\kappa$ = 6.0 $\times$ $10^{13}$ $cm^{2}$ $s^{-1}$, $\gamma$ = 5/3  \\
    \\
    Other thermodynamic quantities \\
    $L$ = 3.846 $\times$ $10^{36}$ $erg$ $s^{-1}$, $C_{p}$ = 3.5 $\times$ $10^{8}$ $erg$ $g^{-1}$ $K^{-1}$  \\
    \bottomrule
    \end{tabular}
    \caption{Parameters for the solar benchmark}
    \label{tab:parameter_Sun}
\end{table}

\subsubsection{Exponential Growth and Nonlinear Saturation}\label{sec:growthsun}
\begin{figure}[!htp]
\centering
\subfigure[]{\label{fig:KE_density_Sun}\includegraphics[width=0.46 \textwidth, angle=0]{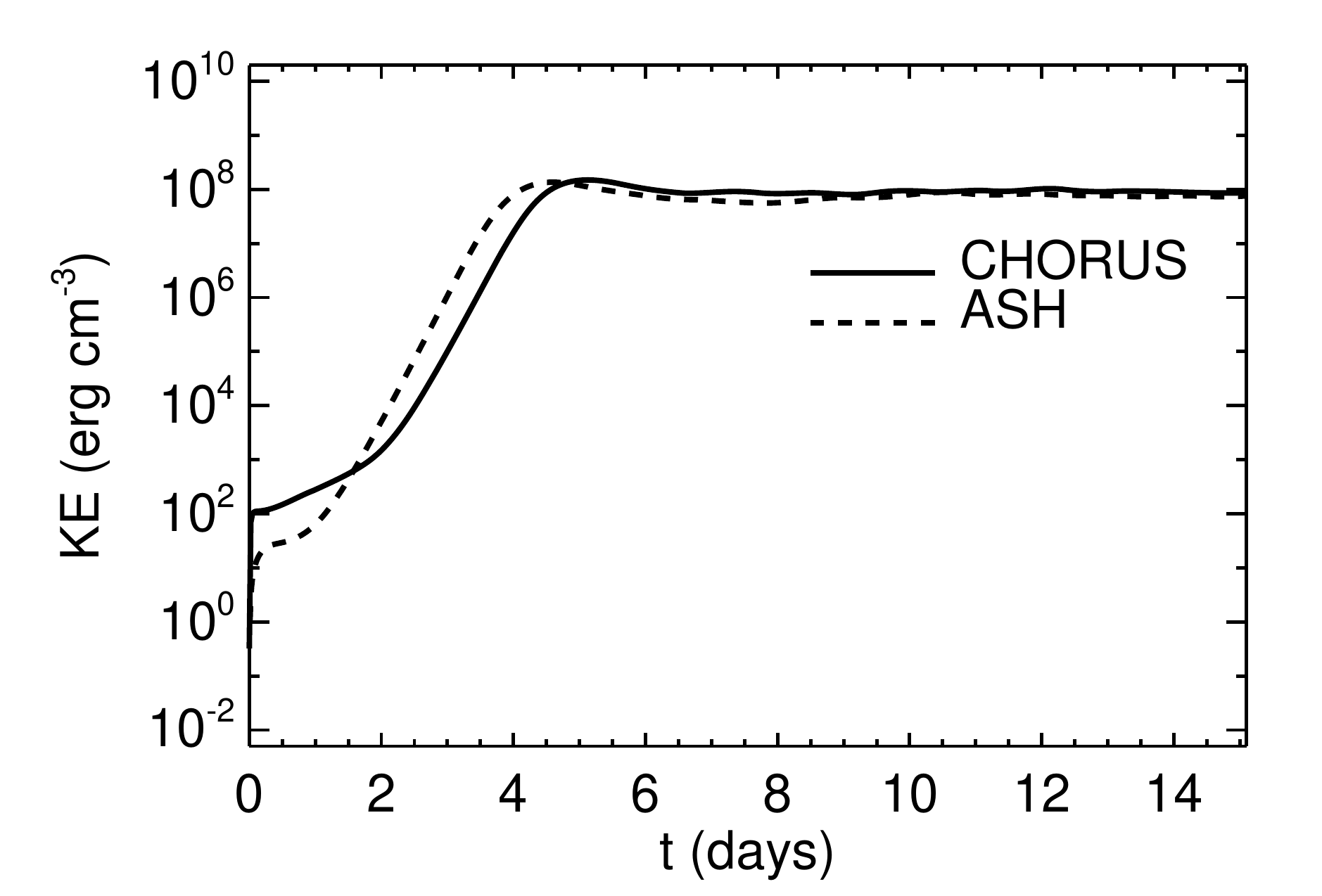}}
\subfigure[]{\label{fig:Growth_Sun}\includegraphics[width=0.47 \textwidth, angle=0]{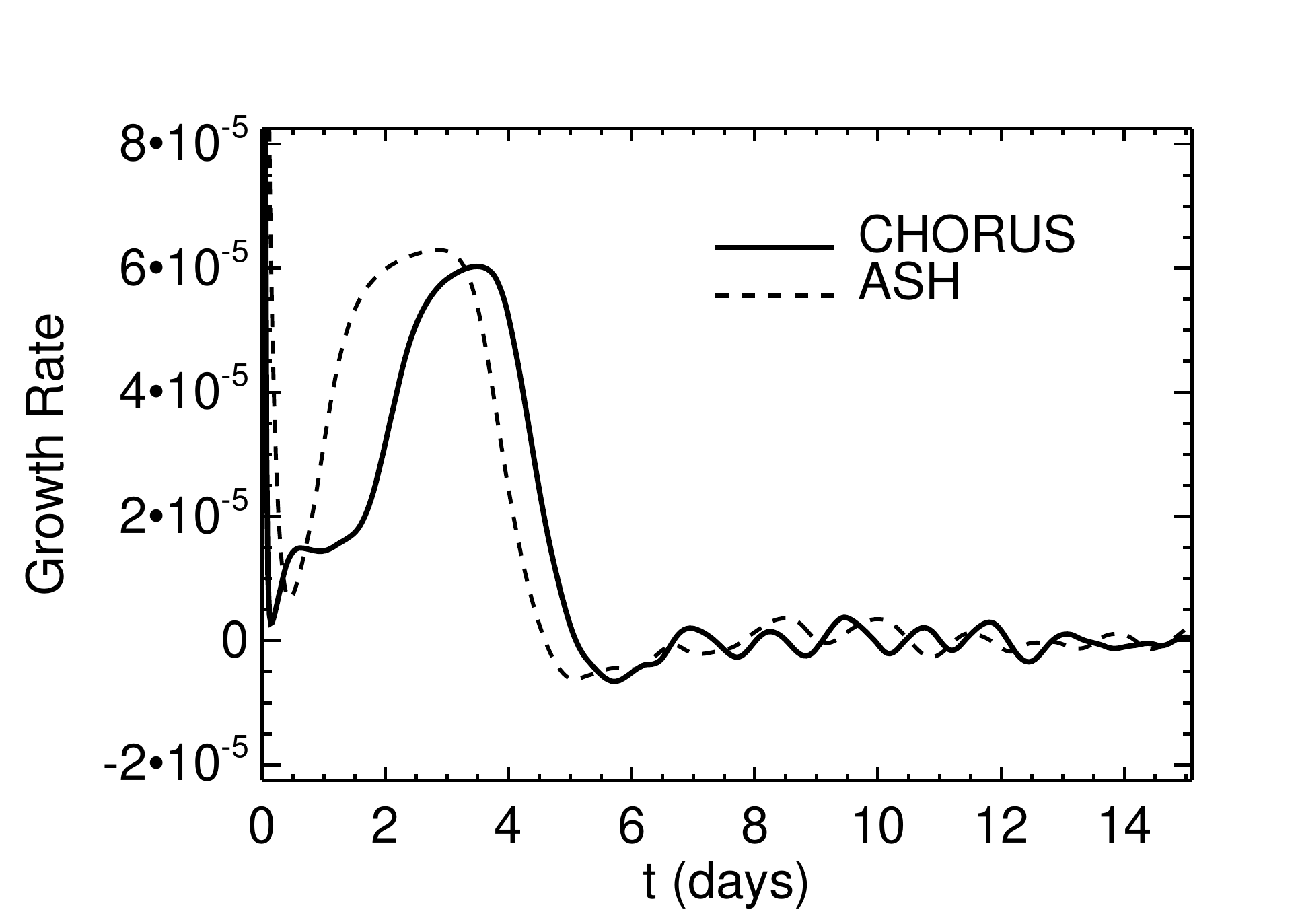}}
\caption{(\textit{a}) Kinetic energy density and (\textit{b}) growth rate for the solar benchmark.}
\label{fig:Sun}.
\end{figure}

The volume-averaged
kinetic energy densities for both the CHORUS and ASH simulations are plotted in Fig.\ref{fig:Sun}(a).
As in the Jupiter benchmark, the random initial conditions lead to differences in the initial transients but once the preferred linear eigenmode begins to grow the two simulations exhibit similar growth rates and nonlinear saturation levels. Averaged between ($T_{s}-1$) and $T_{s}$ days, $KE=9.02 \times 10^{8}$ $erg$ $cm^{-3}$ for the CHORUS simulation and $KE=8.58\times 10^{8}$ $erg$ $cm^{-3}$ for the ASH simulation. The difference is about $5.13\%$. In the linear regime, they grow very fast and the ranges of a well-defined linear unstable regime are narrow as shown in Fig.\ref{fig:Sun}(b) for both CHORUS and ASH simulations. Thus, only their maximum growth rates in that regime are compared. A difference of only $4\%$ is present between $\sigma_{max} = 6.02\times 10^{-5}$ of the CHORUS simulation and $\sigma_{max} = 6.29\times 10^{-5}$ of the ASH simulation. At the sampling time $T_{s}$, DRKE/KE$=4.66\times 10^{-1}$ and MCKE/KE$=9.02\times 10^{-4}$ for the CHORUS simulation while DRKE/KE$=5.08\times 10^{-1}$ and MCKE/KE$=1.07\times 10^{-3}$ for the ASH simulation (a discrepancy of about 9\% and 19\% respectively).  As in the Jupiter benchmark, this relatively large difference in mean flows is likely attributed to the immature state of the simulations, which has not yet achieved flux balance by the sampling time (sec.\ \ref{sec:machsun}).

\subsubsection{Mach Number, $\epsilon$ and Flux Balance}\label{sec:machsun}
\begin{figure}[!htp]
\centering
\subfigure{\label{fig:Ma_SUN}\includegraphics[width=0.49 \textwidth, angle=0]{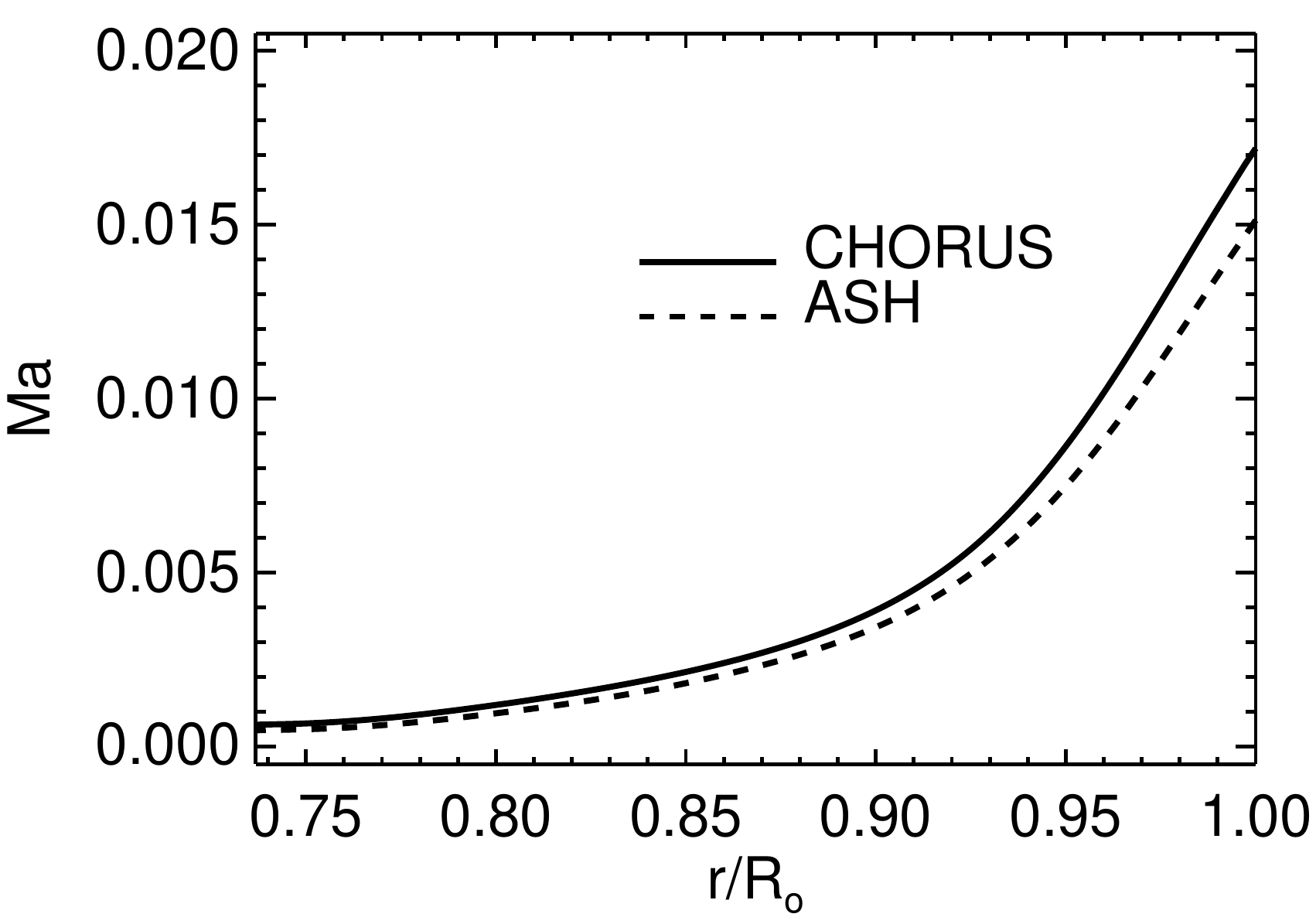}}
\subfigure{\label{fig:epsilon_SUN}\includegraphics[width=0.49 \textwidth, angle=0]{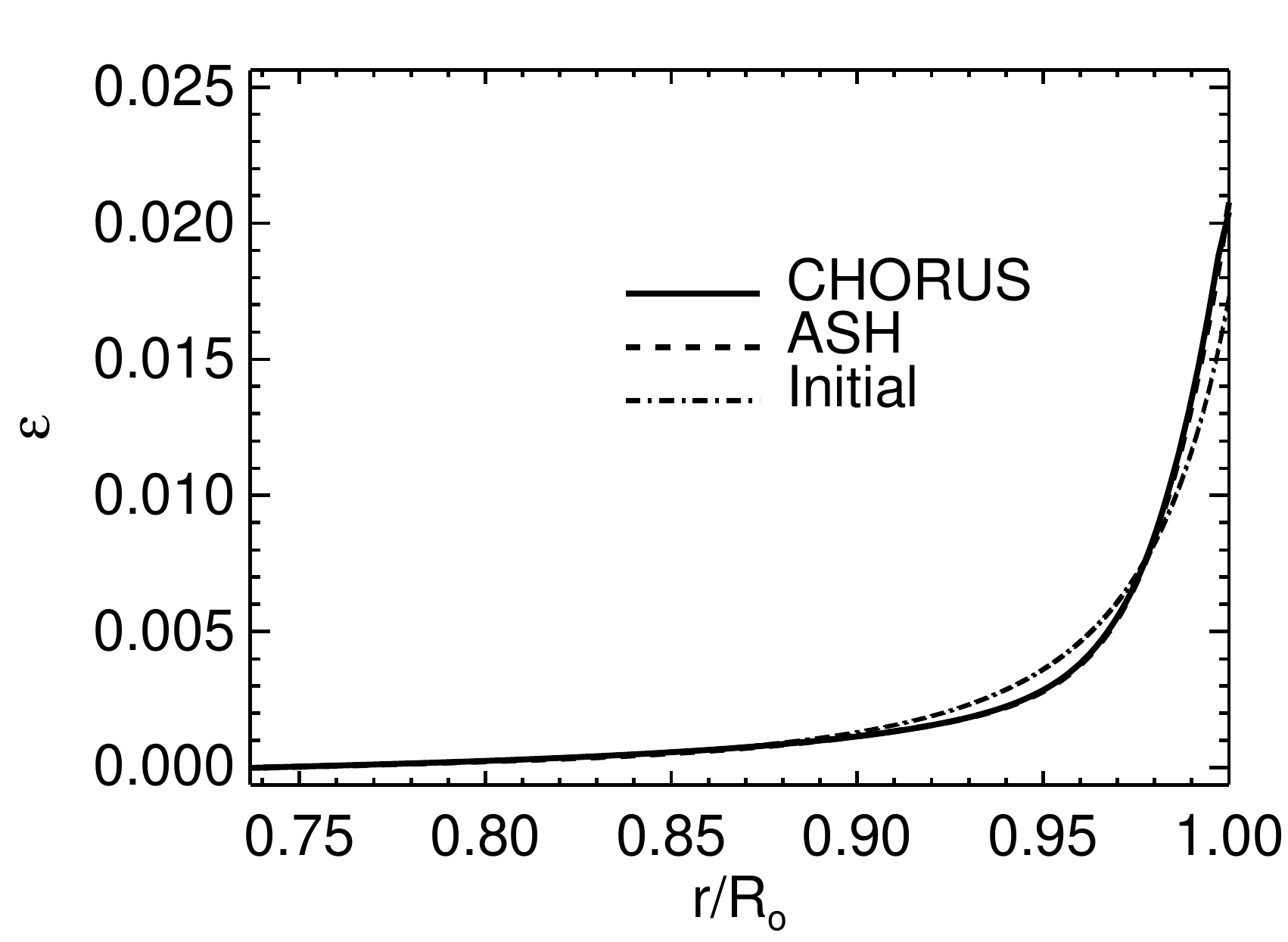}}
\caption{(\textit{a}) Mach number and (\textit{b}) normalized entropy gradient $\epsilon$ for the solar benchmark at the sampling time $T_s$.  As in Fig.\ \ref{fig:epsilon_JUPITER}, the initial profile is included in frame (\textit{b}), where the CHORUS and ASH curves at $T_s$ are slightly steeper and indistinguishable from one another.}
\label{fig:MaEpsilon}.
\end{figure}

As in the Jupiter benchmark, a low Mach number is also achieved in this solar benchmark, with the maximum Mach number around 0.017 and 0.015 for the CHORUS simulation and the ASH simulation, respectively, as shown in Fig.\ref{fig:Ma_SUN}. The value of $\epsilon$ also peaks in the upper convection zone (Fig.\ \ref{fig:epsilon_SUN}).
Reflecting the flatter entropy gradient at the sampling time $T_{s}$, $\epsilon$ becomes smaller where the convective efficiency is largest, near $r=0.95R_{o}$.

\begin{figure}[!htp]
\centering
\subfigure[CHORUS]{\includegraphics[width=0.49 \textwidth, angle=0]{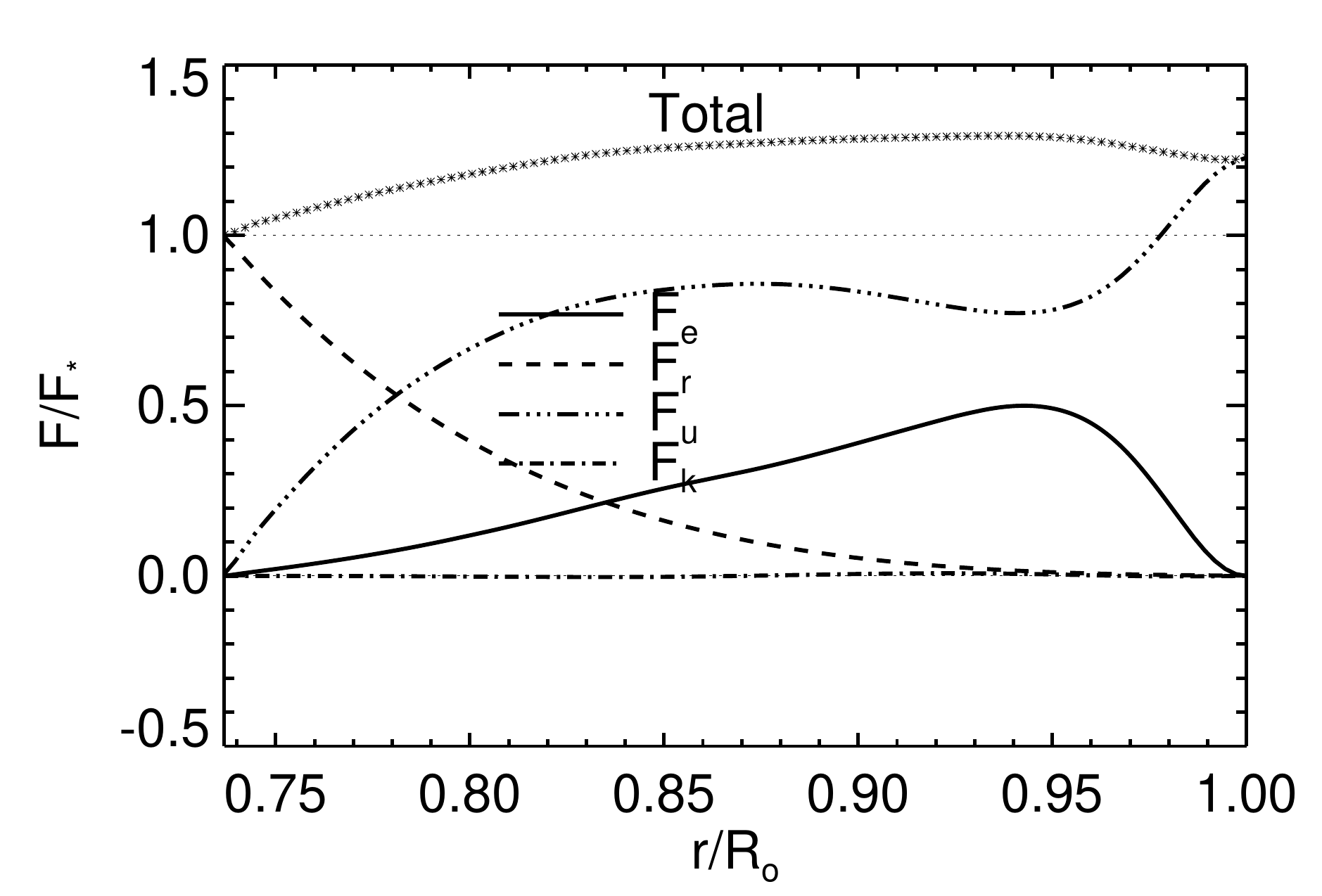}}
\subfigure[ASH]{\includegraphics[width=0.49 \textwidth, angle=0]{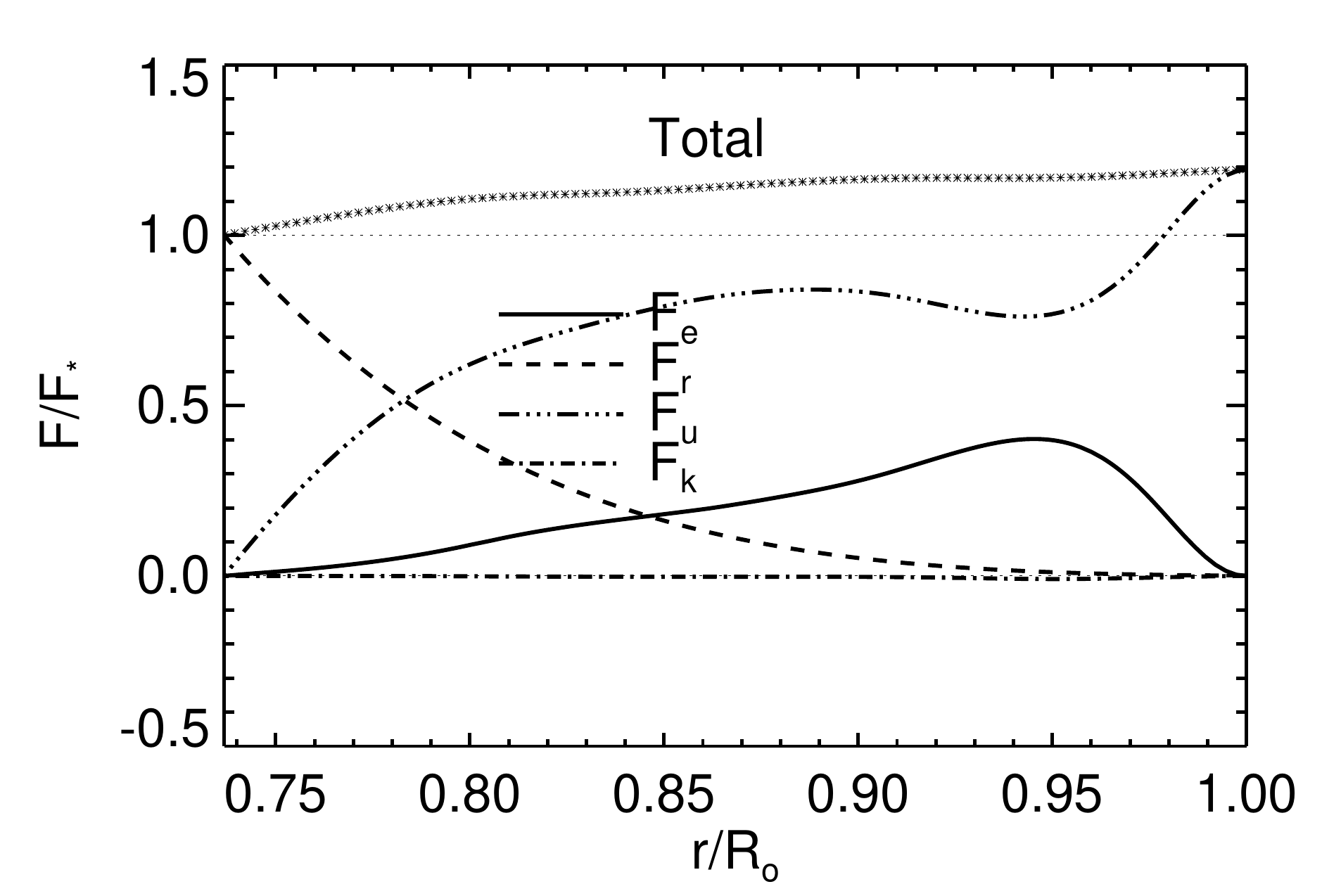}}
\caption{Components of the normalized, horizontally-integrated radial energy flux as in Fig.\ \ref{fig:flux_jupiter} but for the solar benchmark.  Each snapshot corresponds to the sampling time $T_s$.}
\label{fig:flux_Sun}.
\end{figure}

The CHORUS and ASH simulations also have similar flux balances as shown in Fig.\ref{fig:flux_Sun}. Both are over-luminous ($F_e + F_k + F_r + F_u > F_*$) due mainly to the large diffusive and entropy fluxes $F_u$ and $F_e$, which have not yet equilibrated.  This imbalance subsides by $t \sim 540$ days as verified by an extended ASH simulation.  However, as with the Jupiter benchmark, we have not run CHORUS to full equilibration because of the computational expense.  Increasing the luminosity further will mitigate this relaxation time and we plan to exploit this for future production runs.

The radiative flux $F_{r}$ carries energy through the bottom boundary and dominates the heat transport in the lower convection zone while the entropy flux $F_{u}$ carries energy through the top boundary and the upper convection zone. The enthalpy flux $F_{e}$ gradually increases towards the top until peaks near $r =  0.95R_{o}$, and then drops down to zero rapidly as it approaches the impenetrable top boundary. At $r=0.95R_{o}$, $F_{e}=0.50$
and $F_{u}=0.78$ from the CHORUS simulation while $F_{e}=0.42$ and $F_{u}=0.76$ from the ASH simulation.

\subsubsection{Convection Structure}

\begin{figure}[!htp]
\centering
\subfigure[]{\includegraphics[width=0.95 \textwidth, angle=0]{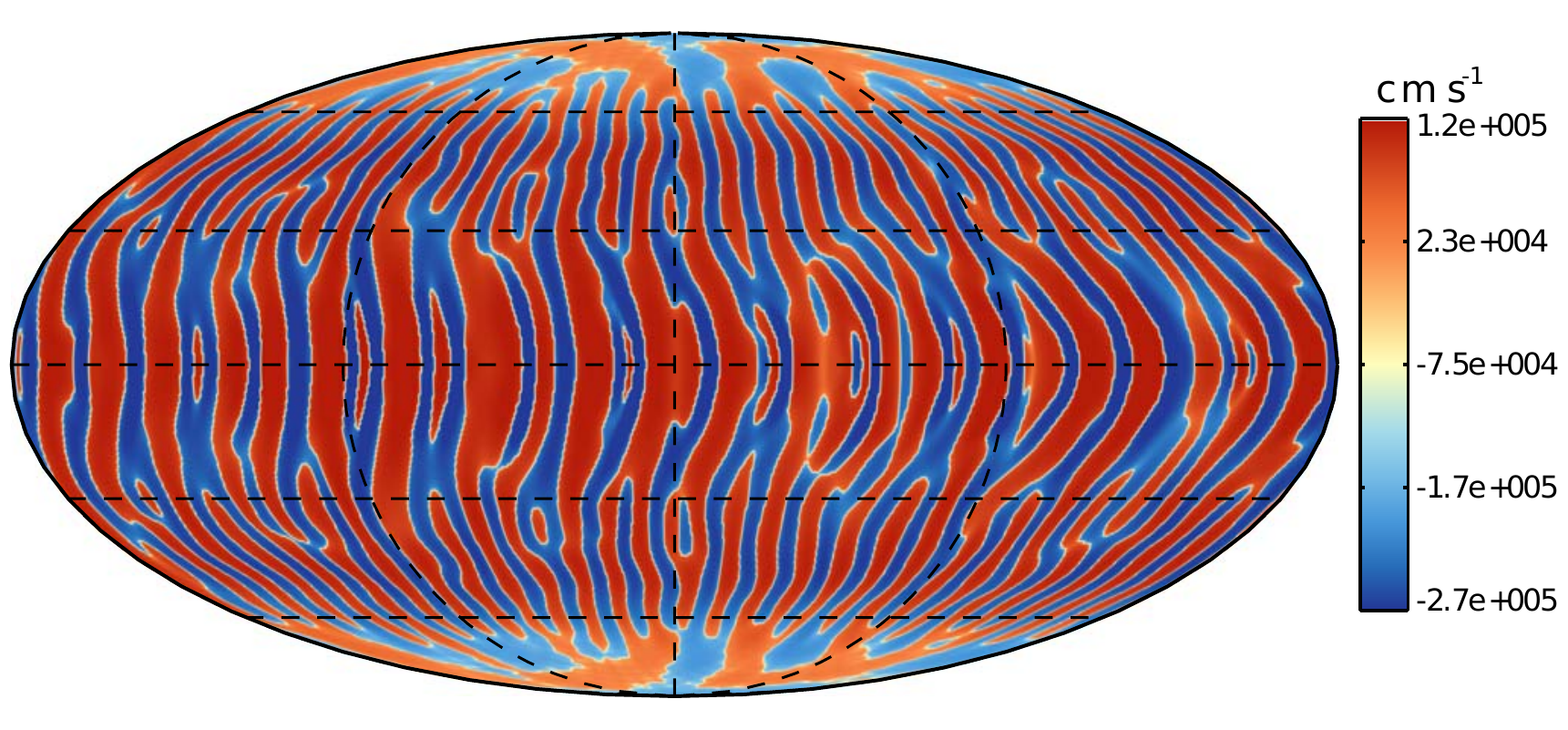}}
\subfigure[]{\includegraphics[width=0.95 \textwidth, angle=0]{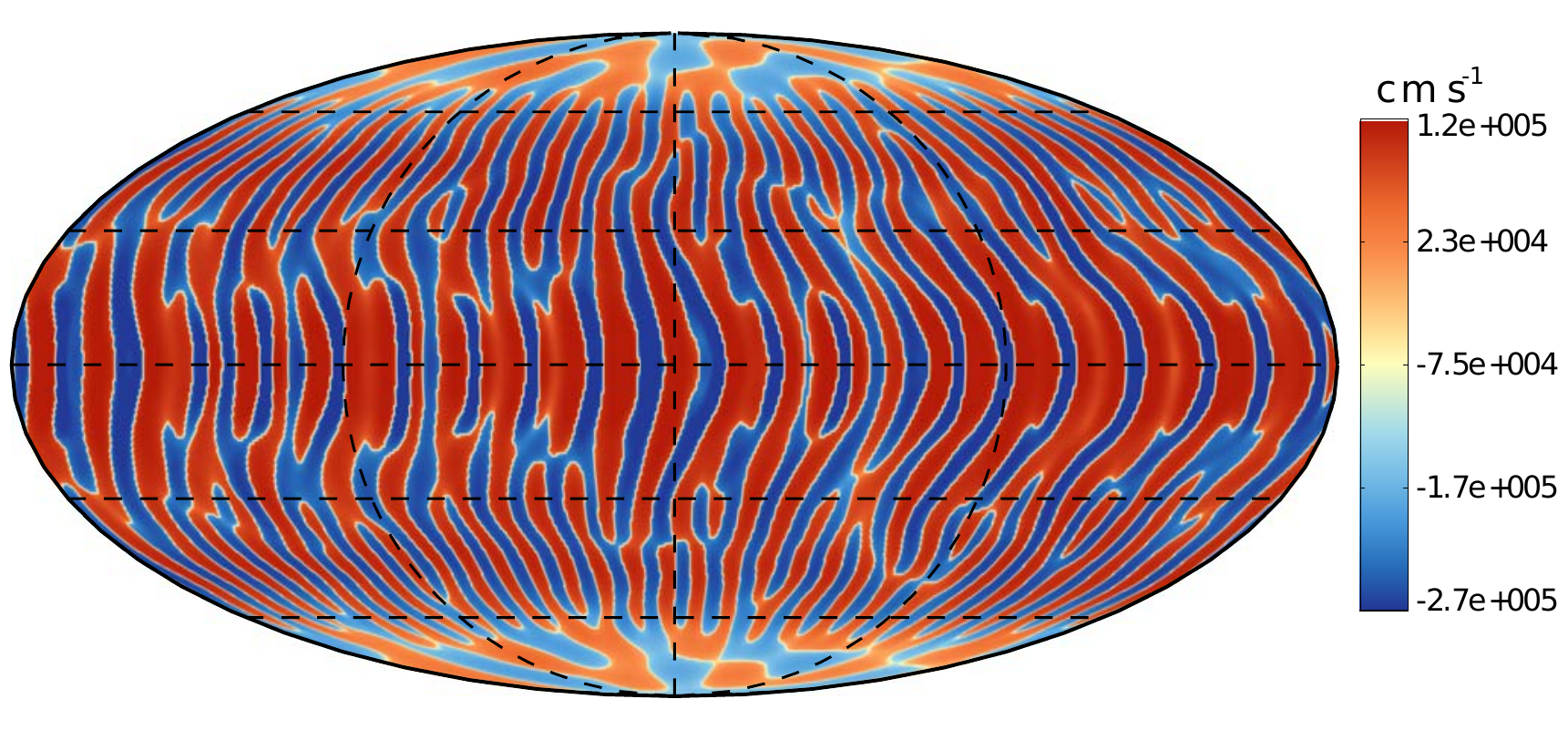}}
\caption{Mollweide projections of the radial velocity $V_{r}$ in the solar benchmark at $r=0.95R_{o}$ and $t = T_s$ for (\textit{a}) CHORUS and (\textit{b}) ASH.  Red and blue tones denote the upflow and downflow as indicated by the color bar.}
\label{fig:Vr_Sun}
\end{figure}

\begin{figure}[!htp]
\centering
\subfigure[]{\label{fig:sun_spectrum_a}\includegraphics[width=0.323 \textwidth, angle=0]{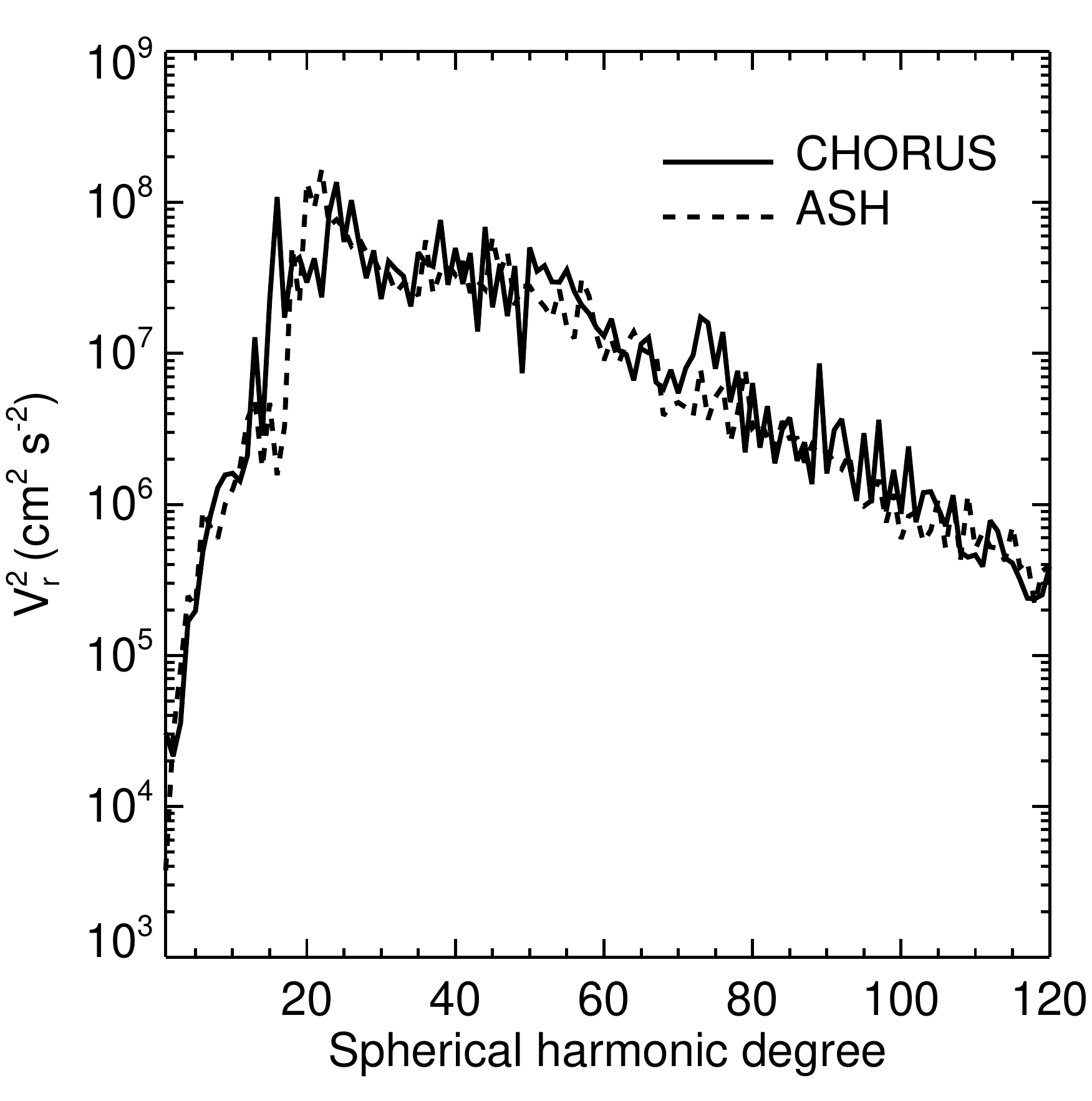}}
\subfigure[]{\label{fig:sun_spectrum_b} \includegraphics[width=0.323 \textwidth, angle=0]{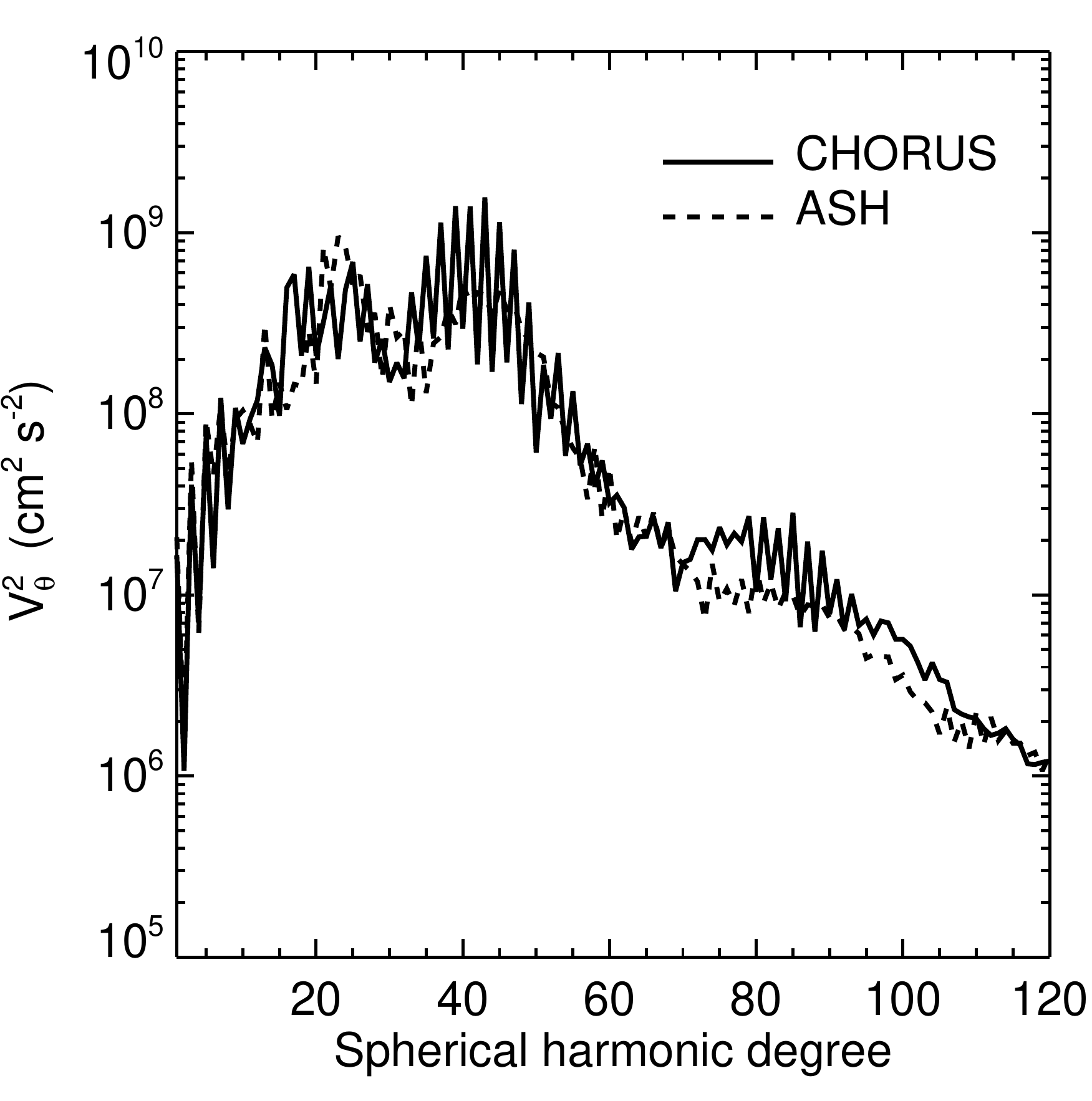}}
\subfigure[]{\label{fig:sun_spectrum_c} \includegraphics[width=0.323 \textwidth, angle=0]{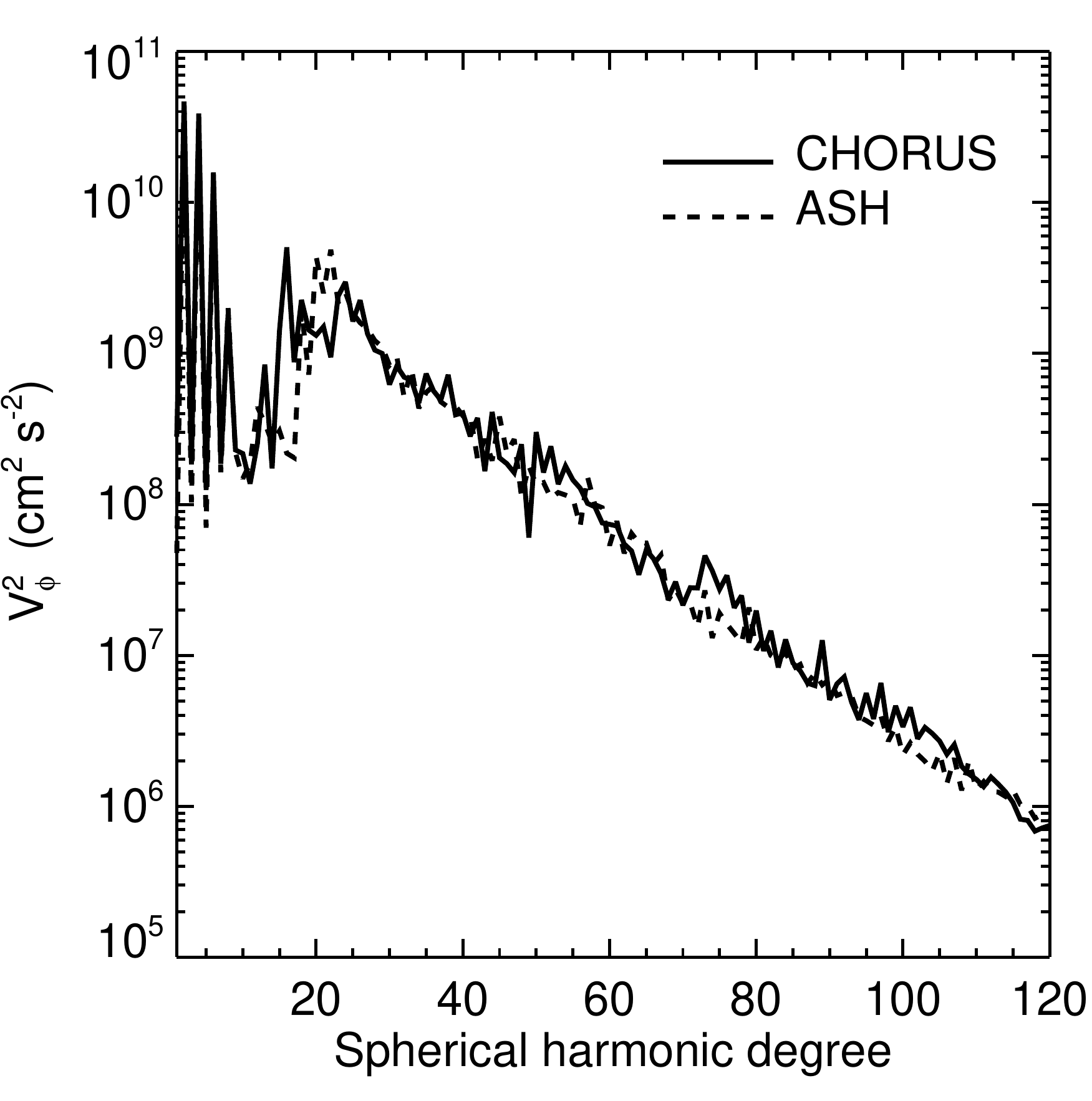}}
\caption{Power spectra of (a) radial velocity $V_{r}$, (b) meridional velocity $V_{\theta}$, and
(c) zonal velocity $V_{\phi}$ as in Fig.\ \ref{fig:Spectrum_Jupiter} but for the solar benchmark at $r = 0.98 R_o$ and $t = T_s$.}
\label{fig:spectrum_Sun}
\end{figure}

The structure of the convection in the solar benchmark is illustrated in Fig.\ref{fig:Vr_Sun}.  One would not expect the instantaneous flow field in a highly nonlinear simulation to correspond exactly between two independent realizations but the qualitative agreement is promising. This qualitative agreement is confirmed quantitatively by comparing the velocity spectra in Fig.\ref{fig:spectrum_Sun}.  The radial velocity spectrum peaks at $\ell = 15-25$ (Fig.\ref{fig:spectrum_Sun}(\textit{a})) and the meridional velocity spectrum exhibits substantial power in two wavenumber bands, namely $\ell = 15-25$ and $\ell = 35-45$ (Fig.\ref{fig:spectrum_Sun}(\textit{b})). For the zonal velocity spectra (Fig.\ref{fig:spectrum_Sun}(\textit{c})), three modes $\ell =2, 4$, and $6$ are prominent in the low $\ell$ ($<10$) range. After peaking at $\ell = 15-25$, the high-$\ell$ power decreases exponentially in amplitude.  Some of the small discrepancies between the two curves in each plot can likely be attributed to random temporal variations that would cancel out with some temporal averaging.

\subsubsection{Mean Flows}
\begin{figure}[!htp]
\centering
\subfigure{\includegraphics[width=0.99 \textwidth, angle=0]{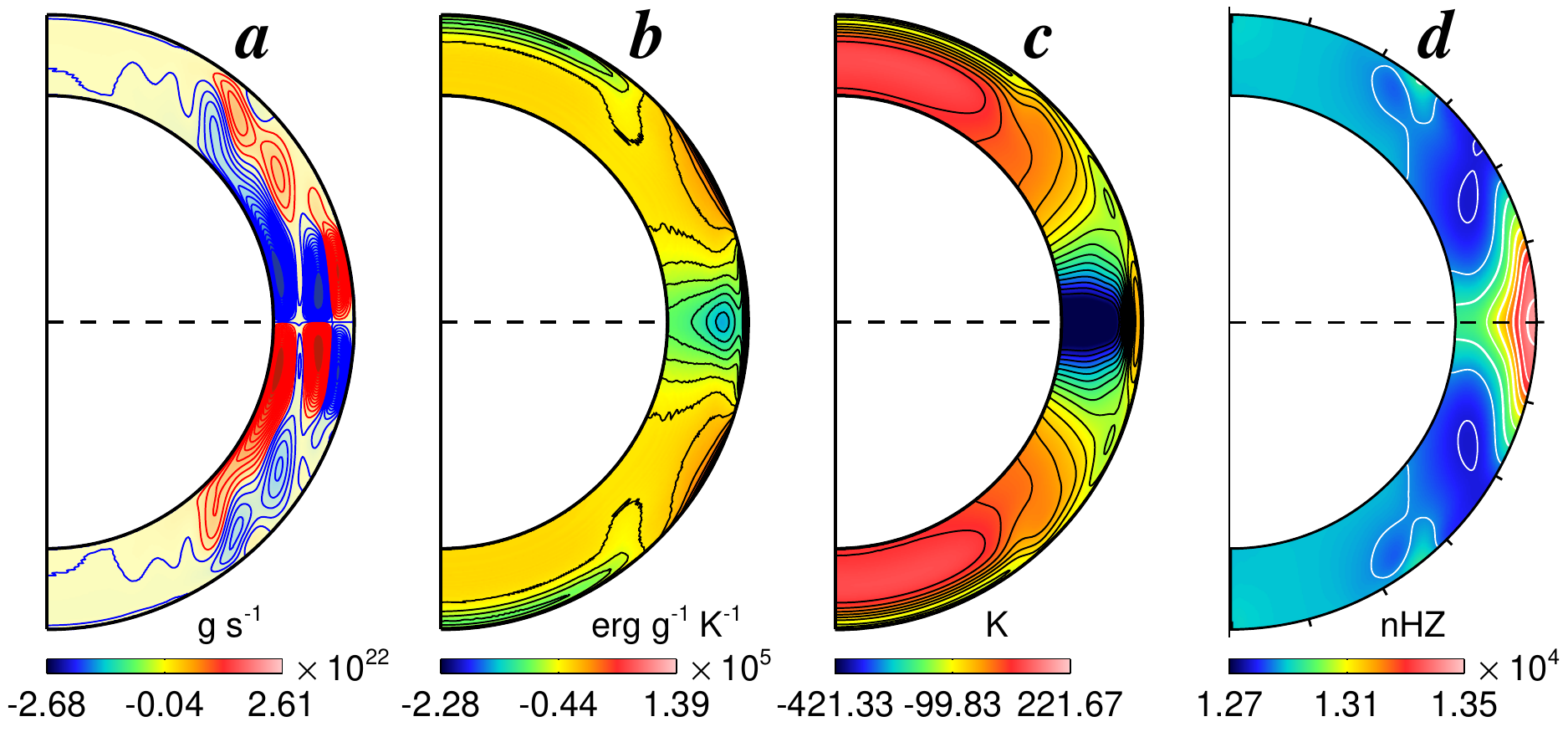}}
\subfigure{\includegraphics[width=0.99 \textwidth, angle=0]{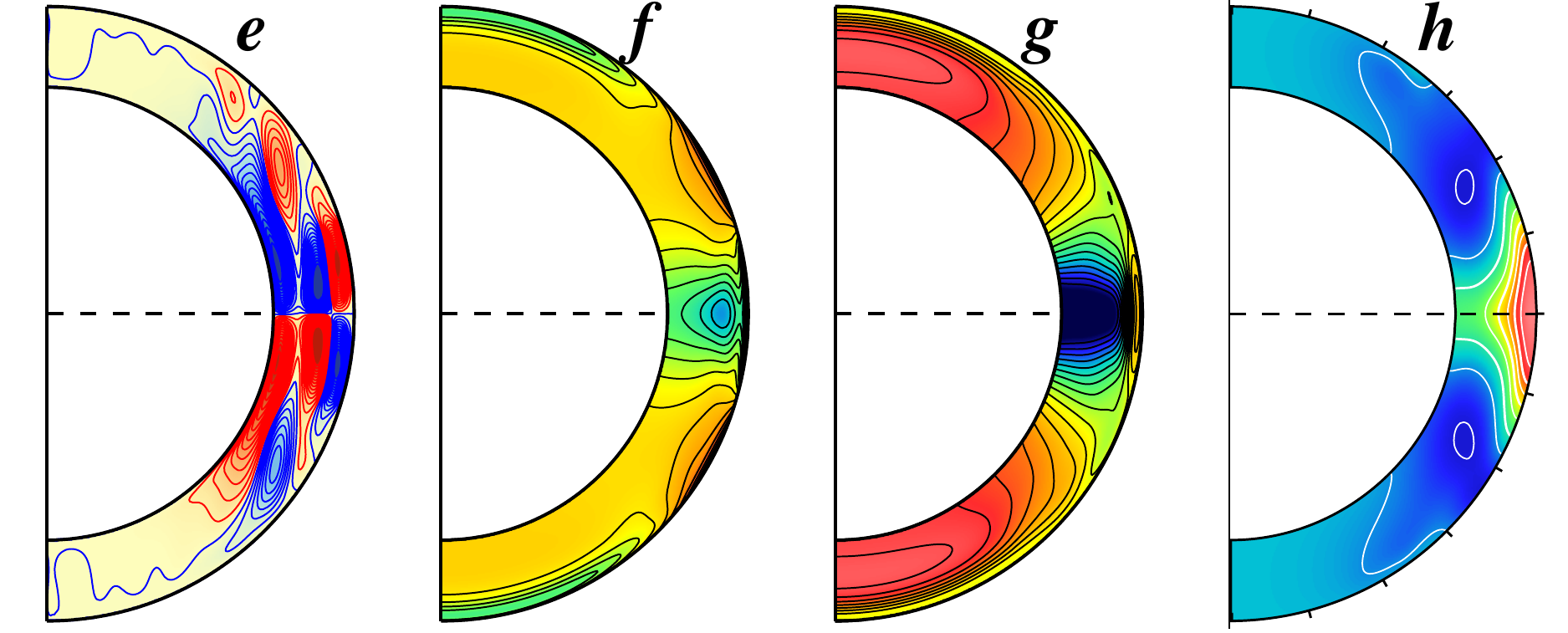}}
\caption{Mean flows as in Fig.\ref{fig:MCDR_Jupiter} but for the solar benchmark.  Top and bottom rows correspond to CHORUS and ASH respectively and all quantities are averaged over longitude and over a two-day time interval beginning at $T_s$ (spanning 2.2 rotation periods). (\textit{a}, \textit{e}) mass flux stream function $\Psi$ with red and blue denoting clockwise and counterclockwise circulations respectively, (\textit{b}, \textit{f}) specific entropy perturbation $S^{'}$, (\textit{c}, \textit{g}) temperature perturbation $T^{'}$, and (\textit{d}, \textit{h}) angular velocity profile. The top and bottom plots in each column share the same color bar.
}
\label{fig:MCDR_Sun}.
\end{figure}

The meridional circulation from the CHORUS simulation (see Fig.\ref{fig:MCDR_Sun}(a)) and the ASH simulation (see Fig.\ref{fig:MCDR_Sun}(e)) exhibit similar flow patterns with most circulations concentrating between the low latitudes and middle latitudes, outside the so-called {\em tangent cylinder}, namely the cylindrical surface aligned with the rotation axis and tangent to the base of the convection zone. The specific entropy perturbation $S^{'}$ are shown in Fig.\ref{fig:MCDR_Sun}(b) for the CHORUS simulation and in Fig.\ref{fig:MCDR_Sun}(f) for the ASH simulation. In both simulations, the contours of $S^{'}$ are symmetric about the equator. By comparing the contours of $T^{'}$ in Fig.\ref{fig:MCDR_Sun}(c) for the CHORUS simulation and Fig.\ref{fig:MCDR_Sun}(g) for the ASH simulation, a good agreement is achieved. For both simulations, they also have similar differential rotation profiles as shown in Fig.\ref{fig:MCDR_Sun}(d) and Fig.\ref{fig:MCDR_Sun}(h). Some of the small discrepancies can likely be attributed to the flux imbalances shown in Fig.\ref{fig:flux_Sun}, causing mean flows to vary slowly as the simulations equilibrate from different random initial conditions and nonlinear saturation states.  Residual random temporal fluctuations may also be present despite the (short) time average, particularly for the meridional circulation which is a relatively weak flow with large fluctuations \citep{miesc05}.

\subsection{Code Performance}\label{sec:performance}

As demonstrated in sec.\ \ref{sec:algorithm}, the CHORUS code achieves excellent scalability out to 12k cores.  This is strong scaling for intermediate-resolution simulations.  We expect higher-resolution simulations to scale even better to tens of thousands of cores.  In this section we consider CHORUS's performance relative to the anelastic code ASH.

The computational efforts are summarized in Table \ref{tab:computational_effort}. The resolution in ASH is expressed as $N_r \times N_\theta \times N_\phi$ where $N_r$, $N_\theta$, and $N_\phi$ are the number of grid points in the radial, latitudinal, and longitudinal directions respectively.  However, due to pseudo-spectral de-aliasing and the symmtery of the spherical harmonics, the effective number of DOFs for ASH is $N_r (\ell_{max} + 1) \ell_{max}$, where $\ell_{max} = (2 N_\theta - 1)/3$ is the maximum spherical harmonic mode.  For both benchmarks, $N_\theta = 256$ and $\ell_{max} = 170$.

The total number of core hours needed to run 10 days is much larger for CHORUS than it is for ASH; by more than three orders of magnitude for the Jupiter benchmark and by a factor of 75 for the solar benchmark.  Much of this is due to the smaller time step required by the compressible scheme and the higher number of degrees of freedom used in running the CHORUS benchmarks.  Furthermore, the ASH simulations were run on only 71 cores whereas the CHORUS runs typically employed several thousand.Thus, imperfect scaling is a factor, as is a difference in the computational platform.  The ASH simulations were run with the Intel Xeon E5-2680v2 (Ivy Bridge) cores on NASA's Plieades machine (2.8 GHz clock speed, 3.2GB/core memory) whereas the CHORUS simulations were run with the Intel Xeon E5-2670 (Sandy Bridge) cores on NCAR's Yellowstone machine (2.6 GHz clock speed, 2 GB/core memory).  Furthermore, CHORUS uses a fourth-order accurate five-stage explicit Runge-Kutta method \citep{spiteri-02} whereas ASH uses a simpler second-order mixed Adams-Bashforth/Crank-Nicolson time stepping.  This also contributes to the larger number of core hours per time step used by CHORUS (Table \ref{tab:computational_effort}).

Though ASH out-performs CHORUS for these simple benchmark problems, it must be remembered that such problems are ideal for pseudo-spectral codes; relatively low resolution, laminar runs dominated by a limited number of spherical harmonic modes.  The real potential of CHORUS will be realized for high-resolution, turbulent, multi-scale convection where its superior scalability and variable mesh refinement will prove invaluable.  It can also be used for studying physical phenomena such as core convection and oblate stars that are challenging or even inaccessible to codes that use structured, spherical grids.  Furthermore, there is much potential for improvement in the efficiency of CHORUS; we intend to implement an implicit time marching scheme and a p-multigrid method \citep{LiangKannan} as well as local time stepping in the future, and to optimize the numerical algorithm for higher performance on heterogeneous (CPU/GPU) architectures that are very suitable for data structures of the SDM.

\begin{table}
    \begin{tabular}{llllc}
    \toprule
     &\multicolumn{2}{c}{Jupiter Case} & \multicolumn{2}{c}{Sun Case} \\
     \hline
     code & CHORUS & ASH & CHORUS & ASH \\
    \hline
    Resolution              & 294,912 elements       & $129\times 256\times 512$ & 307,200 elements & $100\times 256\times 512$ \\
    DOFs                    & 18,874,368          & 3,750,030 & 19,660,800 & 2,907,000 \\
    Time step (s)            & 1.5                 & 533   & 4 & 20    \\ \\
    Core hours per time step &  $7.53\times 10^{-2}$ & $7.03\times 10^{-3}$& $8.03\times 10^{-2}$  & $5.32\times 10^{-3}$ \\ \\
    Number of iteration                                   \\
    required to run 10 days    &  576,000   & 1,621 & 216,000 & 43,200  \\ \\
    Number of core hours                                    \\
    needed to run 10 days      &  43,380     & 11.4  & 17,352 & 230 \\
    \bottomrule
    \end{tabular}
    \caption{Computational efforts for the Jupiter and solar benchmarks.}
    \label{tab:computational_effort}
\end{table}

\section{Summary}

We have developed a novel high-order spectral difference code, CHORUS, to simulate stellar and planetary convection in global spherical geometries.  To our knowledge, the CHORUS code is the first stellar convection code that employs an unstructured grid, giving it unique potential to simulate challenging physical phenomena such as core convection in high and low-mass stars, oblate distortions of rapidly-rotating stars, and multi-scale, hierarchical convection in solar-like stars.

The CHORUS code is fully compressible, which gives it advantages and disadvantages over codes that employ the anelastic approximation.  On the one hand, the hyperbolic nature of the compressible equations promotes more efficient parallel scalability over the (elliptical) anelastic equations.  Indeed, we demonstrated that the CHORUS code does achieve excellent strong scalability for intermediate-size problems extending to 12,000 cores.  We expect even better scalability for higher-resolution problems. Furthermore, the fully compressible equations are required to accurately capture the small-scale surface convection in solar-like and less massive stars where Mach numbers approach unity and where the anelastic approximation breaks down.  On the other hand, the CFL constraint imposed by acoustic waves places strict limits on the allowable time step for simulating deep convection in most stars and planets, where the Mach number is much less than unity.  We intend to address this constraint in the future by implementing implicit and local time stepping schemes.

To verify the CHORUS code, we defined two benchmark simulations designed to bridge the gap between fully compressible and anelastic systems.  This allowed us to compare CHORUS results to the well-established ASH code which employs the anelastic approximation.  The two benchmark cases were formulated to simulate convection in the Jupiter and Sun.
Metrics of physical quantities sensitive to the convective driving and structure such as the linear growth rate and the total KE agree to lowest order in $\epsilon$, the stratification parameter upon which the validity of the anelastic approximation is based.  Mean flows exhibit larger variations (5-20\%) that may be attributed to flux imbalances at the sampling time.  Better agreement between the two codes could likely be achieved by running the simulations longer but full equilibration would require at least an order of magnitude more computing time.  We did not believe this computational expense was warranted given the good agreement at the sampling time.  The level of agreement is remarkable considering that the CHORUS and ASH codes not only solve different equations (compressible versus anelastic) but also they employ dramatically different numerical algorithms.  Thus, we consider the CHORUS code verified.

Future applications of CHORUS will focus on the applications discussed in \S1, namely the interaction of convection, differential rotation, and oblateness in rapidly-rotating stars, core convection in high and low-mass stars, hierarchical convection in solar-like stars, and the excitation of radial and non-radial acoustic pulsations within the context of asteroseismology.   We have already implemented a deformable grid algorithm which we are now using to model oblate stars.  The next step also include further development of CHORUS as a Large-Eddy Simulation (LES) code, with modeling of subgrid-scale (SGS) motions based either on explicit turbulence models or on the implicit numerical dissipation intrinsic to the SDM method.

\section{Acknowledgement}
Junfeng Wang is funded by an Newkirk graduate fellowship by the National Center for Atmospherical Center  (NCAR). Chunlei Liang would like to acknowledge the faculty start up grant from the George Washington University.  NCAR is sponsored by the National Science Foundation.

\newpage
\nocite{*}

\newpage
\begin{thebibliography}{10}
\expandafter\ifx\csname url\endcsname\relax
  \def\url#1{\texttt{#1}}\fi
\expandafter\ifx\csname urlprefix\endcsname\relax\def\urlprefix{URL }\fi
\expandafter\ifx\csname href\endcsname\relax
  \def\href#1#2{#2} \def\path#1{#1}\fi

\bibitem{linsk85}
J.~L. Linsky, Nonradiative activity across the h-r diagram: Which types of
  stars are solar-like?, Solar Phys. 100 (1985) 333--362.

\bibitem{hall08}
J.~C. Hall, Stellar chromospheric activity, Living Reviews in Solar Physics 5,
  http://www.livingreviews.org/lrsp-2008-2.

\bibitem{saar10}
S.~H. Saar, The activity cycles and surface differential rotation of single
  dwarfs, in: M.~Dikpati, T.~Arentoft, I.~G. Hernandez, C.~Lindsey, F.~Hill
  (Eds.), Solar-Stellar Dynamos as Revealed by Helio- and Asteroseismology:
  GONG 2008/SOHO 21, ASP Conference Series,Vol.\ 416, ASP, San Francisco, CA,
  2010, pp. 375--384.

\bibitem{sprui90}
H.~C. Spruit, A.~Nordlund, A.~M. Title, Solar convection, Annu. Rev. Astron.
  Astrophys. 28 (1990) 263--301.

\bibitem{miesc05}
M.~S. Miesch, Large-scale dynamics of the convection zone and tachocline,
  Living Reviews in Solar Physics 2, http://www.livingreviews.org/lrsp-2005-1.

\bibitem{skuma72}
A.~Skumanich, Time scales for ca ii emission decay, rotational braking, and
  lithium depletion, Astrophys.\ J. 171 (1972) 565--567.

\bibitem{charb93}
P.~Charbonneau, K.~B. MacGregor, Angular momentum transport in magnetized
  stellar radiative zones. ii. the solar spin-down, Astrophys.\ J. 417 (1993)
  762--780.

\bibitem{soder01}
D.~R. Soderblom, B.~F. Jones, D.~Fischer, Rotational studies of late-type
  stars. vii. m34 (ngc 1039) and the evolution of angular momentum and activity
  in young solar-type stars, Astrophys.\ J. 563 (2001) 334--340.

\bibitem{mcali05}
H.~A. McAlister, T.~A. {Ten Brummelaar}, D.~R. Gies, W.~Huang, W.~G. {Bagnuolo
  Jr.}, M.~A. Shure, J.~Sturmann, L.~Sturmann, N.~H. Turner, S.~F. Taylor,
  D.~H. Berger, E.~K. Baines, E.~Grundstrom, C.~Ogden, First results from the
  chara array. i. an interferometric and spectroscopic study of the fast
  rotator $\alpha$ leonis (regulus), Astrophys.\ J. 628 (2005) 439--452.

\bibitem{Jones-2011}
C.~A. Jones, P.~Boronski, A.~S. Brun, G.~A. Glatzmaier, T.~Gastine, M.~S.
  Miesch, J.~Wicht, Anelastic convection-driven dynamo benchmarks, Icarus 216
  (2011) 120--135.

\bibitem{Glatzmaier_1984}
G.~A. Glatzmaier, Numerical simulations of stellar convective dynamos. {I} -
  {T}he model and method, J. Comput. Phys. 55 (1984) 461--484.

\bibitem{Miesch-2000}
M.~S. Miesch, J.~R. Elliott, J.~Toomre, T.~C. Clune, G.~A. Glatzmaier, P.~A.
  Gilman, Three-dimensional spherical simulations of solar convection:
  {D}ifferential rotation and pattern evolution achieved with lamiar and
  turbulent states, Astrophys. J. 532 (2000) 593--615.

\bibitem{Kopriva-1996}
D.~A. Kopriva, A conservative staggered-grid chebyshev multidomain method for
  compressible flows. {II} {A} semi-structured method, J. of Comput. Phys. 128
  (1996) 254--265.

\bibitem{Liu-2006}
Y.~Liu, M.~Vinokur, Z.~J. Wang, Spectral difference method for unstructured
  grids.{I}:{B}asic formulation, J. of Comput. Phys. 216 (2006) 780--801.

\bibitem{ZJWang-2007}
Z.~J. Wang, Y.~Liu, G.~May, A.~Jameson, Spectral difference method for
  unstructured grids {II}:{E}xtension to the euler equations, J. of Sci.
  Comput. 32~(1) (2007) 45--71.

\bibitem{wang-sun06}
Z.~J. Wang, Y.~Sun, C.~Liang, Y.~Liu, Extension of the sd method to viscous
  flow on unstructured grids, in: H.~Deconinck, E.~Dick (Eds.), Computational
  Fluid Dynamics 2006, Springer Berlin Heidelberg, 2006, pp. 119--124.

\bibitem{Cockburn-1998}
B.~Cockburn, C.~W. Shu, The runge-kutta discontinuous galerkin method for
  conservation laws {V} - {M}ultidimensional systems, J. of Comput. Phys. 141
  (1998) 199--224.

\bibitem{May-2011}
G.~May, On the connection between the spectral difference method and the
  discontinuous galerkin method, Commun. Comput. Phys. 9 (2011) 1071--1080.

\bibitem{Liang-2009-3D}
C.~Liang, S.~Premasuthan, A.~Jameson, Z.~J. Wang, Large eddy simulation of
  compressible turbulent channel flow with spectral difference method, AIAA
  Paper AIAA-2009-402.

\bibitem{huynh-07}
H.~T. Huynh, A flux reconstruction approach to high-order schemes including
  discontinuous {Galerkin} methods, AIAA Paper AIAA-2007-4079.

\bibitem{Jamson-2010}
A.~Jameson, A proof of the stability of the spectral difference method for all
  orders of accuracy, J. of Sci. Comput. 45 (2010) 348--358.

\bibitem{GAMBIT}
GAMBIT, GAMBIT 2.2 User's Guide, Fluent Inc., Lebanon, New Hampshire, USA
  (2004).

\bibitem{Liang-2009}
C.~Liang, A.~Jameson, Z.~J. Wang, Spectral difference method for
  two-dimensional compressible flow on unstructured grids with mixed elements,
  J. of Comput. Phys. 228 (2009) 2847--2858.

\bibitem{Rusanov-1961}
V.~V. Rusanov, Calculation of interaction of non-steady shock waves with
  obstacles, J. of Comput. Math Phys. USSAR 1 (1961) 261--279.

\bibitem{Roe-1981}
P.~L. Roe, Approximate riemann solvers, parameter vectors and difference
  schemes, J. of Comput. Phys. 34 (1981) 357--372.

\bibitem{Bassi1997}
F.~Bassi, S.~Rebay, A high-order accurate discontinuous finite element method
  for the numerical solution of the compressible {Navier–Stokes} equations,
  Journal of Computational Physics 131 (1997) 267--279.

\bibitem{Bassi05}
F.~Bassi, S.~Rebay, Discontinuous galerkin solution of the reynolds-averaged
  navier-stokes and k-article turbulence model equations, Journal of
  Computational Physics 34 (2005) 507--540.

\bibitem{Karypis-1998}
G.~Karypis, V.~Kumar, A fast and high quality nultilevel scheme for
  partitioning graphs, SIAM J. on Sci. Comput. 20 (1998) 359--392.

\bibitem{chand61}
S.~Chandrasekhar, Hydrodynamic and Hydromagnetic Stability, Oxford University
  Press, Oxford, 1961.

\bibitem{Gilman_and_Glatzmaier}
P.~A. Gilman, G.~A. Glatzmaier, Compressible convection in a rotating spherical
  shell. {I} - {A}nelastic equations., Astrophys. J. Suppl. Ser. 45 (1981)
  335--349.

\bibitem{gough69}
D.~O. Gough, The anelastic approximation for thermal convection, J. Atmos. Sci.
  26 (1969) 448--456.

\bibitem{lantz99}
S.~R. Lantz, Y.~Fan, Anelastic magnetohydrodynamic equations for modeling solar
  and stellar convection zones, Astrophys.\ J. 121 (1999) 247--264.

\bibitem{Clune-1999}
T.~C. Clune, J.~R. Elliott, M.~S. Miesch, J.~Toomre, G.~A. Glatzmaier,
  Computational aspects of a code to study rotating turbulent covection in
  spherical shells, Parallel Computing 25 (1999) 361--380.

\bibitem{Brun-2004}
A.~S. Brun, M.~S. Miesch, J.~Toomre, Global-scale turbulent convection and
  magnetic dynamo action in the solar envelope, Astrophys. J. 614 (2004)
  1073--1098.

\bibitem{miesc07}
M.~S. Miesch, P.~A. Gilman, M.~Dikpati, Nonlinear evolution of global
  magneto-shear instabilities in a three-dimensional thin-shell model of the
  solar tachocline, Astrophys.\ J.\ Suppl.\ Ser. 168 (2007) 337--361.

\bibitem{brun10}
A.~S. Brun, Modeling the sun and stars in 3d, EAS Pub.\ Ser. 44 (2010) 81--95.

\bibitem{strug11}
A.~Strugarek, A.~S. Brun, J.-P. Zahn, Magnetic confinement of the solar
  tachocline: Ii. coupling to a convection zone, Astron.\ Astrophys. 532 (2011)
  A34 (15 pp).

\bibitem{hubba02}
W.~B. Hubbard, A.~Burrows, J.~I. Lunine, Theory of giant planets, Ann.\ Rev.\
  Astron.\ Astrophys. 40 (2002) 103--136.

\bibitem{spiteri-02}
R.~J. Spiteri, S.~J. Ruuth, A new class of optimal high-order
  strong-stability-preserving time-stepping schemes, SIAM Journal on Numerical
  Analysis 40 (2002) 469--491.

\bibitem{LiangKannan}
C.~Liang, R.~Kannan, Z.~J. Wang, A p-multigrid spectral difference method with
  explicit and implicit smoothers on unstructured triangular grids, Computers
  \& Fluids 38 (2009) 254--265.

\end{thebibliography}
\newpage
\listoftables
\newpage

\listoffigures

\end {document}